\documentclass{vldb}
\usepackage{amsmath}
\usepackage{amssymb}
\usepackage{comment}
\usepackage{times}
\usepackage{color}
\usepackage{tikz}
\usepackage{multicol} % Used in the appendix
\usepackage{epstopdf}

\usepackage{subcaption}

\usepackage{float}

\usepackage{enumitem}

\usepackage{epstopdf}

\usetikzlibrary{arrows}

\newtheorem{theorem}{Theorem}[section]
\newtheorem{definition}[theorem]{Definition}
\newtheorem{example}[theorem]{Example}
\newtheorem{proposition}[theorem]{Proposition}

\newcommand{\LIT}{\mathsf{L}}
\newcommand{\U}{\mathsf{U}}
\newcommand{\V}{\mathsf{V}}
\newcommand{\type}{\mathsf{type}}

\newcommand{\sig}{\operatorname{sig}}
\newcommand{\cnt}{\operatorname{count}}
\newcommand{\supp}{\operatorname{supp}}

\newcommand{\Cov}{\operatorname{Cov}}
\newcommand{\Sim}{\operatorname{Sim}}

\newcommand{\Dep}{\operatorname{Dep}}
\newcommand{\SymDep}{\operatorname{SymDep}}
\newcommand{\var}{\operatorname{var}}
\newcommand{\val}{\operatorname{val}}
\newcommand{\subj}{\operatorname{subj}}
\newcommand{\prop}{\operatorname{prop}}

\newcommand{\hash}{\operatorname{hash}}

\newcommand{\total}{\mbox{total}}
\newcommand{\dom}{\text{\rm dom}}

\newcommand{\ws}{\;\wedge\;}

\newcommand{\eits}{\text{\sc ExistsSortRefinement}}

\makeatletter
\renewcommand*\env@matrix[1][*\c@MaxMatrixCols c]{%
  \hskip -\arraycolsep
  \let\@ifnextchar\new@ifnextchar
  \array{#1}}
\makeatother

\begin{document}

\title{A Principled Approach to Bridging the Gap between Graph Data and their Schemas}

\numberofauthors{1}
\author{
\alignauthor
Marcelo Arenas$^{1,2}$,\hspace*{0.1in}Gonzalo D\'iaz$^{1}$,\hspace*{0.1in}Achille Fokoue$^{3}$,\\
\hspace*{0.1in}Anastasios Kementsietsidis$^{3}$,\hspace*{0.1in}Kavitha Srinivas$^{3}$\\
\vspace*{2.0ex}
\begin{tabular}{c c}
\affaddr{$^{1}$Pontificia Universidad Cat\'{o}lica de Chile} & \affaddr{$^{3}$IBM T.J. Watson Research Center}\\
\affaddr{$^{2}$University of Oxford} & \\
%\email{marenas@ing.puc.cl, gdiazc@uc.cl} &
%\email{\{achille, ksrinivs\}@us.ibm.com, tasosk@ca.ibm.com}
marenas@ing.puc.cl, gdiazc@uc.cl & \{achille, ksrinivs\}@us.ibm.com, tasosk@ca.ibm.com
\end{tabular}
}

\maketitle
\thispagestyle{empty}
\pagestyle{empty}

%================================================================
%================================================================

\begin{abstract}

Although RDF graph data often come with an associated schema, recent studies
have proven that real RDF data rarely conform to their perceived schemas.
Since a number of data management decisions, including storage layouts,
indexing, and efficient query processing, use schemas to guide the decision
making, it is imperative to have an accurate description of the
\textit{structuredness} of the data at hand (how well the data conform to the
schema).

In this paper, we have approached the study of the structuredness of an RDF
graph in a principled way: we propose a framework for specifying structuredness
functions, which gauge the degree to which an RDF graph conforms to a schema.
In particular, we first define a formal language for specifying structuredness
functions with expressions we call rules. This language allows a user to state
a rule to which an RDF graph may fully or partially conform. Then we consider
the issue of discovering a refinement of a sort (type) by partitioning the
dataset into subsets whose structuredness is over a specified threshold. In
particular, we prove that the natural decision problem associated to this
refinement problem is NP-complete, and we provide a natural translation of this
problem into Integer Linear Programming (ILP).  Finally, we test this ILP
solution with three real world datasets and three different and intuitive rules,
which gauge the structuredness in different ways. We show that the rules give
meaningful refinements of the datasets, showing that our language can be a
powerful tool for understanding the structure of RDF data, and we show that the
ILP solution is practical for a large fraction of existing data.
\end{abstract}

%================================================================
%================================================================

\section{Introduction}

If there is one thing that is clear from analyzing real RDF data, it is that
the data rarely conform to their assumed schema~\cite{DKSU11}.  One example is
the popular type of DBpedia persons (in this paper, we will use the term
\emph{sort} as a synonym of \emph{type}), which includes all people with an entry in Wikipedia. According to the sort definition, each person in DBpedia
can have 8 properties, namely, a \textit{name}, a \textit{givenName}, a
\textit{surName}, a \textit{birthDate}, a \textit{birthPlace}, a
\textit{deathDate}, a \textit{deathPlace}, and a \textit{description}. There
are 790,703 people listed in DBpedia, and while we expect that a
large portion of them are alive (they do not have a death date or death place)
we do expect that we know at least when and where these people were born. The
statistics however are very revealing: Only 420,242 people have a birthdate, only 323,368 have a birthplace, and for only 241,156 do we have both. There are $\sim$40,000 people for whom we do not even know their
last name. As for death places and death dates, we only know those for 90,246 and 173,507 people, respectively.

There is actually nothing wrong with the DBpedia person data. The data reflect
the simple fact that the information we have about any domain of discourse (in
this case people) is inherently incomplete. But while this is the nature of
things in practice, sorts in general go against this trend since they
favor uniformity, i.e., they require that the data \textit{tightly} conform to
the provided sorts. In our example, this means that we expect to have all 8
properties for every DBpedia person. So the question one needs to address
is how to bridge the gap between these two worlds, the sorts and the respective
data.  In our previous work~\cite{DKSU11}, we considered sorts as being
the unequivocal ground truth, and we devised methods to make the data 
fit these sorts.  Here, we consider a complementary
approach in which we accept the data for what they are and ask ourselves
whether \textit{we can refine the schema to better fit our data} (more precisely, we will seek a \textit{sort refinement} of our data).

Many challenges need to be addressed to achieve our goal. First, we need to
define formally what it means for a dataset to fit a particular sort.
Our own past work has only introduced one such fitness metric, called
\textit{coherence}, but that does not clearly cover every possible
interpretation of fitness. In this work, we propose a new set of alternative
and complementary fitness metrics between a dataset and a sort, and we
also introduce a rule language through which users can define their own metrics.

Second, for a given RDF graph $D$ and a fitness metric $\sigma$,  we study
the problem of determining whether there exists a sort refinement $\mathcal{T}$  of
$D$ with a fitness value above a given threshold $\theta$ that contain at most
$k$ implicit sorts, and we show that the problem is NP-complete. In spite of this negative result,
we present several techniques enabling us to solve this problem
in practice on real datasets as illustrated in our experimental evaluation
section.  Our first attack on the problem is to reduce the size of the input we
have to work with. Given that typical real graph datasets involve millions of
instances, even for a single sort, scalability is definitely a concern. We
address this challenge by introducing views of our input
data that still maintain all the properties of the data in terms of their
fitness characteristics, yet they occupy substantially less space. Using said
view, given any fitness metric expressed as a
rule $r$ in our language, we formulate the previously defined problem as an
Integer Linear Programming (ILP) problem instance.  Although ILP is also known
to be NP-hard in the worst case, in practice, highly optimized commercial
solvers (e.g. IBM ILOG CPLEX) exist to efficiently solve our formulation of the
sort refinement problem (see experimental evaluation for more details). In
particular, we study two complementary formulations of  our problem:  In the
first alternative, we allow the user to specify a desired fitting value
$\theta'$, and we compute a smallest set of implicit sorts, expressed as a partition $\{D_1, D_2, \dots, D_n\}$ of the input dataset $D$, such that the fitness of each $D_i$ is larger
than or equal to $\theta'$. In the second alternative, we allow the user to
specify the desired number $k$ of implicit sorts, and we compute a set of $k$
implicit sorts such that the minimum fitness across all implicit sorts is maximal amongst
all possible decompositions of the sort that involve $k$ implicit sorts. Both
our alternatives are motivated by practical scenarios. In the former
alternative, we allow a user to define a desirable fitness and we try to compute 
a sort refinement with the indicated fitness. However, in other settings,
the user might want to specify a maximum number of sorts to which the data
should be decomposed and let the system figure out the best possible
sort and data decomposition.

A clear indication of the practical value of this work can be found in
the experimental section, where we use different rules over real datasets and
not only provide useful insights about the data themselves, but also
automatically discover sort refinements that, in hindsight, seem natural,
intuitive and easy to understand. We explore the correlations between
alternative rules (and sort refinements) over the same data and show that the
use of \textit{multiple} such rules is important to fully understand the nature
of data. Finally, we study the scalability of the ILP-based solution on a sample
of explicit sorts extracted from the knowledge base YAGO, showing that it is practical
for all but a small minority of sorts.

Our key contributions in this paper are fivefold: (1) we propose a framework
for specifying structuredness functions to measure the degree to which an RDF
graph conforms to a schema; (2) we study the problem of discovering a
refinement of a sort by partitioning the dataset into subsets whose
structuredness is greater than a given threshold and show that the decision
problem associated with this sort refinement problem is NP-complete; (3) we
provide a natural translation of an instance of the sort refinement problem
into an ILP problem instance; (4) we successfully test our ILP approach on two
real world datasets and three different structuredness functions; (5) we study
the scalability of our solution on a sample of sorts extracted from the knowledge
base YAGO, showing that the solution is practical in a large majority of cases.

The remainder of the paper is organized as follows. Section 2 presents a brief
introduction to RDF and sample structuredness functions, while the syntax and
the formal semantics of the language for specifying structuredness functions is
defined in Section 3. Section 4 introduces the key concepts of signatures and
sort refinements. Section 5 presents  the main complexity result of the sort
refinement problem, while Section 6 describes the formulation of the problem as
an ILP problem. In Section 7, we present our experimental evaluation on three
real world datasets. Finally, we review related work in Section 8, and conclude
in Section 9.

An extended version of this paper (with proofs of the theoretical
results) is available at {\small \texttt{http://arxiv.org/abs/1308.5703}}.

%================================================================
%================================================================

\section{Preliminaries}

\subsection{A schema-oriented graph representation}

We assume two countably infinite disjoint sets $\U$ and $\LIT$ of URIs,
Literals, respectively. An RDF triple is a tuple $(s,p,o) \in \U \times \U
\times (\U \cup \LIT)$, and an RDF graph is a finite set of RDF triples. Given
an RDF graph $D$, we define the sets of subjects and properties mentioned in
$D$, respectively denoted by $S(D)$ and $P(D)$, as:

\vspace*{-3.0ex}

\begin{eqnarray*}
S(D) & = & \{ s \in \U \;\mid\; \exists p \exists o \text{ s.t. } (s, p, o) \in D\},\\
P(D) & = & \{ p \in \U \;\mid\; \exists s \exists o \text{ s.t. } (s, p, o) \in D\}.
\end{eqnarray*}

Given an RDF graph $D$  and $s, p \in \U$, we say that $s$ {\em has property} $p$ in $D$ if there exists $o \in \U$ such that $(s,p,o) \in D$.

A natural way of storing RDF data in a relational table, known as the
horizontal database \cite{Pan04}, consists in defining only one relational
table in which each row represents a subject and there is a column for every
property. With this in mind, given an RDF graph $D$, we define an $|S(D)|
\times |P(D)|$ matrix $M(D)$ (or just $M$ if $D$ is clear from the context) as
follows: for every $s \in S(D)$ and $p \in P(D)$,

\vspace*{-3.0ex}

\begin{equation*}
M(D)_{sp} = 
\begin{cases}
  1 &\mbox{if } s \text{ has property } p \text{ in } D\\
  0 &\mbox{otherwise}.
\end{cases}
\end{equation*}
It is important to notice that $M(D)$ corresponds to a view of the RDF graph $D$ in which we have discarded a large amount of information, and only retained information about the structure of the properties in $D$. Thus, in what follows we refer to $M(D)$ as the \emph{property-structure view of $D$.}

In an RDF graph, to indicate that a subject $s$ is of a specific sort $t$ (like
person or country), the following triple must be present: $(s,\type,t)$, where
the constant $\type = $ \texttt{http://www.w3.org/
1999/02/22-rdf-syntax-ns\#type} (note that $\type \in \U$).

Given a URI $t$, we define the following RDF subgraph $D_t \subseteq D$: $D_t =
\{ (s,p,o) \in D \mid (s, \type, t) \in D \}$. This subgraph consists of all 
triples whose subject $s$ is explicitly declared to be of sort $t$ in $D$. With
this subgraph $D_t$, we can mention its set of subjects, $S(D_t)$, which is
also the set of subjects of sort $t$ in $D$, and its set of properties
$P(D_t)$, which is the set of properties set by some subject of sort $t$. We
will use the term \emph{sort} to refer to the constant $t$, the RDF
subgraph $D_t$, and sometimes the set $S(D_t)$.

\subsection{Sample structuredness functions}
\label{sec-sf}

As there are many alternative ways to define the fitness, or
\textit{structuredness}, of a dataset with respect to a schema, it is
convenient to define structuredness initially in the most general way. As such,
we define a \emph{structuredness function} $\sigma$ to be any function which
assigns to every RDF graph $D$ a rational number $\sigma(D)$, such that $0 \leq
\sigma(D) \leq 1$.  Within the context of our framework a structuredness
function will only produce rational numbers. In what follows, we offer concrete
examples of structuredness functions which gauge the structuredness of RDF
graphs in very different ways.

%\begin{definition}\label{def-s-fn}
%{\em A \emph{structuredness function} $\sigma$ is a function which assigns to
%every RDF graph $D$ a rational number $\sigma(D)$, such that $0 \leq \sigma(D)
%\leq 1$. \hfill$\qed$ }
%\end{definition}

\subsubsection{The coverage function}

Duan et.~al.~defined the \textsc{Coverage} function~\cite{DKSU11}
$\sigma_{\Cov}$ to test the fitness of graph data to their respective schemas.
The metric was used to illustrate that though graph benchmark data are very
relational-like and have high fitness (values of $\sigma_{\Cov}(D) $ close to
1)  with respect to their sort, real graph data are fairly unstructured
and have low fitness ($\sigma_{\Cov}(D) $ less than 0.5). Using the property-structure
view $M(D)$,
%of the graph data introduced in previously, 
the coverage
metric of~\cite{DKSU11} can be defined as: $\sigma_{\Cov}(D) =
(\sum_{sp} M(D)_{sp}) / |S(D)||P(D)|$. Intuitively, the metric favors
\textit{conformity}, i.e., if one subject has a property $p$, then the other
subjects of the same sort are expected to also have this property. Therefore,
the metric is not forgiving when it comes to \textit{missing} properties. To
illustrate, consider an RDF graph $D_1$ consisting of $N$ triples: $(s_i, p,
o)$ for $i = 1, \ldots, N$ (i.e.~all $N$ subjects have the same property $p$).
The matrix $M(D_1)$ for $D_1$ is shown in Figure~\ref{fig:sampleM}a. For this
dataset, $\sigma_{\Cov}(D_1) = 1$. If we insert a new triple $(s_1, q,
o)$ for some property $q \neq p$, resulting dataset $D_2 = D_1 \cup \{ (s_1, q,
o) \}$ whose matrix is shown in Figure~\ref{fig:sampleM}b.  Then, the
structuredness of $\sigma_{\Cov}(D_2) \approx 0.5$ (for a large value of $N$). The addition of the single triple generates a new dataset $D_2$ in
which most of the existing subjects are \textit{missing} property $q$, an
indication of unstructureness.

\subsubsection{The similarity function}

The previous behavior motivates the introduction of a structuredness function
that is less sensitive to missing properties. We define the $\sigma_{\Sim}$
structuredness function as the probability that, given two randomly selected
subjects $s$ and $s'$ and a random property $p$ such that $s$ has property $p$
in $D$, $s'$ also has property $p$ in $D$.

To define the function formally, let $\varphi_1^{\Sim}(s,s',p)$ denote the statement ``$s \neq s'$ and $s \text{ has property } p \text{ in } D$'' and let $\varphi_2^{\Sim}(s',p)$ denote ``$s' \text{ has property } p \text{ in } D$''. Next, we define a set of total cases $\total(\varphi_1^{\Sim},D) = \{ (s,s',p) \in S(D) \times S(D) \times P(D) \mid \varphi_1^{\Sim}(s,s',p) \text{ holds} \}$, and a set of favorable cases $\total(\varphi_1^{\Sim} \wedge \varphi_2^{\Sim}, D) = \{ (s,s',p) \in S(D) \times S(D) \times P(D) \mid \varphi_1^{\Sim}(s,s',p) \wedge \varphi_2^{\Sim}(s',p) \text{ holds} \}$. Finally, define:
\begin{equation*}
\sigma_{\Sim}(D) = \frac{ |\total(\varphi_1^{\Sim} \wedge \varphi_2^{\Sim},D)| }{ |\total(\varphi_1^{\Sim}, D)| }.
\end{equation*}
Going back to the example in Figure~\ref{fig:sampleM}, notice that $\sigma_{\Sim}(D_1) = 1$ but also
$\sigma_{\Sim}(D_2)$ is still approx. equal to 1 (for large N).  Unlike
$\sigma_{\Cov}$, function $\sigma_{\Sim}$ allows certain subjects to have
\textit{exotic} properties that either no other subject has, or only a small
fraction of other subjects have (while maintaining high values for
$\sigma_{\Sim}$). As another example, consider the RDF graph $D_3$ in
Figure~\ref{fig:sampleM}c where every subject $s_i$ has only one property $p_i$,
and no two subjects have the same property. This dataset is intuitively very
unstructured.  Indeed, $\sigma_{\Sim}(D_3) = 0$ while $\sigma_{\Cov}(D_3)
\approx 0$ (for a large value of~N).

\begin{figure}[t]
\vspace*{-1.5ex}
\hspace*{-1.5ex}
{\small
\begin{minipage}{3.5in}
\begin{minipage}{0.9in}
\begin{equation*}
\bordermatrix{
 ~   & p \cr
 s_1 & 1 \cr
 s_2 & 1 \cr
 \vdots & \vdots \cr
 s_N & 1 \cr}
\end{equation*}
\centerline{(a) $M(D_1)$}
\end{minipage}
\begin{minipage}{1.0in}
\begin{equation*}
\bordermatrix{
 ~   & p & q \cr
 s_1 & 1 & 1 \cr
 s_2 & 1 & 0 \cr
 \vdots & \vdots & \vdots \cr
 s_N & 1 & 0 \cr}
\end{equation*}
\centerline{(b) $M(D_2)$}
\end{minipage}
\begin{minipage}{1.3in}
\begin{equation*}
\bordermatrix{
  ~   & p_1 & p_2 & \cdots & p_{N} \cr
  s_1 & 1   & 0   & \cdots & 0 \cr
  s_2 & 0   & 1   & \cdots & 0 \cr
  \vdots & \vdots & \vdots & \ddots & \vdots \cr
  s_N & 0 & 0 & \cdots & 1 \cr}
\end{equation*}
\centerline{(c) $M(D_3)$}
\end{minipage}
\end{minipage}
}
\vspace*{-1.0ex}
\caption{Sample matrixes for datasets $D_1$, $D_2$ and $D_3$}
\vspace*{-1.5ex}
\label{fig:sampleM}
\end{figure}

\subsubsection{The dependency functions}

It is also of interest to understand the correlation between different properties in an RDF graph $D$. Let $\mathbf{p}_1, \mathbf{p}_2 \in P(D)$ be two fixed properties we are interested in. Define the $\sigma_{\Dep}$[$\mathbf{p}_1$, $\mathbf{p}_2$] function as the probability that, given a random subject $s\in S(D)$ such that $s$ has $\mathbf{p}_1$, $s$ also has $\mathbf{p}_2$.
In the same way as before, we can define a set of total cases and a set of favorable cases, and we define the value of $\sigma_{\Dep}$[$\mathbf{p}_1$, $\mathbf{p}_2$] to be the ratio of the sizes of both sets.

A closely related structuredness function is the symmetric version of $\sigma_{\Dep}$[$\mathbf{p}_1$, $\mathbf{p}_2$], which we call $\sigma_{\SymDep}$[$\mathbf{p}_1$, $\mathbf{p}_2$]. It is defined as the probability that, given a random subject $s\in S(D)$ such that $s$ has $p_1$ or $s$ has $p_2$, $s$ has both.

%================================================================
%================================================================

%\input{sec_structuredness_functions.tex}

%================================================================
%================================================================

\section{A language for defining structuredness measures}

We have already shown in Section \ref{sec-sf} some intuitive structuredness
measures that give very different results when applied to the same RDF graphs.
As many more natural structuredness functions exist, we do not intend to list
all of them in this article, but instead our goal is to introduce a general
framework to allow users to define their own custom structuredness measures in
a simple way. To this end, we introduce in this section a language for
describing such measures.  This language has a simple syntax and a formal
semantics, which make it appropriate for a formal study, and it is expressive
enough to represent many natural structuredness functions, like the ones
presented in Section \ref{sec-sf}. In general, starting from the matrix $M(D)$
of a dataset $D$, our language can construct statements that involve (i) the
contents of the matrix (the cells of the matrix with 0 or 1 values); (ii) the
\textit{indices} of the matrix, that correspond to the subjects and properties
of the dataset; and (iii) Boolean combinations of these basic building components.
% in the form of conjunctions, disjunctions and negations.

\subsection{Syntax of the language}

To define the syntax of the language, we need to introduce some terminology.
From now on, assume that $\V$ is an infinite set of variables disjoint from
$\U$. We usually use $c$, $c_1$, $c_2$, $\ldots$ to denote the variables in
$\V$, as each one of these variables is used as a pointer to a {\em cell} (or
position) in a matrix associated with an RDF graph. Moreover, assume that $0,
1$ do not belong to $(\U \cup \V)$. Then the set of terms in the language is
defined as follows: (i) $0$, $1$, every $u \in \U$ and every $c \in \V$ is a term, and (ii) if $c \in \V$, then $\val(c)$, $\subj(c)$ and $\prop(c)$ are terms.

%\begin{tabbing}
%X \= $\bullet$ \= \kill
%\> $\bullet$ \> $0$, $1$, every $u \in \U$ and every $c \in \V$ is a term, and\\
%\> $\bullet$ \> if $c \in \V$, then $\val(c)$, $\subj(c)$ and $\prop(c)$ are terms.
%\end{tabbing}

If $c$ is a variable pointing to a particular cell in a matrix, then $\val(c)$
represents the value of the cell, which must be either 0 or 1, $\subj(c)$
denotes the row of the cell, which must be the subject of a triple in $D$, and
$\prop(c)$ denotes the column of the cell, which must be the property of a
triple in $D$. Moreover, the set of formulas in the language is recursively
defined as follows:

\begin{tabbing}
X \= $\bullet$ \= \kill
\> $\bullet$ \> \parbox[t]{3.0in}{If $c \in \V$ and $u \in \U$, then $\val(c) = 0$, $\val(c) = 1$, $\prop(c) = u$ and $\subj(c) = u$ are formulas.}\\
\> $\bullet$ \> \parbox[t]{3.0in}{If $c_1,c_2 \in \V$, then $c_1 = c_2$, $\val(c_1) = \val(c_2)$, $\prop(c_1) = \prop(c_2)$ and $\subj(c_1) = \subj(c_2)$ are formulas.}\\
\> $\bullet$ \> \parbox[t]{3.0in}{If $\varphi_1$ and $\varphi_2$ are formulas, then $(\neg \varphi_1)$, $(\varphi_1 \wedge \varphi_2)$, $(\varphi_1 \vee \varphi_2)$ are formulas.}
\end{tabbing}

If $\varphi$ is a formula, then $\var(\varphi)$ is the set consisting of all
the variables mentioned in $\varphi$. With this notation, we can finally define
the syntax of the rules in the language, which are used to define
structuredness functions. Formally, if $\varphi_1, \varphi_2$ are formulas such
that $\var(\varphi_2) \subseteq \var(\varphi_1)$, then the following is a rule:
\begin{eqnarray}\label{eq-rule}
	\varphi_1 & \mapsto & \varphi_2.
\end{eqnarray}

\subsection{Semantics of the language}

To define how rules of the form \eqref{eq-rule} are evaluated, we need to
define the notion of satisfaction of a formula. In the rest of this section,
assume that $D$ is an RDF graph and $M$ is the $|S(D)| \times |P(D)|$ matrix
associated with $D$. A partial function $\rho: \V \rightarrow S(D) \times P(D)$
is said to be a variable assignment for $M$, whose domain is denoted by
$\dom(\rho)$. Moreover, given a formula $\varphi$ and a variable assignment
$\rho$ for $M$ such that $\var(\varphi) \subseteq \dom(\rho)$, pair $(M, \rho)$
is said to satisfy $\varphi$, denoted by $(M,\rho) \models \varphi$, if:

\begin{tabbing}
X \= $\bullet$ \= \kill
\> $\bullet$ \> \parbox[t]{3.0in}{$\varphi$ is the formula $\val(c) = i$, where $i = 0$ or  $i = 1$, $\rho(c) = (s,p)$ and $M_{sp} = i$.}\\
\> $\bullet$ \> $\varphi$ is the formula $\subj(c) = u$, where $u \in \U$, and $\rho(c) = (u, p)$.\\
\> $\bullet$ \> $\varphi$ is the formula $\prop(c) = u$, where $u \in \U$, and $\rho(c) = (s, u)$.\\
\> $\bullet$ \> $\varphi$ is the formula $c_1 = c_2$, and $\rho(c_1) = \rho(c_2)$.\\
\> $\bullet$ \> \parbox[t]{3.0in}{$\varphi$ is the formula $\val(c_1) =
\val(c_2)$,  $\rho(c_1) = (s_1, p_1)$, $\rho(c_2) = (s_2, p_2)$ and $M_{s_1
p_1} = M_{s_2 p_2}$.}\\
\> $\bullet$ \> \parbox[t]{3.0in}{$\varphi$ is the formula $\subj(c_1) =
\subj(c_2)$, $\rho(c_1) = (s_1, p_1)$, $\rho(c_2) = (s_2, p_2)$ and $s_1 =
s_2$.}\\
\> $\bullet$ \> \parbox[t]{3.0in}{$\varphi$ is the formula $\prop(c_1) =
\prop(c_2)$, $\rho(c_1) = (s_1, p_1)$, $\rho(c_2) = (s_2, p_2)$ and $p_1 =
p_2$.}\\
\> $\bullet$ \> $\varphi$ is the formula $(\neg \varphi_1)$ and $(M, \rho) \models \varphi_1$ does not hold.\\
\> $\bullet$ \> $\varphi$ is the formula $(\varphi_1 \wedge \varphi_2)$, $(M, \rho) \models \varphi_1$ and $(M, \rho) \models \varphi_2$.\\
\> $\bullet$ \> $\varphi$ is the formula $(\varphi_1 \vee \varphi_2)$, and $(M, \rho) \models \varphi_1$ or~$(M, \rho) \models \varphi_2$. %(or both).
\end{tabbing}

Moreover, the set of satisfying assignments for a formula $\varphi$ w.r.t. $M$, denoted by $\total(\varphi,M)$, is defined as follows:
\begin{multline*}
\big\{ \rho \mid \rho \text{ is a variable assignment for } M \text{ such that }\\ \dom(\rho) = \var(\varphi) \text{ and } (M,\rho) \models \varphi \big\}.
\end{multline*}

We now have the necessary ingredients to define the semantics of rules. Assume that $r$ is the rule \eqref{eq-rule}. Then the structuredness function given by rule $r$ is defined as a function $\sigma_r$ that assigns to every matrix $M$ the value
\begin{eqnarray*}
  \sigma_r(M) & = &  \frac{ |\total(\varphi_1 \wedge \varphi_2, M)| }{ |\total(\varphi_1, M)| }
\end{eqnarray*}

\noindent if $ |\total(\varphi_1, M)| > 0$, and 1 otherwise (notice that $0 \leq \sigma_r(M) \leq 1$, as we assume that $\var(\varphi_2) \subseteq \var(\varphi_1)$). Thus, $\sigma_r(M)$ is defined as the probability that a variable assignment $\rho$ satisfies $\varphi_2$ given that $\rho$ satisfies $\varphi_1$.

The functions presented in Section~\ref{sec-sf} can be expressed in our language as follows. The $\sigma_{\Cov}$ structuredness measure can be expressed with the rule $c = c  \mapsto \val(c) = 1$. In this case, given a matrix $M$, $\total(c=c,M)$ is the set of all cells of $M$ and $\total(c=c \wedge \val(c) =1, M)$ is the set of all cells of $M$ containing a value 1 (which is represented by the condition $\val(c) = 1$).

In some cases, it is desirable to compute a structuredness functions without considering some predicate (or set of predicates), which can be easily done in our language. For instance, a modified $\sigma_{\Cov}$ structuredness measure which ignores a specific column called $\mathbf{p}$ is defined  by the following rule:
\begin{eqnarray*}
c = c \wedge \neg(\prop(c) = \mathbf{p})  & \mapsto & \val(c) = 1.
\end{eqnarray*}
The $\sigma_{\Sim}$ structuredness measure can be expressed with the rule
\begin{multline*}
	\neg(c_1 = c_2) \wedge \prop(c_1) = \prop(c_2) \wedge \val(c_1) = 1 \ \mapsto \\ \val(c_2) = 1,
\end{multline*}
where $\neg(c_1 = c_2)$ considers two variables $c_1$ and $c_2$ that should
point to different cells, and $ \prop(c_1) = \prop(c_2)$ requires that the
two variables range over the same property column, say property $p$.
Taken together, the first two formulas iterate over all pairs of subjects for
each property $p$.  The last part of the formula $\val(c_1) = 1$ requires that
the value of the first cell be 1, i.e., the first subject actually has
property $p$. If the consequence formula is satisfied, then the rule
considers the cases where the value of the second cell is also 1, which translates to the
second subject also having property $p$. Notice that this is exactly the
definition of the function $\sigma_{\Sim}$.

Finally, for fixed $\mathbf{p}_1, \mathbf{p}_2 \in \U$, 
%we can also express 
the dependency measures. $\sigma_{\Dep}$[$\mathbf{p}_1$, $\mathbf{p}_2$] can be expressed with the rule
\begin{multline*}
  \subj(c_1) = \subj(c_2) \wedge \prop(c_1)=\mathbf{p}_1 \wedge \prop(c_2)=\mathbf{p}_2 \\
\wedge \val(c_1) = 1 \mapsto \val(c_2) = 1,
\end{multline*}

\noindent while $\sigma_{\SymDep}$[$\mathbf{p}_1$, $\mathbf{p}_2$] can be expressed with the rule
\begin{multline*}
  \subj(c_1) = \subj(c_2) \wedge \prop(c_1)=\mathbf{p}_1 \wedge \prop(c_2)=\mathbf{p}_2 \\
\wedge (\val(c_1) = 1 \vee \val(c_2) = 1) \\ \mapsto \val(c_1) = 1 \wedge \val(c_2) = 1.
\end{multline*}

A variant of the dependency rule uses disjunction in the consequent and corresponds to the probability that a random subject $s$ satisfies that: if $s$ has $\mathbf{p}_1$, then $s$ also has $\mathbf{p}_2$:
\begin{multline*}
  \subj(c_1) = \subj(c_2) \wedge \prop(c_1)=\mathbf{p}_1 \wedge \prop(c_2)=\mathbf{p}_2 \\
\mapsto \val(c_1) = 0 \vee \val(c_2) = 1.
\end{multline*}

%================================================================
%================================================================

\section{Sort refinements and signatures}

We can use the language from the previous section to
define a structuredness measure for a dataset. If the value of the measure for
the dataset is high, say 0.9 or even 1.0, then this is probably a positive
indication for the current state of the data, and the measure computation can
be repeated at a later stage, as the data change. Of most interest, however, is
what happens if this value is relatively low, say, 0.5 or even 0.1? Then, we
know that the dataset does not have the desired characteristic, as expressed
by the measure, and the question is whether there is anything we can do about
it. In particular, it is interesting to investigate if there is a way to convert
the existing dataset into one whose measure is high.

In our previous work, we tried to change the data themselves to fit the
measure, by introducing new triples or removing existing ones. The approach
made sense in the context of benchmarking for which it was introduced, but in
any practical setting one does not want to contaminate their data with dummy
triples, or even worse lose real data by deleting triples just so that the data
fit some desired measure. So a more pragmatic solution is to leave data as they
are and try to figure out whether we can refine the sort that the data
is supposed to fit, in an effort to improve structuredness.

To this end, we consider the situation in which one wishes to partition the
dataset into $k$ implicit sorts such that each implicit sort has a high structuredness (as
defined by a rule in our language).  For a certain subject $s_0 \in S(D)$ we
are interested in keeping all triples of the form $(s_0,p,o)$ (for some $p, o
\in \U$) together. We refer to these triples collectively as the \emph{entity}
$s_0$.

We define an \emph{entity preserving} partition of size $k$ of an RDF graph $D$ to be a set of non-empty RDF graphs $\{ D_1, \ldots, D_k \}$ where (i) $D_i \subseteq D$ for every $i \in \{1, \ldots, k\}$, (ii) $D_i \cap D_j = \emptyset$ for every $i,j \in \{1, \ldots, k\}$ such that $i \neq j$, (iii) $\bigcup_{i=1}^k D_i = D$, and (iv) for all $s, p_1, p_2, o_1, o_2 \in \U$, we have that:
%
%\begin{equation*}
\begin{center}
if $(s,p_1,o_1) \in D_i$ and $(s,p_2,o_2) \in D_j$, then $i = j$.
\end{center}
%\end{equation*}
%
While the first three items specify a partition of $D$, the last item indicates
that for every entity $s$, we include the full entity in a sort. 

A second consideration we shall make is concerned with the grouping of subjects
which have the same properties in $D$. For this, we define the concept of
signature:

\begin{definition}\label{def-sig}
{\em Given an RDF graph $D$ and a subject $s \in S(D)$, the signature of $s$ in $D$ is a function $\sig(s,D): P(D) \rightarrow \{ 0, 1 \}$, which assigns to every property $p \in P(D)$ a 1 if $s$ has property $p$ in $D$, and a 0 otherwise.
\hfill $\qed$
}
\end{definition}

Similarly, a signature set is the set of all subjects in $S(D)$ which share the same signature. The \emph{size} of a signature set is the number of subjects sharing that signature.

We are now ready to define our main objects of study. For the following definition, let $D$ be a fixed RDF graph and $\theta$ be a rational number such that $0 \leq \theta \leq 1$ ($\theta$ is required to be a rational number for compatibility with the reduction to the Integer Linear Programming instance).

\begin{definition}
{\em Given a structuredness function $\sigma$, a $\sigma$-\emph{sort refinement $\mathcal{T}$ of $D$ with threshold $\theta$} is an entity preserving partition $\{ D_1, \ldots, D_n \}$ of $D$ such that:
\begin{itemize}
\item[i)] $\sigma(D_i) \geq \theta$ for $i = 1, \ldots, n$, and
\item[ii)] each $D_i$ ($1 \leq i \leq n$) is closed under signatures. That is, for every pair of subjects $s_1, s_2 \in S(D)$, if $\sig(s_1,D) = \sig(s_2,D)$ and $s_1 \in S(D_i)$, then $s_2 \in S(D_i)$. \hfill $\qed$
\end{itemize}
}
\end{definition}

In the rest of this paper, we will refer to the elements of the sort refinement (i.e.~the elements of the partition of $D$) as implicit sorts.

The requirement that each implicit sort be closed under signatures is due to the fact that two subjects with equal signatures are structurally identical, and thus it would not be reasonable to place them in different subsets. This has the added benefit of reducing the difficulty of partitioning the dataset, as the basic units to be moved will be signatures sets, and not individual entities.

In what follows we will be concerned with discovering sort refinements in RDF data.
%We use the term signature in two ways: (i) to refer to the binary function described in Definition \ref{def-sig}, and (ii) to refer to the set of all entities in an RDF graph $D$ which share a common signature.

\begin{figure}[t]
\begin{minipage}{3.5in}
\begin{minipage}{1.6in}
\centering
\includegraphics[trim=6.8cm 11.6cm 7.2cm 4.5cm, clip=true, width=0.8\textwidth]{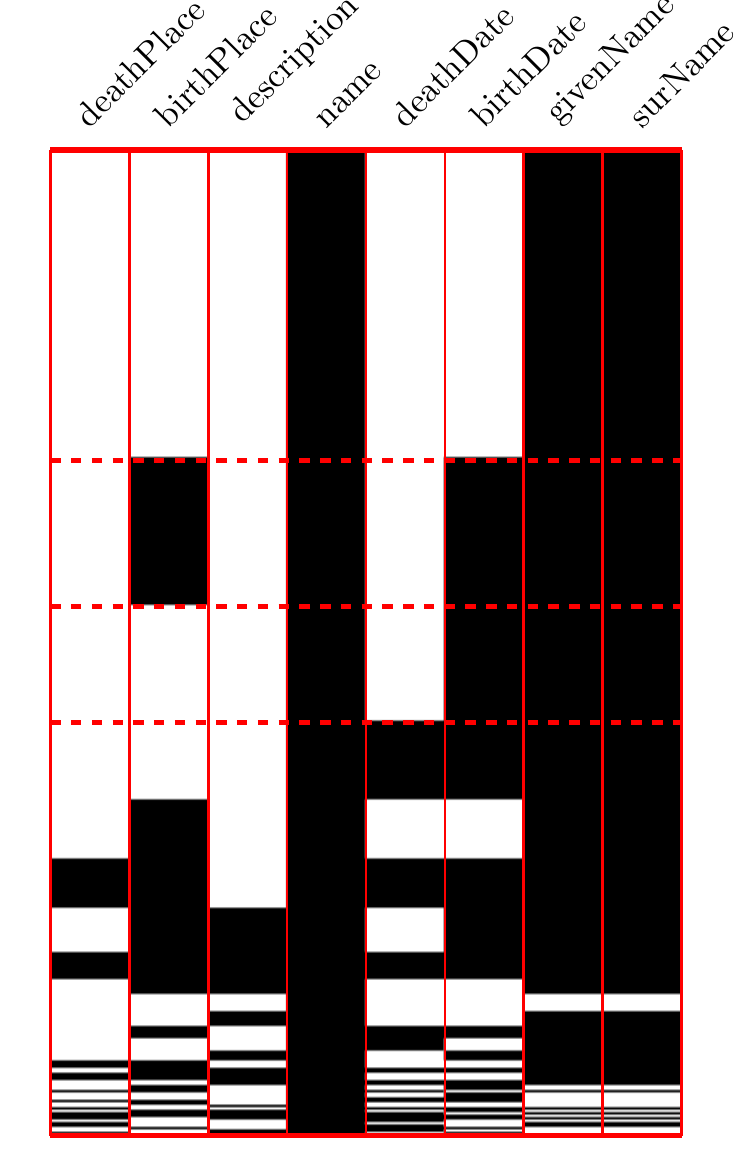}
\hspace*{-3.0ex}\parbox[t]{1.5in}{
\caption{DBPedia Persons has 790,703 subjects, 8 properties and 64 signature sets. Its $\sigma_{\Cov} = 0.54$, while $\sigma_{\Sim} = 0.77$.}
\label{fig:full_sigs_a}
}
\end{minipage}
\begin{minipage}{1.8in}
\vspace*{-3.0ex}
\centering
\includegraphics[trim=5.3cm 10.4cm 5.5cm 4.1cm, clip=true, width=0.9\textwidth]{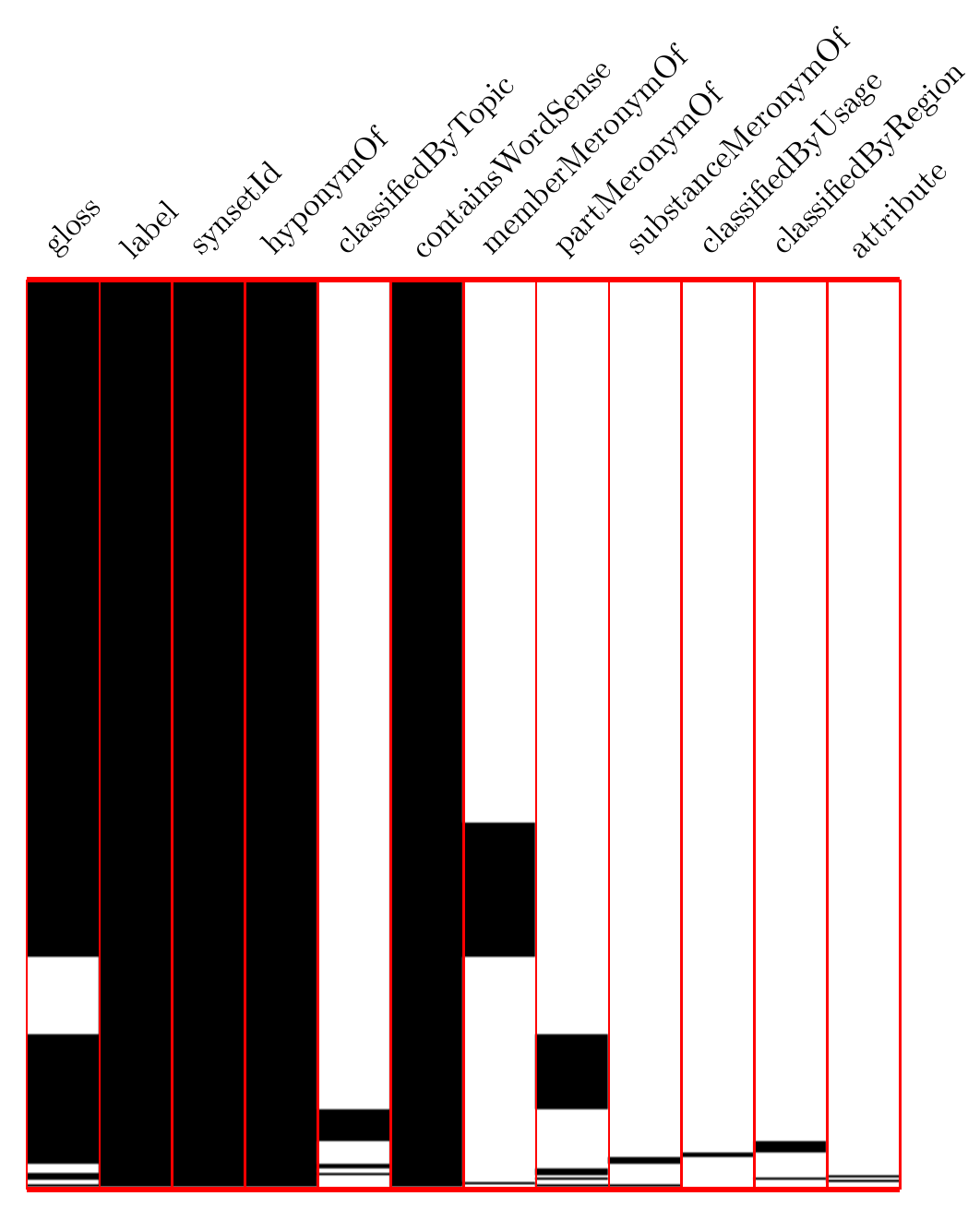}
\parbox[t]{1.7in}{
\caption{WordNet Nouns has 79,689 subjects, 12 properties and 53 signature sets. Its $\sigma_{\Cov} = 0.44$, while $\sigma_{\Sim} = 0.93$.}\label{fig:full_sigs_b}
}
\end{minipage}
\end{minipage}
\vspace*{-3.0ex}
\end{figure}

Figures~\ref{fig:full_sigs_a} and \ref{fig:full_sigs_b} present a visual representation of an
RDF graph's horizontal table. Every column represents a property and the rows
have been grouped into signature sets, in descending order of signature set size. The first 3 signature sets in Figure \ref{fig:full_sigs_a} have been delimited with a dashed line, for clarity. The subsequent signature sets can be visually separated by searching for the change in pattern. The
black zones represent data (i.e.~non-\texttt{null} values) whereas the white
regions represent \texttt{null} cells. The difference between DBpedia Persons
(Fig.~\ref{fig:full_sigs_a}) and WordNet Nouns (Fig.~\ref{fig:full_sigs_b}) is immediately visible. DBpedia Persons is a relatively unstructured dataset, with only 3 clearly common properties: \texttt{name}, \texttt{givenName}, and \texttt{surName} (these three
attributes are usually extractable directly from the URL of a Wikipedia
article). On the other hand, WordNet Nouns has 5 clearly common properties, and
the rest of the properties are realtively rare (very few subjects have them).
The values of the structuredness functions show how they differ in judging the
structuredness of an RDF graph.

We shall use this visual representation of the horizontal table of an RDF graph
to present the results of the experimental settings. In this context, a
sort refinement corresponds loosely to a partitioning of the rows of the
horizontal table into subtables (in all figures for a given dataset, we depict the same number of columns for easy comparison, even if some columns are not present in a given implicit sort of the sort refinement).

%\begin{figure}[t]
%\centering
%\begin{subfigure}[b]{\columnwidth}
%    \centering
%    \includegraphics[trim=6.8cm 11.6cm 7.2cm 4.5cm, clip=true, width=0.40\textwidth]{images/dbpedia_persons_sigs2.pdf}
%
%    \caption{The DBPedia Persons dataset has 790,703 subjects, 8 properties and 64 signature sets. For this RDF graph, $\sigma_{\Cov} = 0.54$, $\sigma_{\Sim} = 0.77$.}
%  \label{fig:full_sigs_a}\end{subfigure}
%
%  \begin{subfigure}[b]{\columnwidth}
%    \centering
%    \includegraphics[trim=5.3cm 10.4cm 5.5cm 4.1cm, clip=true, width=0.5\textwidth]{images/wordnet_nounsynsets_nice.pdf}
%
%    \caption{The WordNet Nouns dataset has 79,689 subjects, 12 properties and 53 signature sets. For this RDF graph, $\sigma_{\Cov} = 0.44$, $\sigma_{\Sim} = 0.93$.}
%  \label{fig:full_sigs_b}\end{subfigure}
%
%  \caption{Signature set view of the two real world datasets: (a) DBpedia Persons, and (b) WordNet Nouns. The datasets are depicted as horizontal tables, where the columns correspond to the properties of a dataset, the black regions correspond to data (non-\texttt{null} cells), and the white regions correspond to \texttt{null} cells.}
%\label{fig:full_sigs}\end{figure}

%================================================================
%================================================================

\section{Formal definition of the decision problem}

Fix a rule $r$. The main problem that we address in this paper can be formalized as follows.
%We will now formally define the relevant decision problem. Let $r$ denote a rule, let $\theta$ denote a rational number such that $0 \leq \theta \leq 1$, and let $k$ be a positive integer.

%We will now formally define the relevant decision problem. Let $r$ denote a rule, let $\theta$ denote a rational number such that $0 \leq \theta \leq 1$, and let $k$ be a positive integer.

\begin{center}
\framebox{
\begin{tabular}{rp{6cm}}
{\bf Problem}: & $\eits(r)$\\
{\bf Input}:   &  An RDF graph $D$, a rational number $\theta$ such that $0 \leq \theta \leq 1$, and a positive integer $k$.\\
{\bf Output}:  & \textbf{true} if there exists an $\sigma_{r}$-sort refinement $\mathcal{T}$ of $D$ with threshold $\theta$ that contains at most $k$ implicit sorts, and \textbf{false} otherwise.
\end{tabular}}
\end{center}

In the following theorem, we pinpoint the complexity of the problem $\eits(r)$.
\begin{theorem}\hfill
{\em 
%The following holds:
\begin{itemize}
  \item $\eits(r)$ is in NP for every rule $r$.
  \item There is a rule $r_0$ for which $\eits(r_0)$ is NP-complete. Moreover, this result holds even if we fix $k = 3$ and $\theta = 1$. \hfill \qed
\end{itemize}
}
\label{the:eits_in_np}
\end{theorem}

The first part of Theorem \ref{the:eits_in_np} is a corollary of the fact that one can efficiently check if a sort refinement is an entity preserving partition of an RDF graph and has the correct threshold, as for every (fixed) rule $r$, function $\sigma_r$ can be computed in polynomial time. The second statement in Theorem \ref{the:eits_in_np} shows that there exists a (fixed) rule $r_0$ for which $\eits(r_0)$ is NP-hard, even if the structuredness threshold $\theta$ and the maximum amount of implicit sorts $k$ are fixed. The proof of this part of the theorem relies on a reduction from the graph 3-coloring problem to $\eits(r_0)$ with $\theta = 1$ and $k = 3$. In this reduction, a graph $G$ (the input to the 3-coloring problem) is used to construct an RDF graph $D_G$ in such a way that a partition of the nodes of $G$ can be represented by an entity preserving partitioning of the RDF graph. Although the rule $r_0$ will not be shown explicitly, it is designed to calculate the probability that 2 subjects in a subset of the entity preserving partitioning of $D_G$ represent 2 nodes of $G$ which are not adjacent. This probability will be 1 only when said subset represents an independent set of $G$. Therefore, setting the threshold $\theta = 1$ ensures that each subset of $D_G$ will represent an independent set of $G$. Finally, setting $k = 3$ ensures that at most 3 subsets will be generated. If the graph $G$ is 3-colorable, then it will be possible to generate the sort refinement of $D_G$ in which each subset represents an independent set of $G$, and thus will have a structuredness value of 1. Conversly, if there is a sort refinement of at most 3 subsets, then it is possible to partition the nodes of $G$ into 3 or less independent sets, and thus, is 3-colorable.

Note that the fixed rule $r_0$ used in the reduction does not contain statements of the form $\subj(c) = a$ (where $a$ is a constant URI), although it does use statements of the form $\prop(c) = a$ and other equalities. It is natural to exclude rules which mention specific subjects, as the structuredness of an RDF graph should not depend on the presence of a particular subject, but rather on the general uniformity of all entities in the RDF graph.

The decision problem presented in this section is theoretically intractable, which immediately reduces the prospects of finding reasonable algorithms for its solution. Nevertheless, the inclusion of the problem in NP points us to three NP-complete problems for which much work has been done to produce efficient solvers: the travelling salesman problem, the boolean satisfiability problem, and the integer linear programming problem.

An algorithm for our problem must choose a subset for each signature set, producing a series of decisions which could in principle be expressed as boolean variables, suggesting the boolean satisfiability problem. However, for a candidate sort refinement the function $\sigma_r$ must be computed for every subset, requiring non-trivial arithmetics which cannot be naturally formulated as a boolean formula. Instead, and as one of the key contributions of this paper, we have successfully expressed the previous decision problem in a natural way as an instance of Integer Linear Programming. It is to this reduction that we turn to in the next section.

%================================================================
%================================================================

\section{Reducing to Integer Linear Programming}

We start by describing the general structure of the Integer Linear Programming
(ILP) instance which, given a fixed rule $r$, solves the problem $\eits(r)$.
Given an RDF graph $D$, a rational number $\theta$ such that $0 \leq \theta
\leq 1$ and a positive integer $k$, we define in this section an instance of
integer linear programing, which can be represented as a pair
$(A_{(D,k,\theta)}, \vec b_{(D,k,\theta)})$, where $A_{(D,k,\theta)}$ is a
matrix of integer values, $\vec b_{(D,k,\theta)}$ is a vector of integer
values, and the problem is to find a vector $\vec d$ of integer values
(i.e.~the values assigned to the variables of the system of equations) such
that $A_{(D,k,\theta)} \vec d \leq \vec b_{(D,k,\theta)}$. Moreover, we prove
that $(D,k,\theta) \in \eits(r)$ if and only if the instance
$(A_{(D,k,\theta)}, \vec b_{(D,k,\theta)})$ has a solution.

Intuitively, our ILP instance works as follows: the integer variables decide
which signature sets are to be included in which subsets, and they keep track
of which properties are used in each subset.  Also, we group variable
assignments into objects we call \emph{rough} variable assignments, which
instead of assigning each variable to a subject and a property, they assign
each variable to a \textit{signature set} and a property. In this way, another
set of variables keeps track of which rough assignments are valid in a given
subset (i.e.~the rough assignment mentions only signature sets and properties
which are present in the subset). With the previous, we are able to count the
total and favorable cases of the rule for each subset.

For the following, fix a rule $r = \varphi_1 \mapsto \varphi_2$ and assume that $\var(\varphi_1) = \{ c_1, \ldots, c_n \}$ (recall that $\var(\varphi_2) \subseteq \var(\varphi_1)$). Also, fix a rational number $\theta \in [0,1]$, a positive integer $k$, and an RDF graph $D$, with the matrix $M = M(D)$.

\subsection{Variable definitions}

We begin by defining the ILP instance variables. Recall that our goal when
solving $\eits(r)$ is to find a $\sigma_r$-sort refinement of $D$ with
threshold $\theta$ and at most $k$ implicit sorts. 

All the variables used in the ILP instance take only integer values. To introduce these variables, we begin by
defining the set of signatures of $D$ as $\Lambda(D) = \{ \sig(s,D) \mid s \in
S(D) \}$, and for every $\mu \in \Lambda(D)$, by defining the support of $\mu$,
denoted by $\supp(\mu)$, as the set $\{ p \in P(D) \mid \mu(p) = 1\}$. Then for
each $i \in \{ 1, \ldots, k \}$ and each $\mu \in \Lambda(D)$, we define the
variable:
\begin{equation*}
X_{i,\mu} = \begin{cases}
1 & \text{if signature $\mu$ is placed in implicit sort $i$} \\
0 & \text{otherwise}.
\end{cases}
\end{equation*}

\noindent These are the primary variables, as they encode the generated sort
refinement. Notice that it could be the case that for some $i \in \{1, \ldots,
k\}$ value 0 is assigned to every variable $X_{i,\mu}$ ($\mu \in \Lambda(D)$),
in which case we have that the $i$-th implicit sort is empty.

For each $i \in \{ 1, \ldots, k \}$ and each $p \in P(D)$ define the variable:
\begin{equation*}
U_{i,p} = \begin{cases}
1 & \text{if implicit sort $i$ uses property $p$} \\
0 & \text{otherwise}.
\end{cases}
\end{equation*}

Each variable $U_{i, p}$ is used to indicate whether the $i$-th implicit sort uses property $p$, that is, if implicit sort $i$ includes a signature $\mu \in \Lambda(D)$ such that $\mu(p) = 1$ ($p \in \supp(\mu)$).

For the last set of variables, we consider a \emph{rough} assignment of
variables in $\varphi_1$ to be a mapping of each variable to a signature and a
property. Recall that $\var(\varphi_1) = \{ c_1, \ldots, c_n \}$. Then we denote rough assignments with $\tau = ((\mu_1, p_1), \ldots,$
$(\mu_n, p_n)) \in (\Lambda(D) \times P(D))^n$, and for each $i \in \{ 1,
\ldots, k \}$ and each $\tau \in (\Lambda(D) \times P(D))^n$, we define the
variable:
\begin{equation*}
T_{i,\tau} = \begin{cases}
1 & \text{if $\tau$ is consistent in the $i$-th implicit sort} \\
0 & \text{otherwise}.
\end{cases}
\end{equation*}

\noindent The rough assignment $\tau = ((\mu_1, p_n), \ldots, (\mu_n, p_n))$ is consistent in the $i$-th implicit sort if it only mentions signatures and properties which are present in it, that is, if for each $j \in \{ 1, \ldots, n \}$ we have that $\mu_j$ is included in the $i$-th implicit sort and said implicit sort uses $p_j$.

\subsection{Constraint definitions}

Define function $\cnt(\varphi, \tau, M)$ to be the number of variable
assigments for rule $r$ which are restricted by the rough assignment $\tau$ and
which satisfy the formula $\varphi$. Formally, if $\tau = ((\mu_1, p_n)$,
$\ldots$, $(\mu_n, p_n))$, then $\cnt(\varphi, \tau, M)$ is defined as the
cardinality of the following set:

\vspace*{-3.0ex}

\begin{multline*}
\{ \rho \mid \rho \text{ is a variable assignment for $D$ s.t. } \dom(\rho) = \var(\varphi),\\
(M,\rho) \models \varphi \text{ and for every } i \in \{1, \ldots, n\},\\
 \text{ if } \rho(c_i) = (s,p) \text{ then } \sig(s,D) = \mu_i \text{ and } p = p_i\}.
\end{multline*}

Note that the value of $\cnt(\varphi, \tau, M)$ is calculated offline and is
used as a constant in the ILP instance.  We now present the inequalities that
constrain the acceptable values of the defined variables.

\vspace*{-2.0ex}

\begin{tabbing}
X \= $\bullet$ \= \kill
\> $\bullet$ \> \parbox[t]{3.0in}{The following inequalities specify the
obvious lower and upper bounds of all variables:

\vspace*{-3.0ex}

\begin{align*}
0 \leq X_{i, \mu} \leq 1  & \quad i \in \{1, \ldots, k\} \text{ and } \mu  \in \Lambda(D) \\
0 \leq U_{i, p} \leq 1    & \quad i \in \{1, \ldots, k\} \text{ and } p \in P(D) \\
0 \leq T_{i, \tau} \leq 1 & \quad i \in \{1, \ldots, k\} \text{ and } \tau \in (\Lambda(D) \times P(D))^n
\end{align*}
}\\

\> $\bullet$ \>  \parbox[t]{3.0in}{For every $\mu \in \Lambda(D)$, the
following equation indicates that the signature $\mu$ must be assigned to
exactly one implicit sort:

\vspace*{-3.0ex}

\begin{equation*}
\sum_{i = 1}^k X_{i,\mu} = 1.
\end{equation*}
}\\

\> $\bullet$ \> \parbox[t]{3.0in}{For every $i \in \{1, \ldots, k\}$ and $p \in
P(D)$,  we include the following equations to ensure that $U_{i, p}$ is
assigned to $1$ if and only if the $i$-th implicit sort includes a signature
$\mu \in \Lambda(D)$ such that $\mu(p) = 1$ ($p \in \supp(\mu)$):

\vspace*{-3.0ex}

\begin{eqnarray*}
X_{i,\mu} & \leq & U_{i,p} \quad\quad \text{if } p \in \supp(\mu) \\
U_{i,p} & \leq &  \sum_{\mu' \in \Lambda(D) \,:\, p \in \supp(\mu')} X_{i,\mu'}
\end{eqnarray*}

The first equation indicates that if signature $\mu$ has been assigned to the
$i$-th implicit sort and $p \in \supp(\mu)$, then $p$ is one of the properties
that must be considered when computing $\sigma_r$ in this implicit sort. The
second equation indicates that if $p$ is used in the computation of $\sigma_r$
in the $i$-th implicit sort, then this implicit sort must include a signature
$\mu' \in \Lambda(D)$ such that $p \in \supp(\mu')$.
}\\

\> $\bullet$ \> \parbox[t]{3.0in}{For $i \in \{ 1, \ldots, k \}$, and $\tau =
((\mu_1, p_1), \ldots, (\mu_n,p_n)) \in (\Lambda(D) \times P(D))^n$, recall
that $T_{i, \tau} = 1$ if and only if for every $ j \in \{1, \ldots, n \}$, it
holds that $ X_{i, \mu_j} = 1$ and  $U_{i, p_j} = 1$. This is expressed as
integer linear equations as follows:

\vspace*{-3.0ex}

\begin{align*}
\sum_{j=1}^n (X_{i,\mu_j} + U_{i,p_j}) & \leq T_{i, \tau} + 2 \cdot n - 1 \\
2 \cdot n \cdot T_{i, \tau} & \leq \sum_{j=1}^n (X_{i,\mu_j} + U_{i,p_j})
\end{align*}

}\\

\>  \> \parbox[t]{3.0in}{The first equation indicates that if the signatures $\mu_1$, $\ldots$, $\mu_n$
are all included in the $i$-th implicit sort (each variable $X_{i,\mu_j}$ is
assigned value 1), and said implicit sort uses the properties $p_1$, $\ldots$,
$p_n$ (each variable $U_{i,p_j}$ is assigned value 1), then $\tau$ is a valid
combination when computing favorable and total cases (variable $T_{i,\tau}$ has
to be assigned value 1).  Notice that if any of the variables $X_{1,\mu_1}$,
$U_{1,p_1}$, $\ldots$, $X_{n,\mu_n}$, $U_{n,p_n}$ is assigned value 0 in the
first equation, then $\sum_{j=1}^n (X_{i,\mu_j} + U_{i,p_j}) \leq 2 \cdot n -
1$ and, therefore, no restriction is imposed on $T_{i, \tau}$ by this equation,
as we already have that $0 \leq T_{i, \tau}$.  The second equation indicates
that if variable $T_{i, \tau}$ is assigned value 1, meaning that $\tau$ is
considered to be

}\\

\> \> \parbox[t]{3.0in}{

 a valid combination when computing $\sigma_r$ over the $i$-th
implicit sort, then each signature mentioned in $\tau$ must be included in this
implicit sort (each variable $X_{i,\mu_j}$ has to be assigned value 1), and
each property mentioned in $\tau$ is used in this implicit sort (each variable
$U_{i,p_j}$ has to be assigned value~1).

}\\

\> $\bullet$ \> \parbox[t]{3.0in}{Finally, assuming that $\theta =
\theta_1/\theta_2$, where $\theta_1, \theta_2$ are natural numbers, we include
the following equation for each $i \in \{ 1, \ldots, k \}$:

\vspace*{-2.0ex}

\begin{align*}
\theta_2 \cdot \bigg(\sum_{\tau  \in (\Lambda(D) \times P(D))^n} \cnt(\varphi_1 \wedge \varphi_2, \tau, M) \cdot T_{i,\tau}\bigg) \\
\geq \theta_1 \cdot \bigg(\sum_{\tau  \in (\Lambda(D) \times P(D))^n} \cnt(\varphi_1, \tau,M) \cdot T_{i,\tau}\bigg)
\end{align*}

To compute the numbers of favorable and total cases for $\sigma_r$ over the $i$-th implicit sort, we consider each rough assignment $\tau$ in turn. The term $\sum_{\tau  \in (\Lambda(D) \times P(D))^n} \cnt(\varphi_1 \wedge \varphi_2, \tau, M) \cdot T_{i,\tau}$ evaluates to the amount of favorable cases (i.e.~variable assignments which satisfy the antecedent and the consequent of the rule), while the term $\sum_{\tau \in (\Lambda(D) \times P(D))^n}$ $\cnt(\varphi_1, \tau, M) \cdot T_{i,\tau}$ evaluates to the number of total cases (i.e.~variable assignments which satisfy the antecedent of the rule). Consider the former term as an example: for each rough variable assignment $\tau$, if $\tau$ is a valid combination in the $i$-th implicit sort, then the amount of variable assignments which are compatible with $\tau$ and which satisfy the full rule are added.

}

%If $\tau$ is a valid rough assignment in the $i$-subset, we count all the variable assignments which conform to $\tau$ and also satisfy the appropriate formula.

%we split $M$ according to the tuples $\tau \in (\Lambda(D) \times P(D))^n$. More precisely, first we consider each $\tau = ((\mu_1,p_1),$ $\ldots$, $(\mu_n,p_n))$ such that value 1 has been assigned to $T_{i,\tau}$, and compute the number of favorable cases for $\sigma_r$ that can be found by restricting the possible values of variable $c_j$ ($1 \leq j \leq n$) to the tuples $(s_j,p_j)$ such that $s_j \in S$ and $\lambda_{s_j} = \mu_j$ (notice that this value corresponds to $\cnt(\varphi_1 \wedge \varphi_2, \tau, M)$). Then we compute the total number of favorable cases by adding all the values computed for the variables $T_{i,\tau}$ with value 1 (and likewise for the total cases).

%The term $\sum_{\tau  \in (\Lambda(M) \times P)^n} \cnt(\varphi_1 \wedge \varphi_2, \tau, M) \cdot T_{i,\tau}$ represents the number of favorable cases for $\sigma_r$ over the $i$-th subset, while the term $\sum_{\tau  \in (\Lambda(M) \times P)^n} \cnt(\varphi_1, \tau, M) \cdot T_{i,\tau}$ represents the number of total cases for $\sigma_r$ over the $i$-th subset. Thus, the above equation checks whether, over this subset, the value of $\sigma_r$ is at least $\theta_1/\theta_2 = \theta$.
\end{tabbing}

It is now easy to see that the following result holds. 
\begin{proposition}
{\em There exists a $\sigma_r$-sort refinement of $D$ with threshold $\theta$ that contains at most $k$ implicit sorts if and only if the instance of ILP defined in this section has a solution. \hfill\qed}
\end{proposition}

\subsection{Implementation details}

Although the previously defined constraints suffice to solve the decision
problem, in practice the search space is still too large due to the presence of
sets of solutions which are \textit{equivalent}, in the sense that the
variables describe the same partitioning of the input RDF graph $D$. More
precisely, if there is a solution of the ILP instance where for each $i \in \{
1, \ldots, k \}$, $\mu \in \Lambda(D)$, $p \in P(D)$, and $\tau \in (\Lambda(D)
\times P(D))^n$, $X_{i,\mu} = a_{i,\mu}$, $U_{i,p} = b_{i,p}$, and $T_{i,\tau}
= c_{i,\tau}$, then for any permutation $(l_1, \ldots, l_k)$ of $(1, \ldots,
k)$, the following is also a solution: $X_{i,\mu} = a_{l_i, \mu}$, $U_{i,p} =
b_{l_i, p}$, and $T_{i,\tau} = c_{l_i, \tau}$.

In order to break the symmetry between these equivalent solutions, we define the following hash function for the $i$-th implicit sort. For this, consider $\ell = |\Lambda(D)|$ and consider any (fixed) ordering $\mu_1, \ldots, \mu_{\ell}$ of the signatures in $\Lambda(D)$. Then:
\begin{equation*}
\hash(i) = \sum_{j = 0}^{\ell} 2^{j} X_{i,\mu_j},
\end{equation*}

With the previous hash function defined, the following constraint is added, for $i = 1, \ldots, k-1$:
\begin{equation*}
\hash(i) \leq \hash(i+1).
\end{equation*}

The $\hash$ function as defined above uniquely identifies a subset of
signatures, and thus the previous constraints eliminate the presence of
multiple solutions due to permutations of the $i$ index. Care must be taken,
however, if the amount of signatures in the RDF graph is large (64 for DBpedia Persons) as large exponent values cause numerical instability in
commercial ILP solvers. This issue may be addressed on a case by case basis.
One alternative is to limit the maximum exponent in the term $2^{j}$, which has
the drawback of increasing the amount of collisions of the hash function, and
therefore permitting the existence of more equivalent solutions.

%================================================================
%================================================================

\section{Experimental results}

For our first experiments, we consider two real datasets--DBpedia Persons and
WordNet Nouns--and two settings:

\begin{itemize}
\item \textbf{A highest $\theta$ sort refinement for $k = 2$}: This setup
can be used to obtain an intuitive understanding of the dataset at hand. We fix
$k = 2$ to force at most 2 implicit sorts.

\item \textbf{A lowest $k$ sort refinement for $\theta = 0.9$}: As a
complementary approach, we specify $\theta = 0.9$ as the threshold, and we
search for the lowest $k$ such that an sort refinement with threshold $\theta$
and $k$ implicit sorts exists. This approach allows a user to refine their
current sort by discovering sub-sorts. In some cases the structuredness of the
original dataset under some structuredness function is higher than 0.9, in
which case we increase the threshold to a higher value.
\end{itemize}

In the first case the search for the optimum value of $\theta$ is done in the
following way: starting from the initial structuredness value $\theta =
\sigma_r(D)$ (for which a solution is guaranteed) and for values of $\theta$
incremented in steps of $0.01$, an ILP instance is generated with $k = 2$ and
the current value of $\theta$. If a solution is found by the ILP solver, then
said solution is stored. If the ILP instance is found to be infeasible, then
the last stored solution is used (this is the solution with the highest
threshold). This sequential search is preferred over a binary search because
the latter will generate more infeasible ILP instances on average, and it has
proven to be much slower to find an instance infeasible than to find a solution
to a feasible instance. A similar strategy is used for the second case (the
search for the lowest $k$), with the following difference: for some setups it
is more efficient to search downwards, starting from $k = |\Lambda(D)|$
(i.e.~as many available sorts as signatures in the dataset), and yet for others
it is preferrable to search upwards starting from $k = 1$, thus dealing with a
series of infeasible ILP instances, before discovering the first value of $k$
such that a solution is found. Which of the two directions is to be used has
been decided on a case by case basis.

The amount of variables and constraints in each ILP instance depends on the amount of
variables of the rules, on the degrees of freedom given to the variables in the
rules (e.g.~the two variables in $\sigma_{\Dep}$[$\mathbf{p}_1$,
$\mathbf{p}_2$] lose a degree of freedom when considering the restriction
$\subj(c_1) = \subj(c_2)$ in the antecedent), and on the characteristics of the
dataset. Here, the enourmous reduction in size offered by the signature
representation of a dataset has proven crucial for the efficiency of solving
the ILP instances.

The previous two settings are applied both to the DBpedia Persons and WordNet
Nouns datasets. Furthermore, they are repeated for the 
$\sigma_{\Cov}$, $\sigma_{\Sim}$, and $\sigma_{\Dep}$ functions (the last function is
only used on DBpedia Persons). All experiments are conducted on dual
2.3GHz processor machine (with 6 cores per processor), and 64GB of RAM. The ILP solver used is
IBM ILOG CPLEX version 12.5.

In section 7.3 we study how our solution scales with larger and more complex datasets by extracting a representative sample of explicit sorts from the knowledge base YAGO and solving a highest $\theta$ sort refinement for fixed $k$ for each explicit sort. Finally, in section 7.4 we challenge the solution to recover two different explicit sorts from a mixed dataset, providing a practical test of the results.

%================================================================
%  DBPEDIA PERSONS
%================================================================
\subsection{DBpedia Persons}

DBpedia corresponds to RDF data extracted from Wikipedia. DBpedia Persons
refers to the following subgraph (where $\mathsf{Person}$ is a shorthand for
{\small \texttt{http://xmlns.com/foaf/0.1/Person}}):
\begin{align*}
D_{\text{DBpedia Persons}} = \{ &(s,p,o) \in D_{\text{DBpedia}} \;\mid\;\\
& (s, \type, \mathsf{Person}) \in D_{\text{DBpedia}} \}.
\end{align*}

This dataset is 534 MB in size, and contains 4,504,173 triples, 790,703
subjects, and 8 properties (excluding the $\type$ property). It consists of 64
signatures, requiring only 3 KB of storage. The list of properties is as
follows: {\small \texttt{deathPlace}}, {\small \texttt{birthPlace}}, {\small
\texttt{description}}, {\small \texttt{name}}, {\small \texttt{deathDate}},
{\small \texttt{birthDate}}, {\small \texttt{givenName}}, and {\small
\texttt{surName}} (note that these names are abbreviated versions of the full
URIs).

For this sort, $\sigma_{\Cov} = 0.54$, and $\sigma_{\Sim} = 0.77$. We are also
interested in studying the dependency functions for different properties
$\mathbf{p}_1$ and $\mathbf{p}_2$. If $\mathbf{p}_1 =$ {\small
\texttt{deathPlace}} and $\mathbf{p}_2 =$ {\small \texttt{deathDate}}, for
example, then the value of the function $\sigma_{\SymDep}$ [{\small
\texttt{deathPlace}}, {\small \texttt{deathDate}}] is $0.39$. This specific
choice of $\mathbf{p}_1$ and $\mathbf{p}_2$ is especially interesting because
it might be temping to predict that a death date and a death place are equally
obtainable for a person. However, the value 0.39 reveals the contrary. The
generally low values for the three structuredness functions discussed make
DBpedia Persons interesting to study.

%================================================================
%  Subsubsection: dbpedia k2
%================================================================
\subsubsection{A highest $\theta$ sort refinement for $k = 2$}

\begin{figure}[t]
\centering

  \begin{subfigure}[b]{\columnwidth}
    \centering
    \includegraphics[trim=6cm 11.5cm 6cm 4.5cm, clip=true, width=3.0cm]{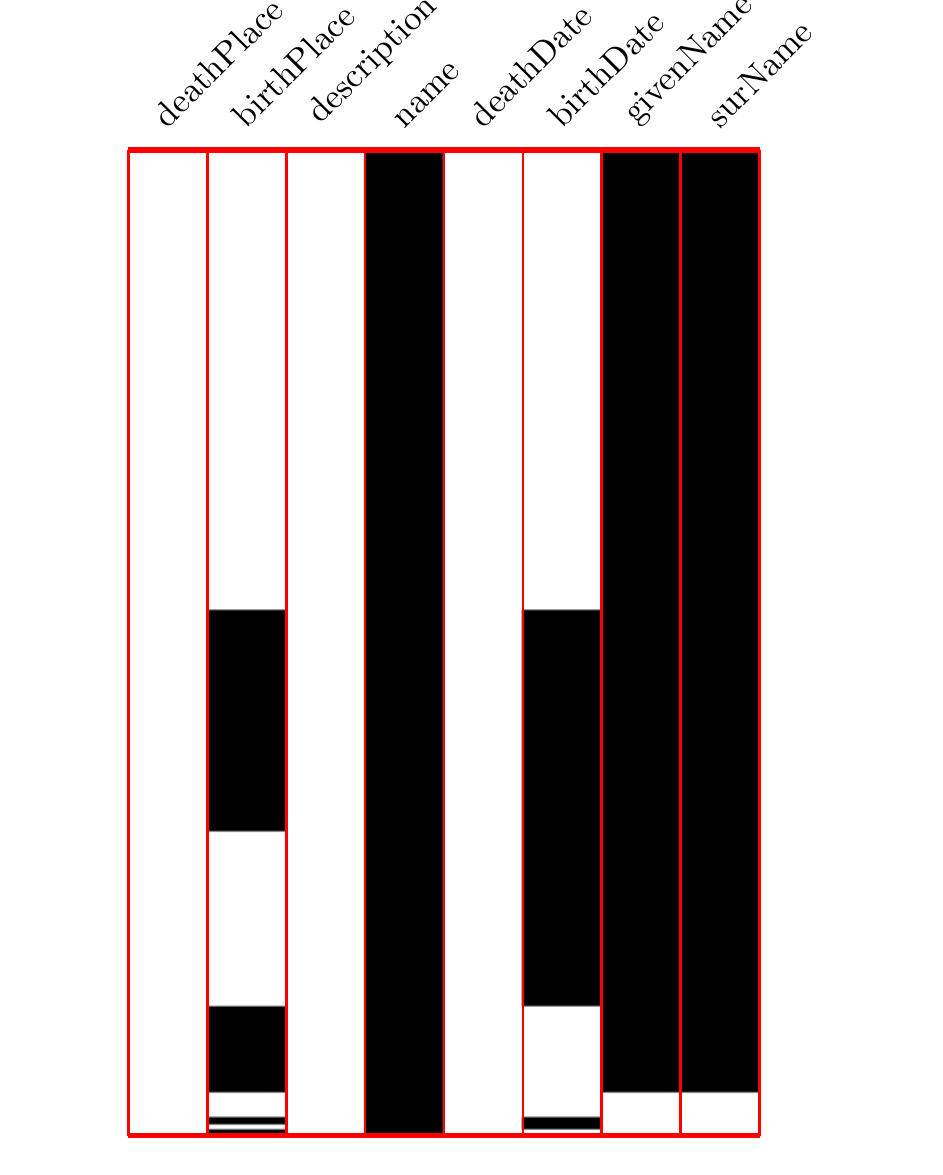}
    \includegraphics[trim=6cm 11.5cm 6cm 4.5cm, clip=true, width=3.0cm]{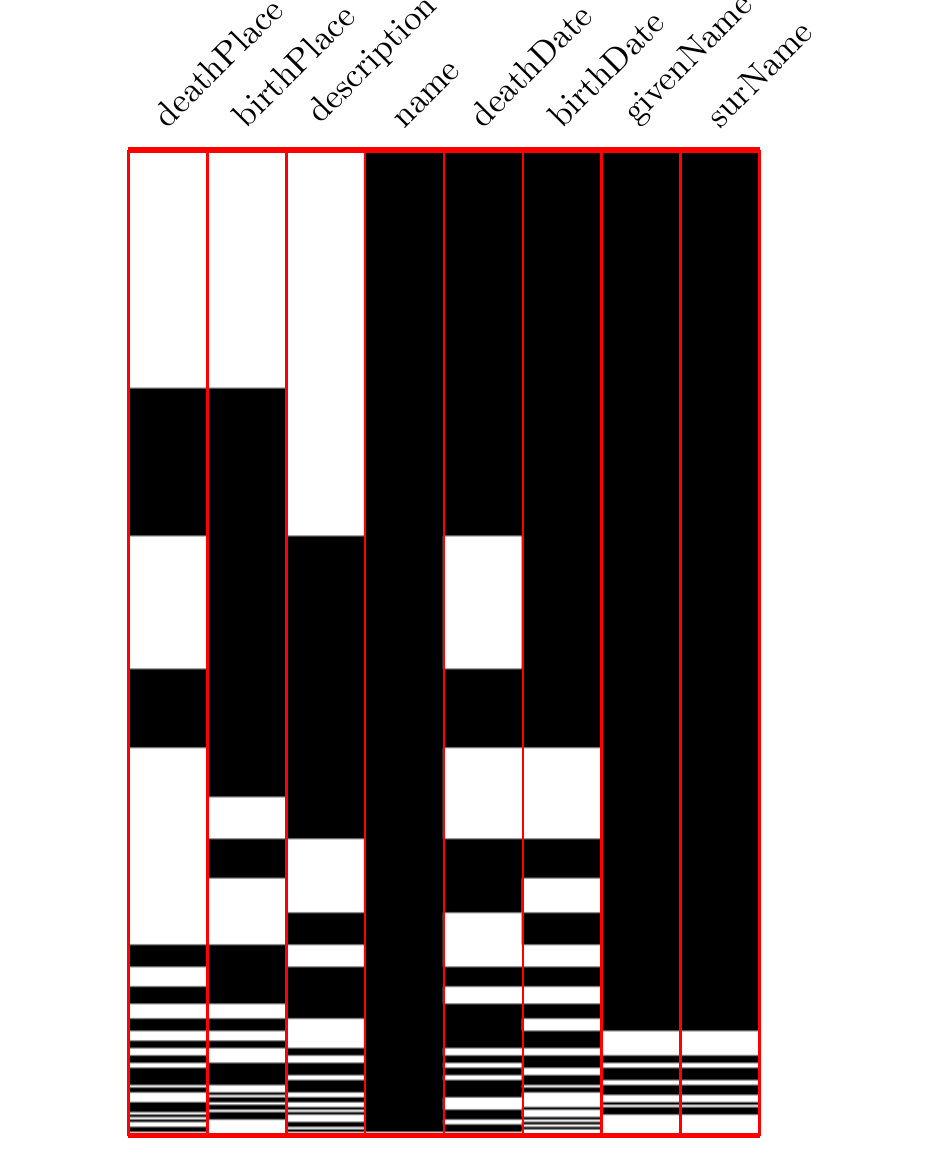}
    \caption{Using the $\sigma_{\Cov}$ function, the left sort has 528,593 subjects and 8 signatures, $\sigma_{\Cov} = 0.73$, and $\sigma_{\Sim} = 0.85$. The right sort has 262,110 subjects and 56 signatures, $\sigma_{\Cov} = 0.71$, and $\sigma_{\Sim} = 0.78$.}

  % Finished quickly in 12 s. Has logfile.
  \label{fig:dbpedia_persons_k2_bestTheta_cov}\end{subfigure}

  \begin{subfigure}[b]{\columnwidth}
    \centering
    \includegraphics[trim=6cm 13.4cm 6cm 3.5cm, clip=true, width=3.0cm]{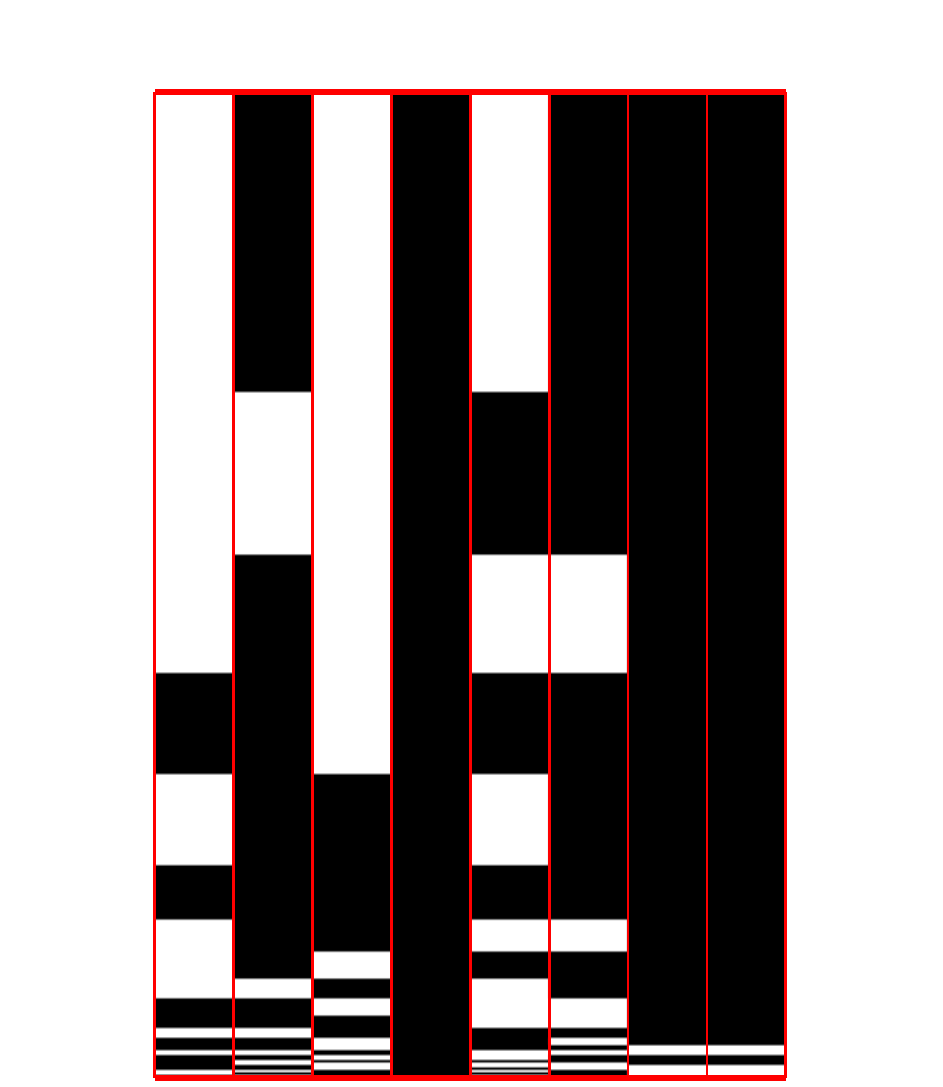}
    \includegraphics[trim=6cm 13.4cm 6cm 3.5cm, clip=true, width=3.0cm]{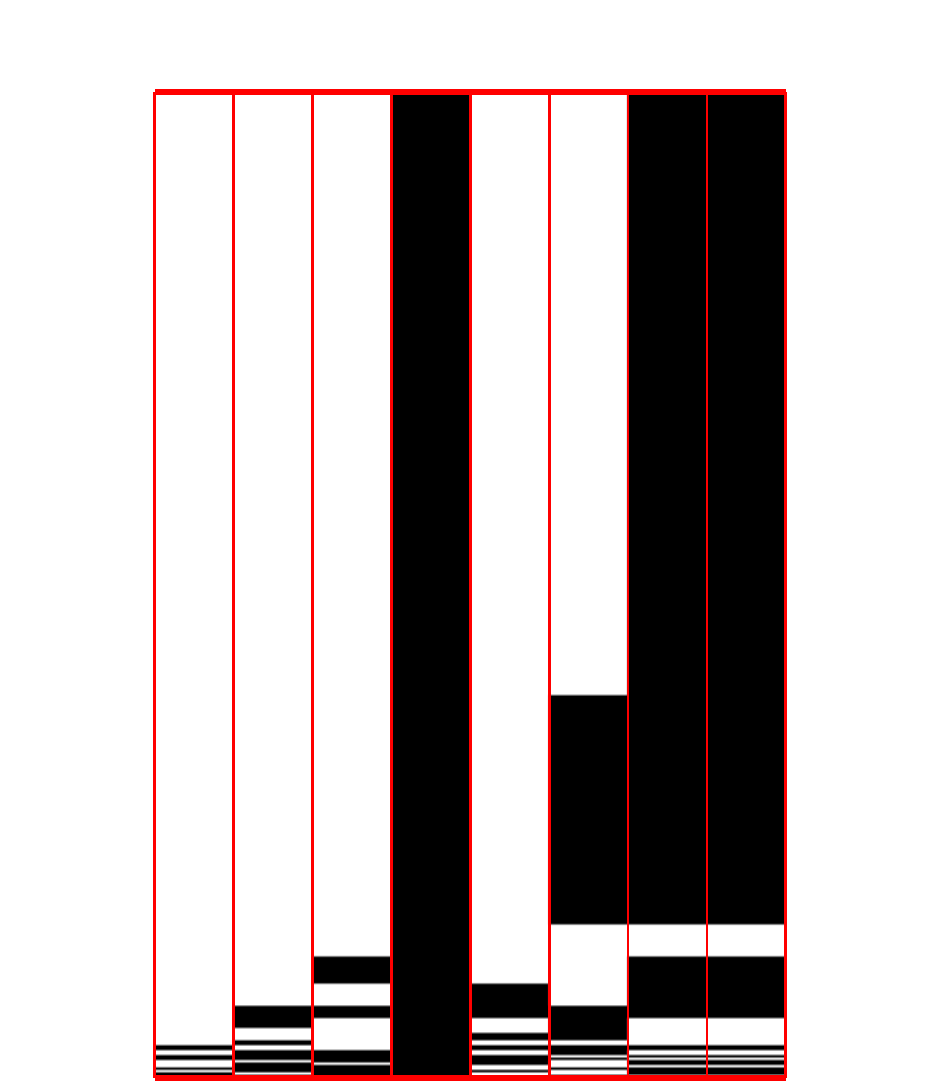}
    \caption{Using the $\sigma_{\Sim}$ function, the left sort has 387,297 subjects and 37 signatures, $\sigma_{\Cov} = 0.67$, and $\sigma_{\Sim} = 0.82$. The right sort has 403,406 subjects and 27 signatures, $\sigma_{\Cov} = 0.42$ and $\sigma_{\Sim} = 0.85$.}

  %Finished in 3871 s (timedout). Has logfile with old Dif!

  \label{fig:dbpedia_persons_k2_bestTheta_sim}\end{subfigure}

  \begin{subfigure}[b]{\columnwidth}
    \centering
    \includegraphics[trim=6cm 13.4cm 6cm 3.5cm, clip=true, width=3.0cm]{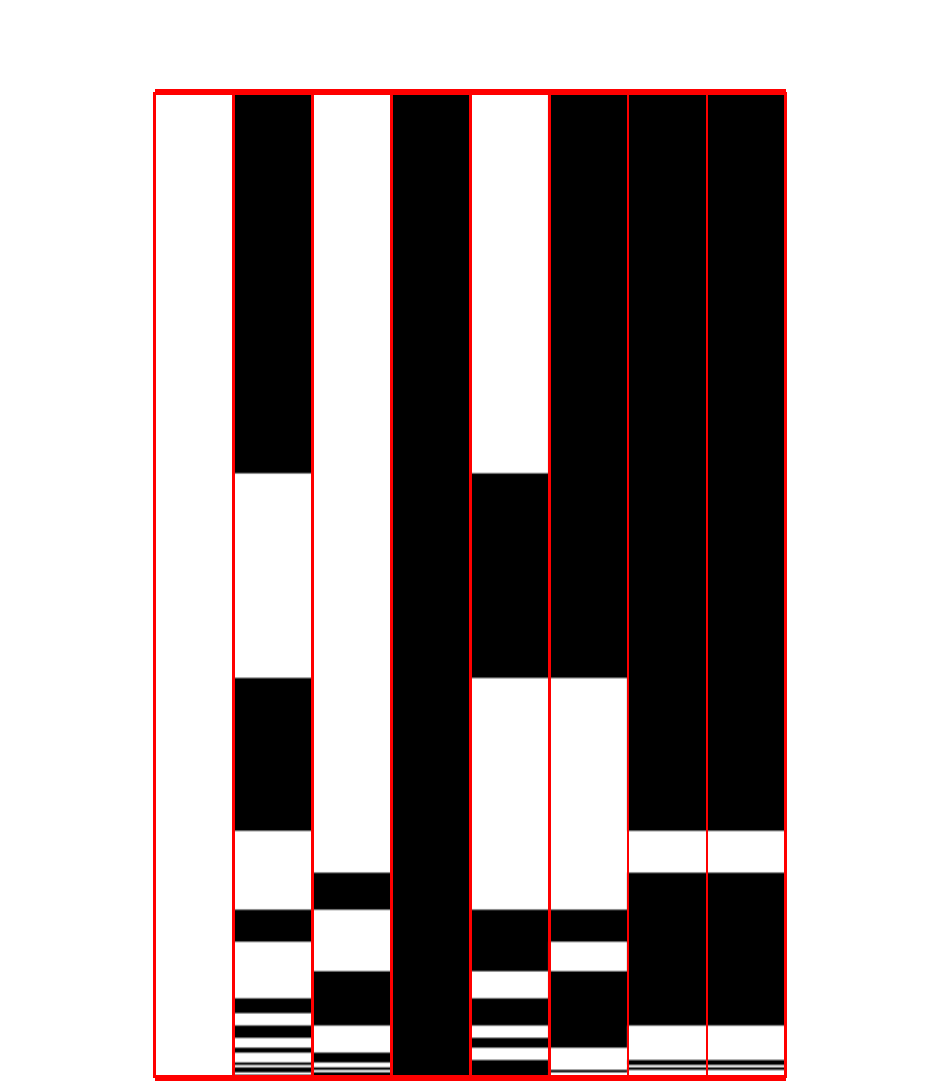}
    \includegraphics[trim=6cm 13.4cm 6cm 3.5cm, clip=true, width=3.0cm]{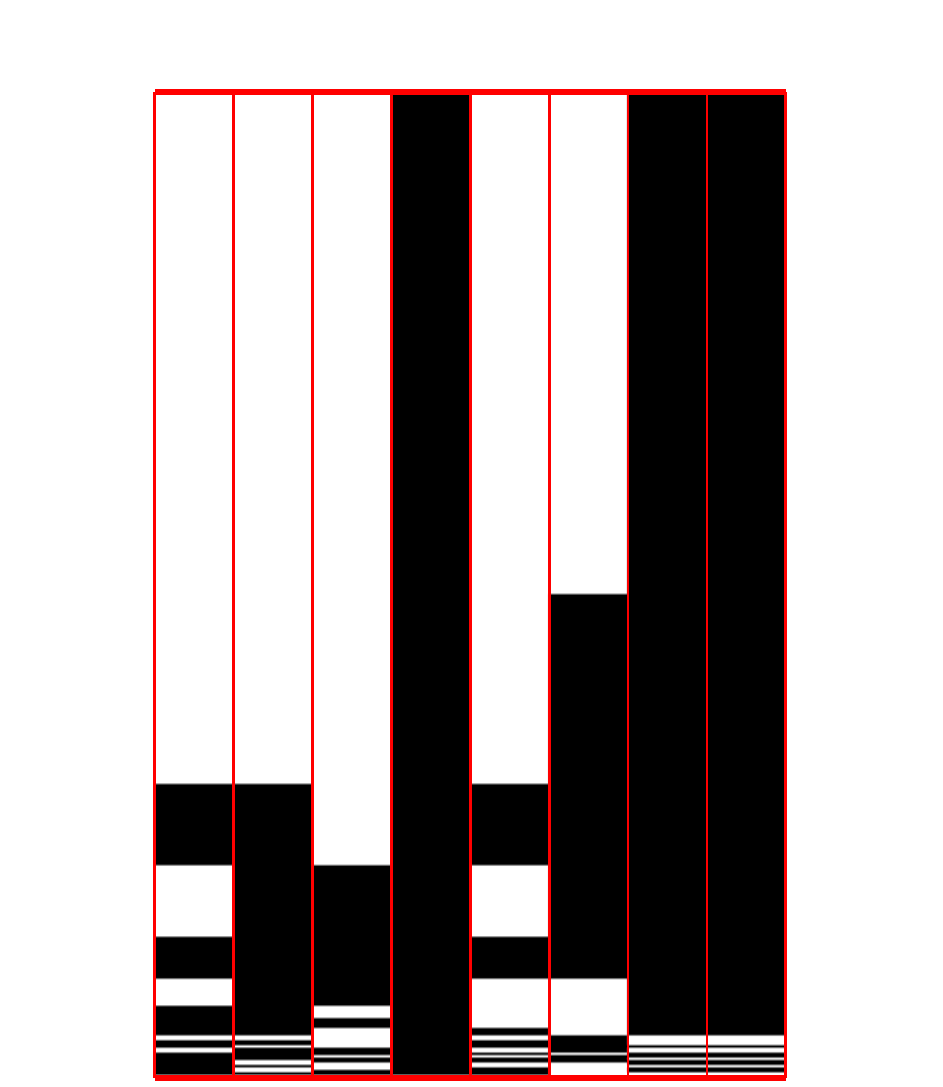}
    \caption{Using the $\sigma_1 = \sigma_{\SymDep}[\texttt{deathPlace}, \texttt{deathDate}]$ function, the left sort has 305,610 subjects and 25 signatures, $\sigma_{\Cov} = 0.66$, $\sigma_{\Sim} = 0.80$, and $\sigma_1 = 1.0$. The right sort has 485,093 subjects and 39 signatures, $\sigma_{\Cov} = 0.52$, $\sigma_{\Sim} = 0.78$, and $\sigma_1 = 0.82$.}
  \label{fig:dbpedia_persons_k2_bestTheta_dep}\end{subfigure}

  \caption{DBpedia Persons split into $k = 2$ implicit sorts, using the structuredness functions (a) $\sigma_{\Cov}$, (b) $\sigma_{\Sim}$, and (c) $\sigma_{\Dep}$.}
\label{fig:dbpedia_persons_k2_bestTheta}
\vspace*{-2.0ex}
\end{figure}

We set $k = 2$ in order to find a two-sort sort refinement with the best
threshold $\theta$.  Figure~\ref{fig:dbpedia_persons_k2_bestTheta_cov} shows
the result for the $\sigma_{\Cov}$ function. The left sort, which is also the
largest (having 528,593 subjects) has a very clear characteristic: no subject
has a {\small \texttt{deathDate}} or a {\small \texttt{deathPlace}}, i.e.~it
represents the sort for people that are alive! Note that without encoding any
explicit schema semantics in the generic rule of $\sigma_{\Cov}$, our ILP
formulation is able to discover a very intuitive decomposition of the initial
sort. In the next section, we show that this is the case even if we consider
larger values of $k$. In this experiment, each ILP instance is solved in under
800 ms.

Figure~\ref{fig:dbpedia_persons_k2_bestTheta_sim} shows the results for the
$\sigma_{\Sim}$ function. Here, the second sort accumulates subjects for which
very little data is known (other than a person's name). Notice that whereas
$\Cov$ has excluded the columns {\small \texttt{deathPlace}}, {\small
\texttt{description}}, and {\small \texttt{deathDate}} from its first sort,
$\Sim$ does not for its second sort, since it does not \textit{penalize} the
largely missing properties in these columns (which was what motivated us to
introduce $\sigma_{\Sim}$ in the first place). Also, notice that unlike the
$\sigma_{\Cov}$ function, the cardinality of the generated sorts from
$\sigma_{\Sim}$ is more \textit{balanced}. In this experiment, each ILP
instance is solved in under 2 minutes, except the infeasible (last) instance
which was completed in 2 hrs.

Finally, Figure~\ref{fig:dbpedia_persons_k2_bestTheta_dep} shows the results
for $\sigma_{\SymDep}$[{\small \texttt{deathPlace}}, {\small
\texttt{deathDate}}], a structuredness function in which we measure the
probability that, if a subject has a {\small \texttt{deathPlace}} or a {\small
\texttt{deathDate}}, it has both. In the resulting sort refinement, the second
sort to the right has a high value of 0.82. It is easy to see
that indeed our ILP solution does the right thing. In the sort on the right,
the {\small \texttt{deathDate}} and {\small \texttt{deathPlace}} columns look
almost identical which implies that indeed whenever a subject has one property
it also has the other.  As far as the sort on the left is concerned, this
includes all subjects that do not have a {\small \texttt{deathPlace}} column.
This causes the sort to have a structuredness value of 1.0 for
$\sigma_{\SymDep}$[{\small \texttt{deathPlace}}, {\small \texttt{deathDate}}]
since the rule is trivially satisfied; the absence of said
column eliminates all total cases (i.e.~there are no assignments
in the rule that represents $\sigma_{\SymDep}$[{\small \texttt{deathPlace}},
{\small \texttt{deathDate}}] for which the antecedent is true, as it is
never true that $\prop(c_1)=$ {\small \texttt{deathPlace}}). This setting is
completed in under 1 minute.

%================================================================
%  Figure: dbpedia k2
%================================================================

%================================================================
%  Subsubsection: dbpedia theta09
%================================================================
\subsubsection{A lowest $k$ sort refinement for $\theta = 0.9$}

We now consider a fixed threshold $\theta = 0.9$. We seek the smallest sort
refinement for DBpedia persons with this threshold.  Figure
\ref{fig:dbpedia_persons_theta09_bestK_cov} shows the result for
$\sigma_{\Cov}$, where the optimum value found is for $k = 9$. As in the
previous setting, the $\Cov$ function shows a clear tendency to produce sorts
which do not use all the columns (i.e.~sorts which exclude certain properties).
People that are alive can now be found in the first, second, third, fourth, and
sixth sorts. The first sort considers living people who have a description (and
not even a birth place or date). The second sort shows living people who are
even missing the description field. The third sort considers living people who
have a description and a birth date or a birth place (or both).  The fourth
sort considers living people with a birth place or birth date but no
description. Finally, the sixth sort considers living people with a birth place
only. It is easy to see that similarly dead people are separated into different
sorts, based on the properties that are known for them. The eighth sort is
particularly interesting since it contains people for which we mostly have all
the properties. This whole experiment was completed in 30 minutes.

Figure \ref{fig:dbpedia_persons_theta09_bestK_sim} shows the result for
$\sigma_{\Sim}$, where the optimum value found is for $k = 4$. Again, the
function is more lenient when properties appear for only a small amount of
subjects (hence the smaller $k$). This is evident in the first sort for
this function, which corresponds roughly to the second sort generated for the
$\sigma_{\Cov}$ function (Fig.~\ref{fig:dbpedia_persons_theta09_bestK_cov}) but
also includes a few subjects with birth/death places/dates. This is
confirmed by the relative sizes of the two sorts, with the sort for
$\sigma_{\Cov}$ having 260,585 subjects, while the sort for $\sigma_{\Sim}$
having 292,880 subjects. This experiment is more difficult as the
running time of individual ILP instances is apx.~8 hrs.

%\begin{figure}
%\centering
%
%
%Set 0: 
%  260585 subjects
%  2 signatures
%  cov = 0.9667542900269266
%  sim = 0.9673258079423555
%Set 1: 
%  13253 subjects
%  2 signatures
%  cov = 0.9280162981966347
%  sim = 0.9335948677124786
%Set 2: 
%  169248 subjects
%  3 signatures
%  cov = 0.9363005766685574
%  sim = 0.9486189701908235
%Set 3: 
%  6236 subjects
%  5 signatures
%  cov = 0.9012721830233056
%  sim = 0.9085289517806974
%Set 4: 
%  76402 subjects
%  4 signatures
%  cov = 0.9549383523991518
%  sim = 0.9573430974556094
%Set 5:
%  60852 subjects
%  6 signatures
%  cov = 0.9033228160126208
%  sim = 0.9113630084602864
%Set 6: 
%  96652 subjects
%  2 signatures
%  cov = 0.9854270992840293
%  sim = 0.9856424607769418
%Set 7: 
%  69312 subjects
%  19 signatures
%  cov = 0.9040817504287033
%  sim = 0.9085240299006502
%Set 8: 
%  38163 subjects
%  21 signatures
%  cov = 0.9002469669575243
%  sim = 0.9112465738649611
%
%\caption{DBpedia Persons split into $k = 9$ implicit sorts, using the $\sigma_{\Cov}$ function. The threshold of this sort system is $\theta = 0.9$, (i.e.~every sort $D_i$ has $\sigma_{\Cov}(D_i) \geq 0.9$. The sizes of the sorts range from 260,585 subjects (the first sort) to 6,236 subjects (the fourth sort).}
%\label{fig:dbpedia_persons_cov_k9}\end{figure}

%================================================================
%  Figure: dbpedia theta09
%================================================================
\newcommand{\tinyWidth}{1.4cm}
\newcommand{\tinySpacing}{2mm}
\begin{figure}[t]
\centering
  \begin{subfigure}[b]{\columnwidth}
    \centering
    \includegraphics[trim=7.5cm 13.5cm 7.5cm 4.3cm, clip=true, width=\tinyWidth]{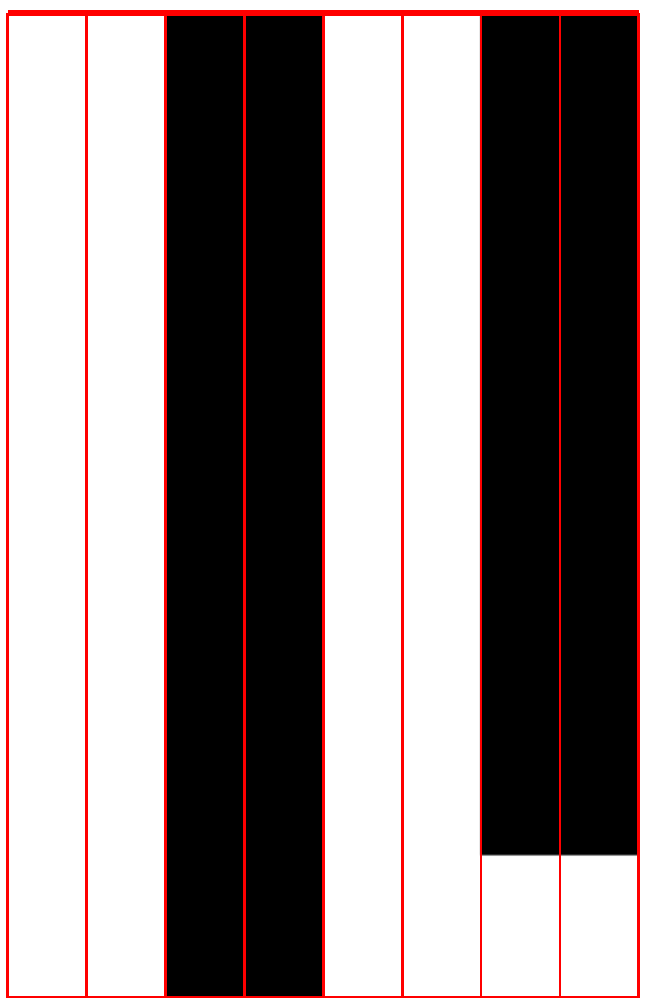}\hspace{\tinySpacing}
    \includegraphics[trim=7.5cm 13.5cm 7.5cm 4.3cm, clip=true, width=\tinyWidth]{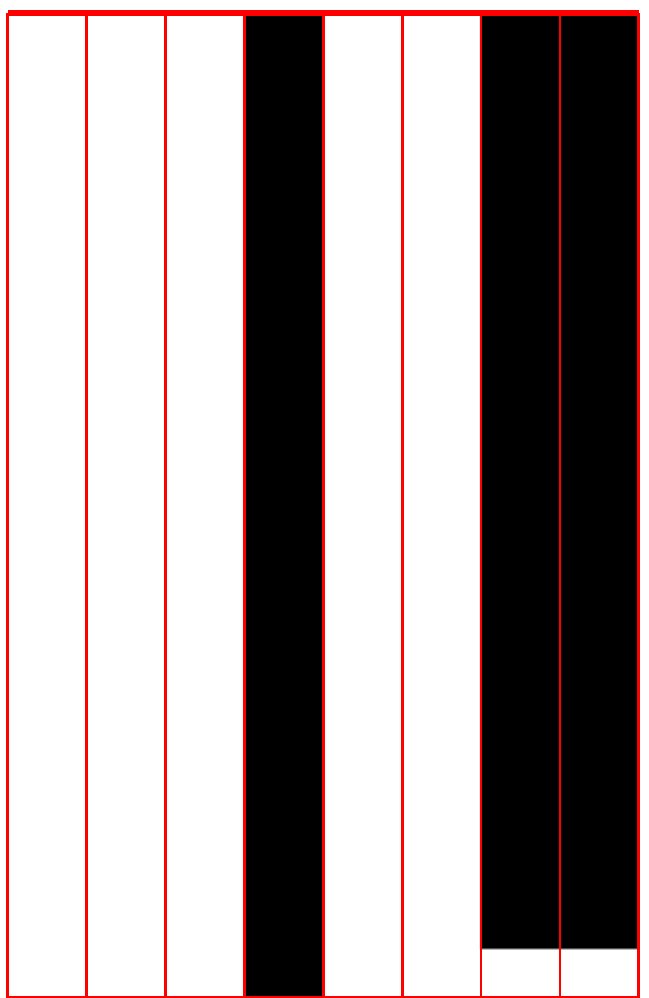}\hspace{\tinySpacing}
    \includegraphics[trim=7.5cm 13.5cm 7.5cm 4.3cm, clip=true, width=\tinyWidth]{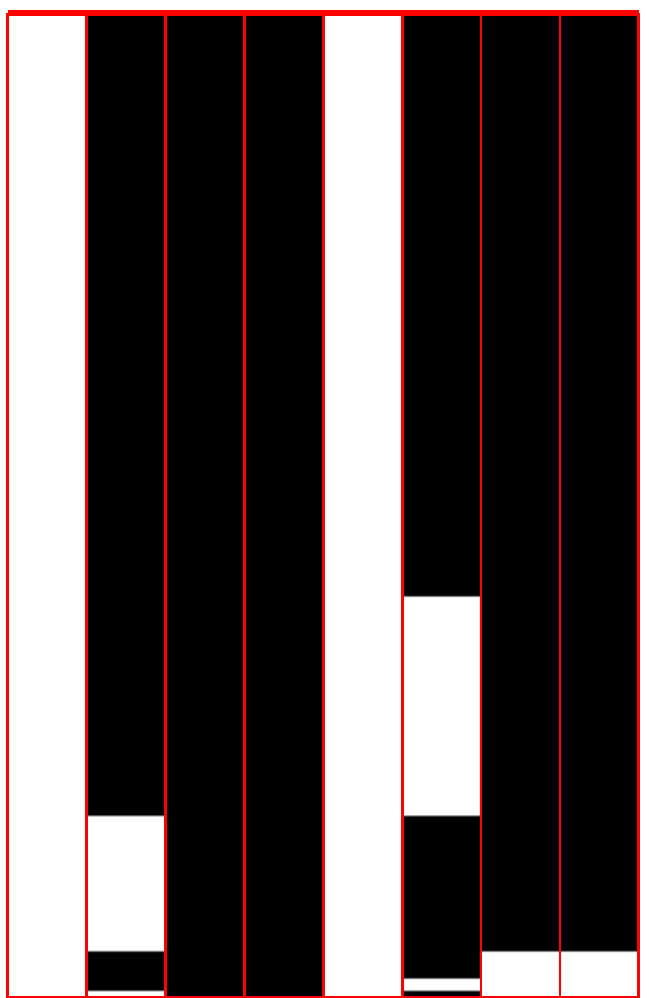}\hspace{\tinySpacing}
    \includegraphics[trim=7.5cm 13.5cm 7.5cm 4.3cm, clip=true, width=\tinyWidth]{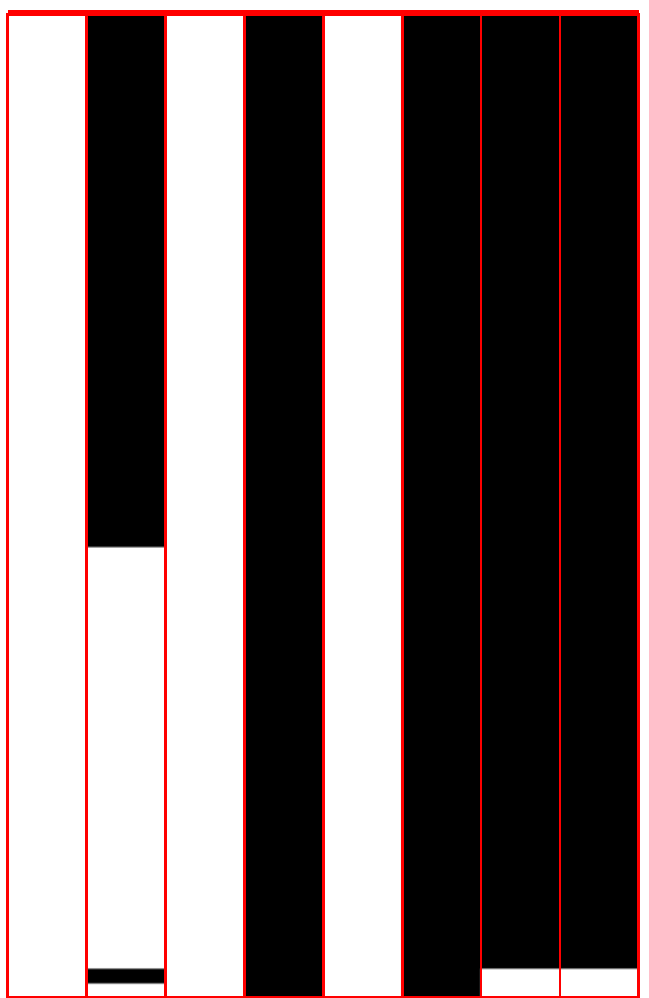}\hspace{\tinySpacing}
    \includegraphics[trim=7.5cm 13.5cm 7.5cm 4.3cm, clip=true, width=\tinyWidth]{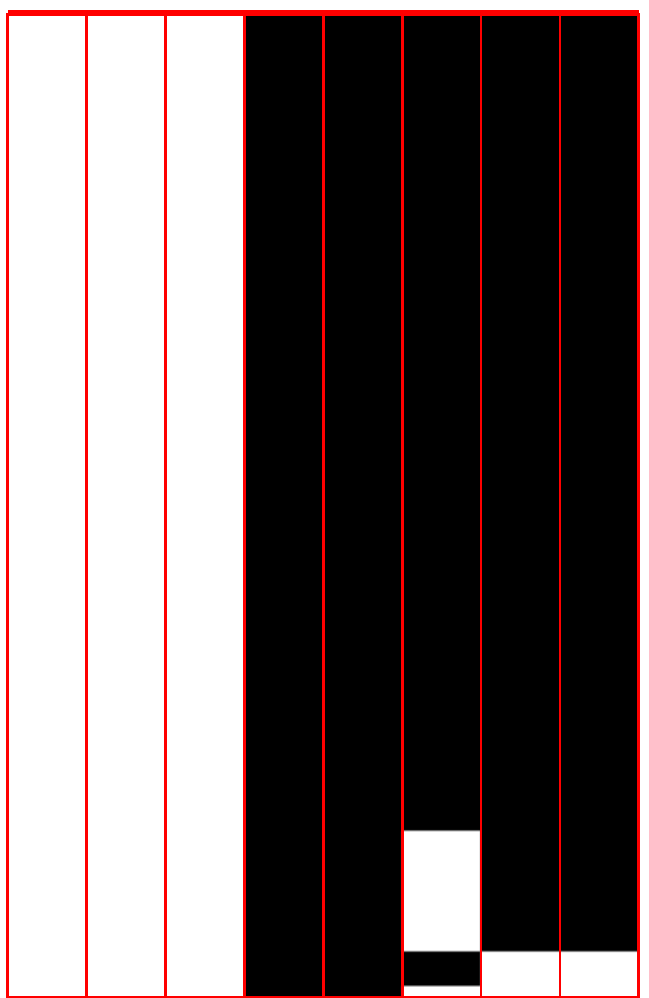}\\\vspace{\tinySpacing}

    \includegraphics[trim=7.5cm 13.5cm 7.5cm 4.3cm, clip=true, width=\tinyWidth]{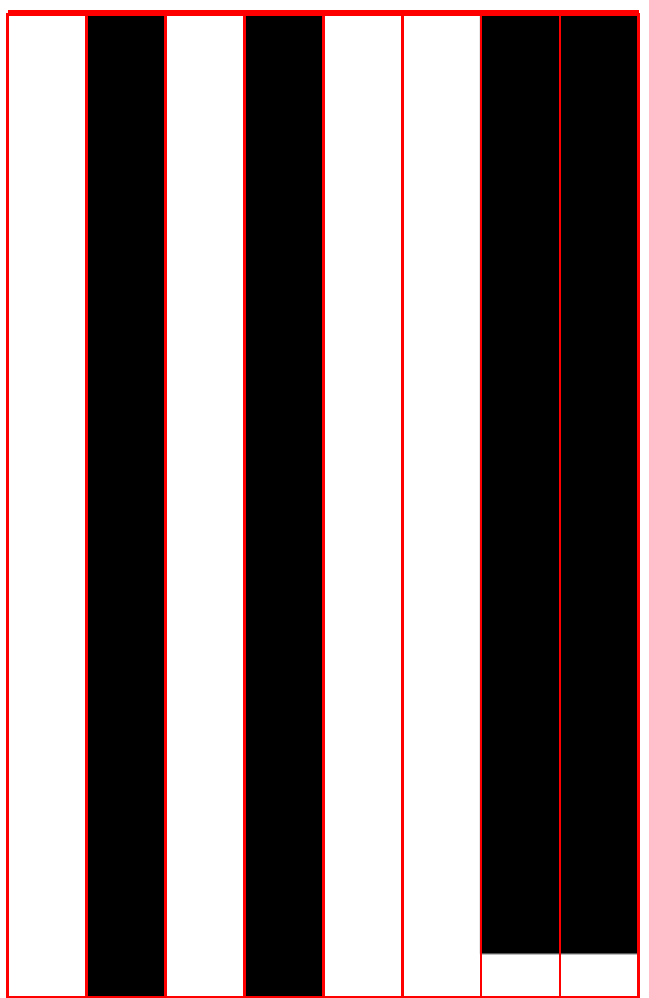}\hspace{\tinySpacing}
    \includegraphics[trim=7.5cm 13.5cm 7.5cm 4.3cm, clip=true, width=\tinyWidth]{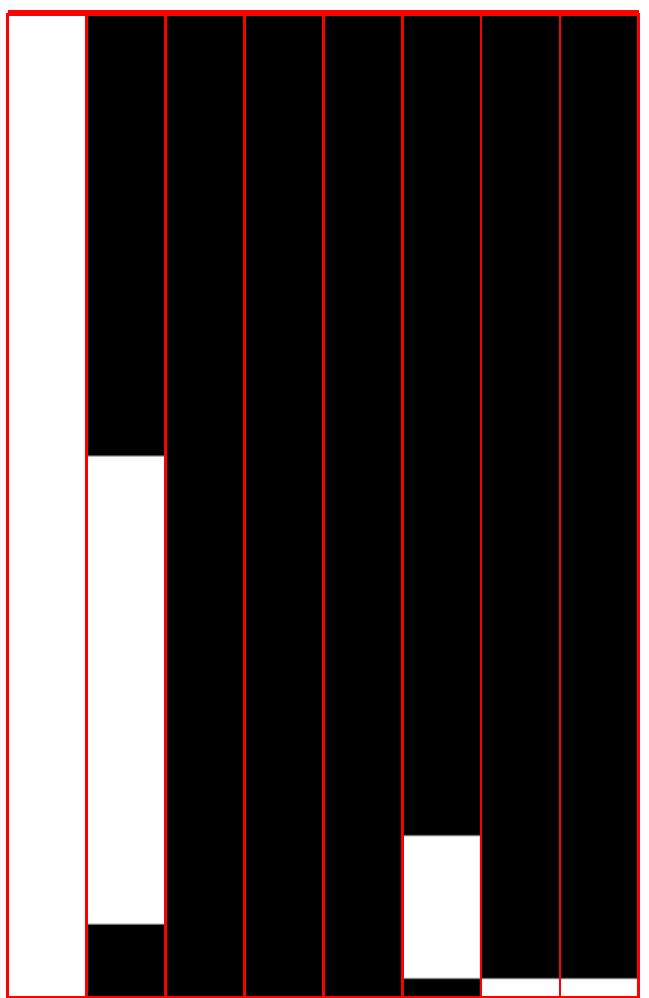}\hspace{\tinySpacing}
    \includegraphics[trim=7.5cm 13.5cm 7.5cm 4.3cm, clip=true, width=\tinyWidth]{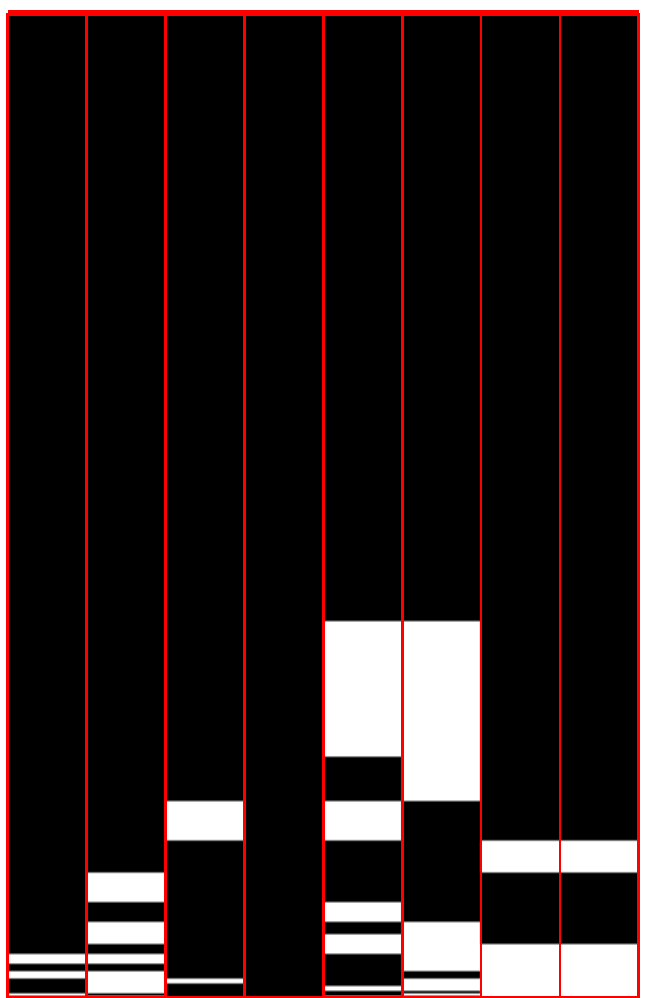}\hspace{\tinySpacing}
    \includegraphics[trim=7.5cm 13.5cm 7.5cm 4.3cm, clip=true, width=\tinyWidth]{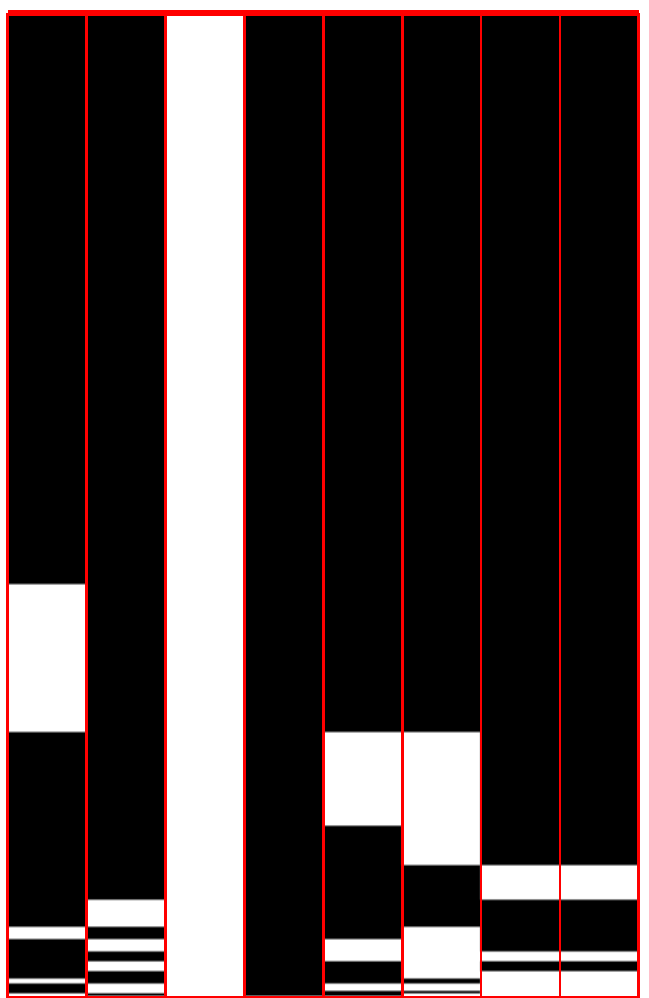}
    \caption{DBpedia Persons split into $k = 9$ implicit sorts, using the $\sigma_{\Cov}$ function. The threshold of this sort refinement is $\theta = 0.9$, (i.e.~every sort $D_i$ has $\sigma_{\Cov}(D_i) \geq 0.9$. The sizes of the sorts range from 260,585 subjects (the second sort) to 10,748 subjects (the seventh sort).}
  \label{fig:dbpedia_persons_theta09_bestK_cov}\end{subfigure}

  \begin{subfigure}[b]{\columnwidth}
    \centering
    \includegraphics[trim=7.5cm 13.5cm 7.5cm 3cm, clip=true, width=\tinyWidth]{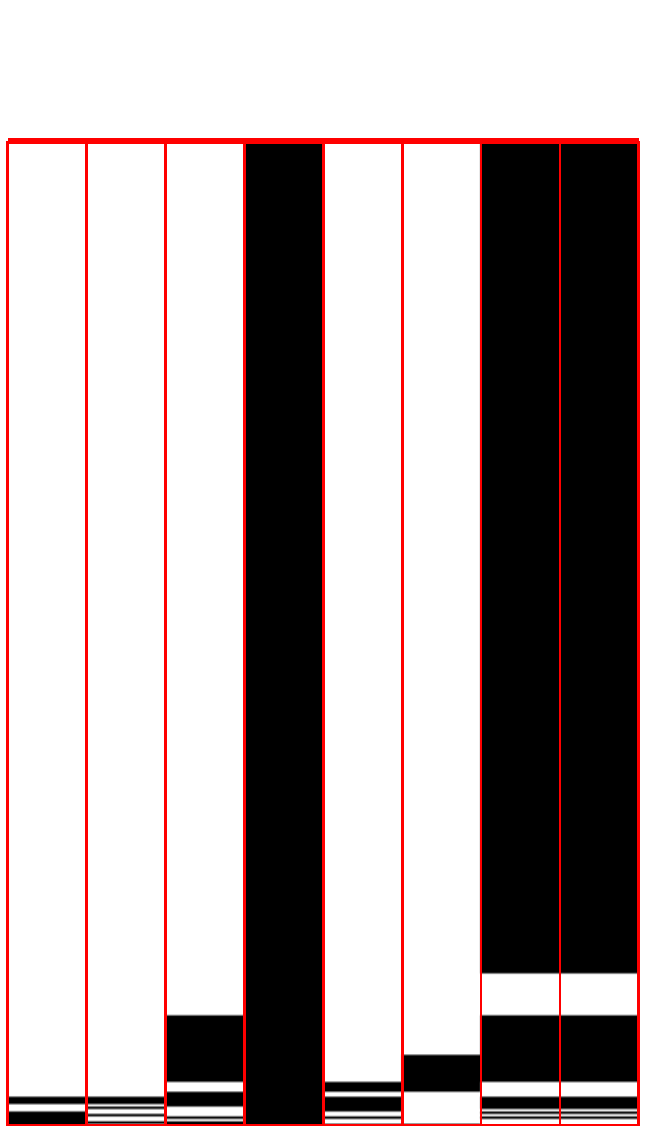}\hspace{\tinySpacing}
    \includegraphics[trim=7.5cm 13.5cm 7.5cm 3cm, clip=true, width=\tinyWidth]{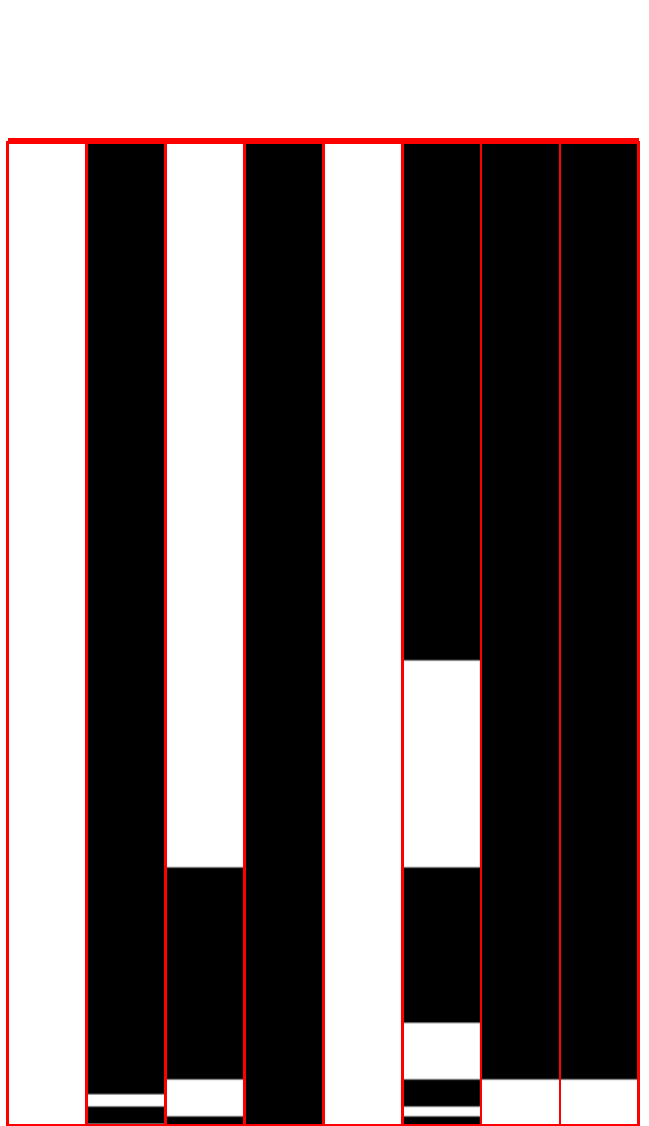}\hspace{\tinySpacing}
    \includegraphics[trim=7.5cm 13.5cm 7.5cm 3cm, clip=true, width=\tinyWidth]{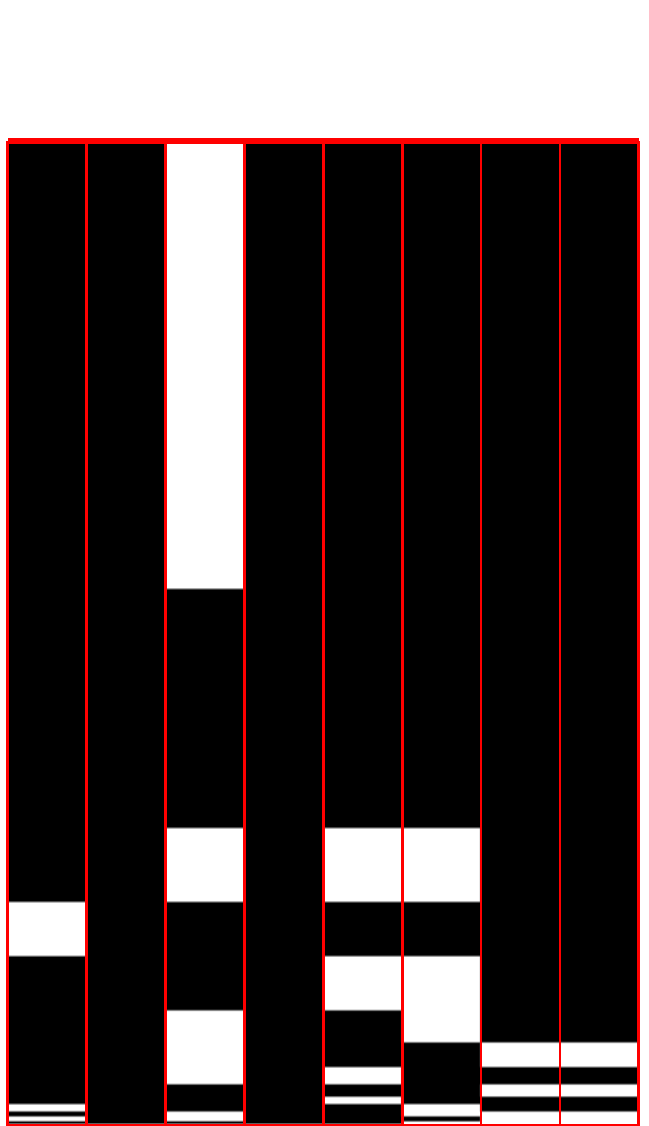}\hspace{\tinySpacing}
    \includegraphics[trim=7.5cm 13.5cm 7.5cm 3cm, clip=true, width=\tinyWidth]{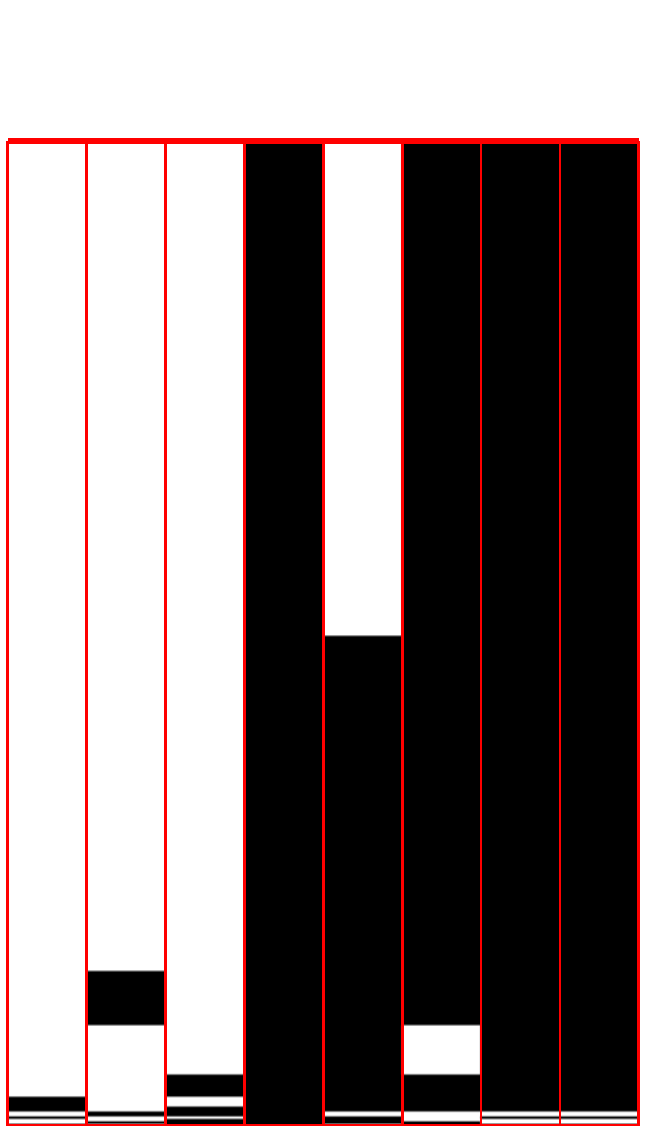}

    \caption{A $k = 4$ $\sigma_{\Sim}$-sort refinement with threshold 0.9 for DBpedia Persons. The sizes of the sorts range from 292,880 subjects (the first sort) to 87,117 subjects (the third sort).}
  \label{fig:dbpedia_persons_theta09_bestK_sim}\end{subfigure}

  \caption{DBpedia Persons split into the lowest $k$ such that the threshold is $\theta = 0.9$, using the structuredness functions (a) $\sigma_{\Cov}$, and (b) $\sigma_{\Sim}$.}
\label{fig:dbpedia_persons_theta09_bestK}\end{figure}

\subsubsection{Dependency functions in DBpedia Persons}

We now turn our attention to the dependency functions. In terms of creating a
new sort refinement using the  function $\sigma_{\Dep}[\mathbf{p}_1,
\mathbf{p}_2]$, for any constants $\mathbf{p}_1, \mathbf{p}_2 \in \U$, we can
generate a sort refinement with $\theta = 1.0$ for $k = 2$, consisting of the
following two sorts: (i) all entities which do not have $\mathbf{p}_1$, and
(ii) all entities which do have $\mathbf{p}_2$.  The sort (i) will have
structuredness 1.0 because there are no assignments that satisfy the antecedent
(no assigments satisfy $\prop(c_2) = \mathbf{p}_1$), and sort (ii) has
structuredness 1.0 because every assigment which satisfies the antecedent will
also satisfy the consequent ($\val(c_2) = 1$ because all entities have
$\mathbf{p}_2$). On the other hand, $\sigma_{\SymDep}$ with constants
$\mathbf{p}_1, \mathbf{p}_2 \in \U$ can generate an sort refinement with
$\theta = 1.0$ for $k = 3$, consisting of the following three sorts: (i)
entities which have $\mathbf{p}_1$ but not $\mathbf{p}_2$, (ii) entities which
have $\mathbf{p}_2$ but not $\mathbf{p}_1$, and (iii) entites which have both
$\mathbf{p}_1$ and $\mathbf{p}_2$ or have neither. The first two sorts will not
have any total cases, and for the third sort every total case is also a
favorable case.

The dependency functions, as shown, are not very well suited to the task of
finding the lowest $k$ such that the threshold $\theta$ is met, which is why
these functions were not included in the previous results. The dependency
functions are useful, however, for characterizing an RDF graph or a sort
refinement which was generated with a different structuredness function, such
as $\sigma_{\Cov}$ or $\sigma_{\Sim}$, since they can help analyze the
relationship between the properties in an RDF graph. To illustrate, we consider
the $\sigma_{\Dep}$[$\mathbf{p}_1$, $\mathbf{p}_2$] function, and we tabulate
(in Table.~\ref{table:dbpedia_persons_dep_table}) the structuredness value of
DBpedia Persons when replacing the parameters $\mathbf{p}_1$ and $\mathbf{p}_2$
by all possible combinations of {\small \texttt{deathPlace}}, {\small
\texttt{birthPlace}}, {\small \texttt{deathDate}}, and {\small
\texttt{birthDate}}. Recall that $\sigma_{\Dep}$ with parameters $\mathbf{p}_1
=$ {\small \texttt{death\-Place}} and $\mathbf{p}_2 =$ {\small \texttt{birthPlace}} measures the
probability that a subject which has {\small \texttt{deathPlace}} also has
{\small \texttt{birthPlace}}.

\begin{table}[t]
  \centering
  {\small
  \begin{tabular}{ | r || c | c | c | c | }
    \hline
                        & \texttt{dP} & \texttt{bP} & \texttt{dD} & \texttt{bD} \\\hline\hline
    \texttt{deathPlace} & 1.0         & .93         & .82         & .77 \\
    \texttt{birthPlace} & .26         & 1.0         & .27         & .75 \\
    \texttt{deathDate}  & .43         & .50         & 1.0         & .89 \\
    \texttt{birthDate}  & .17         & .57         & .37         & 1.0 \\
  \hline
  \end{tabular}}
  \caption{DBpedia Persons structuredness according to $\sigma_{\Dep}$ with different combinations of parameters $\mathbf{p}_1$ and $\mathbf{p}_2$. The property names are abbreviated in the column headers.}
\label{table:dbpedia_persons_dep_table}
\vspace*{-2.0ex}
\end{table}

The table reveals a very surprising aspect of the dataset. Namely, the first
row shows high structuredness values when $\mathbf{p}_1 =$ {\small
\texttt{death\-Place}}.  This implies that if we somehow know the {\small
\texttt{deathPlace}} for a particular person, there is a very high probability
that we also know all the other properties for her. Or, to put it another way,
knowing the death place of a person implies that we know a lot about the
person. This is also an indication that it is somehow the hardest fact to
acquire, or the fact that is least known among persons in DBpedia. Notice that
none of the other rows have a similar characteristic. For example, in the
second row we see that given the {\small \texttt{birthPlace}} of a person there
is a small chance (0.27) that we know her {\small \texttt{deathDate}}.
Similarly, given the {\small \texttt{deathDate}} of a person there is only a
small chance (0.43) that we know the {\small \texttt{deathPlace}}.

We can do a similar analysis with the $\sigma_{\SymDep}$[$\mathbf{p}_1$,
$\mathbf{p}_2$] function. In Table~\ref{table:dbpedia_persons_symdep_table} we
show the pairs of properties with the highest and lowest values of
$\sigma_{\SymDep}$. Given that the {\small \texttt{name}} property in DBpedia
persons is the only property that every subject has, one would expect that the
most correlated pair of properties would include {\small \texttt{name}}.
Surprisingly, this is not the case. Properties {\small \texttt{givenName}} and
{\small \texttt{surName}} are actually the most correlated properties, probably
stemming from the fact that these two properties are extracted from the same
source. The least correlated properties all involve {\small
\texttt{deathPlace}} and the properties of {\small \texttt{name}}, {\small
\texttt{givenName}} and {\small \texttt{surName}}.

\begin{table}[t]
  \centering
  {\small
  \begin{tabular}{ | c | c | c | }
    \hline
    $\mathbf{p}_1$     & $\mathbf{p}_2$   & $\sigma_{\SymDep}$ \\\hline
    \texttt{givenName} & \texttt{surName}     & 1.0 \\
    \texttt{name} & \texttt{givenName}        & .95 \\
    \texttt{name} & \texttt{surName}          & .95 \\
    \texttt{name} & \texttt{birthDate}        & .53 \\
    $\ldots$      & $\ldots$                  & $\ldots$ \\
    \texttt{description} & \texttt{givenName} & .14 \\
    \texttt{deathPlace} & \texttt{name}       & .11 \\
    \texttt{deathPlace} & \texttt{givenName}  & .11 \\
    \texttt{deathPlace} & \texttt{surName}    & .11 \\
  \hline
  \end{tabular}
  }
  \caption{A ranking of DBpedia Persons structuredness according to $\sigma_{\SymDep}$ with different combinations of the 8 properties in $P(D_{\text{DBpedia Persons}})$. Only the highest and lowest entries are shown.}
\label{table:dbpedia_persons_symdep_table}\end{table}

%================================================================
%  WORDNET NOUNS
%================================================================
\subsection{For WordNet Nouns}

WordNet is a lexical database for the english language. WordNet Nouns refers to
the following subgraph (where $\mathsf{Noun}$ stands for {\small
\texttt{http://www.w3.org/2006/03/wn/wn20/schema/NounSynset}}):
\begin{align*}
D_{\text{WordNet Nouns}} = \{ &(s,p,o) \in D_{\text{WordNet}} \;\mid\;\\
& (s, \type, \mathsf{Noun}) \in D_{\text{WordNet}} \}.
\end{align*}

This dataset is 101 MB in size, and contains 416,338 triples, 79,689 subjects,
and 12 properties (excluding the $\type$ property). Its signature
representation consists of 53 signatures, stored in 3 KB. The
properties are the following: {\small \texttt{gloss}}, {\small \texttt{label}},
{\small \texttt{synsetId}}, {\small \texttt{hyponymOf}}, {\small
\texttt{classifiedByTopic}}, {\small \texttt{containsWordSense}}, \\
{\small \texttt{memberMeronymOf}}, {\small \texttt{partMeronymOf}}, {\small
\texttt{substanceMeronymOf}}, \\ {\small \texttt{classifiedByUsage}}, {\small
\texttt{classifiedByRegion}}, and {\small \texttt{attribute}}.

For this sort, $\sigma_{\Cov} = 0.44$, and $\sigma_{\Sim} = 0.93$. There is a
significant difference in the structuredness of WordNet Nouns as measured by
the two functions. This difference is clearly visible in the signature view of
this dataset (fig.~\ref{fig:full_sigs_b}); the presence of nearly empty
properties (i.e.~properties which relatively few subjects have) is highly
penalized by the $\Cov$ rule, though mostly ignored by $\Sim$.

%================================================================
%  Subsubsection: wordnet k2
%================================================================
\subsubsection{A highest $\theta$ sort refinement for $k = 2$}

As mentioned, the WordNet case proves to be very different from the previous, partly because this dataset has roughly 5 dominant signatures representing a most subjects, and yet only using 8 of the 12 properties, causing difficulties when partitioning the dataset.

\begin{figure}[t]
\centering
  \begin{subfigure}[b]{\columnwidth}
    \centering
    \includegraphics[trim=5cm 10.6cm 5.5cm 4.2cm, clip=true, width=4cm]{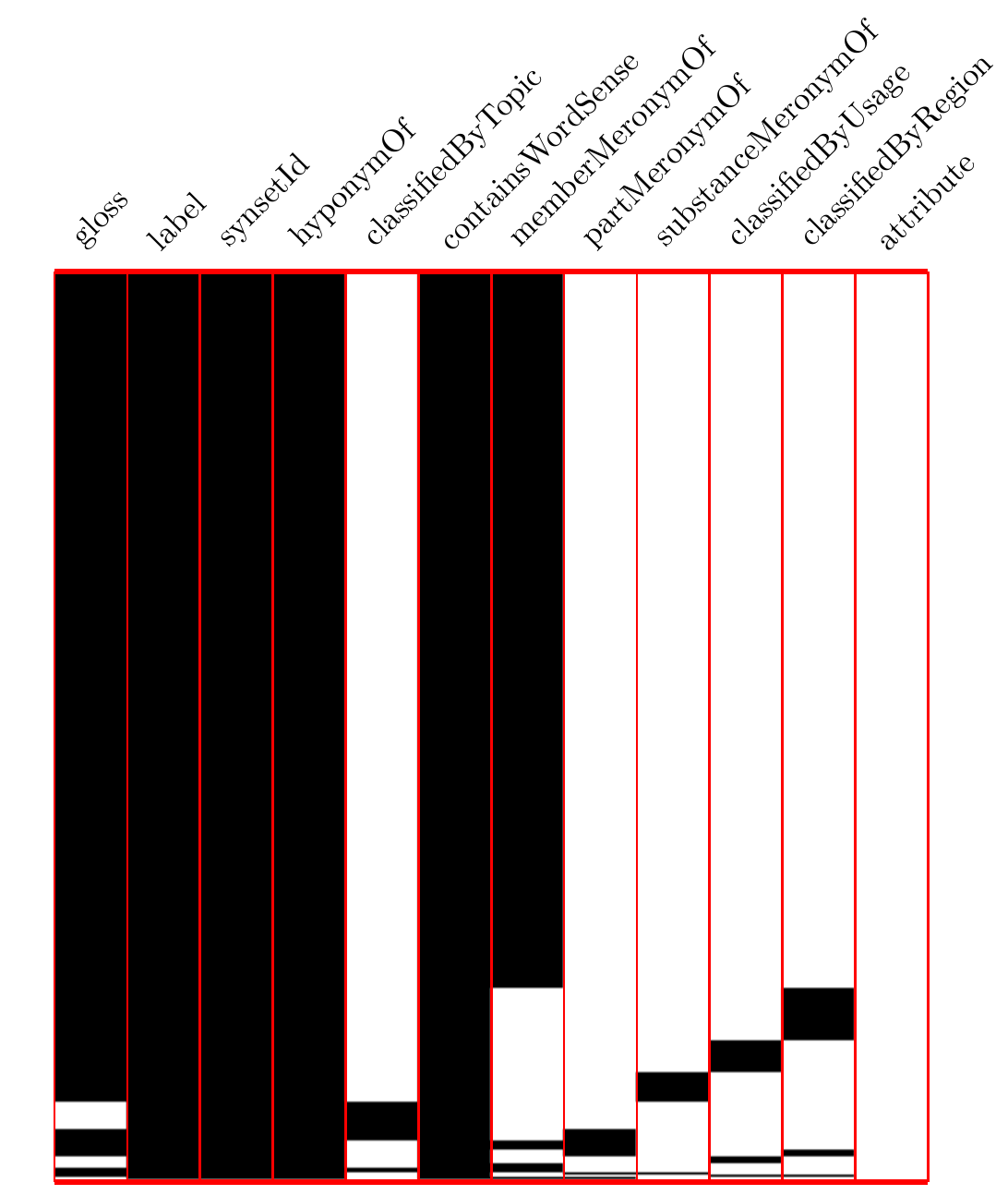}
    \includegraphics[trim=5cm 10.6cm 5.5cm 4.2cm, clip=true, width=4cm]{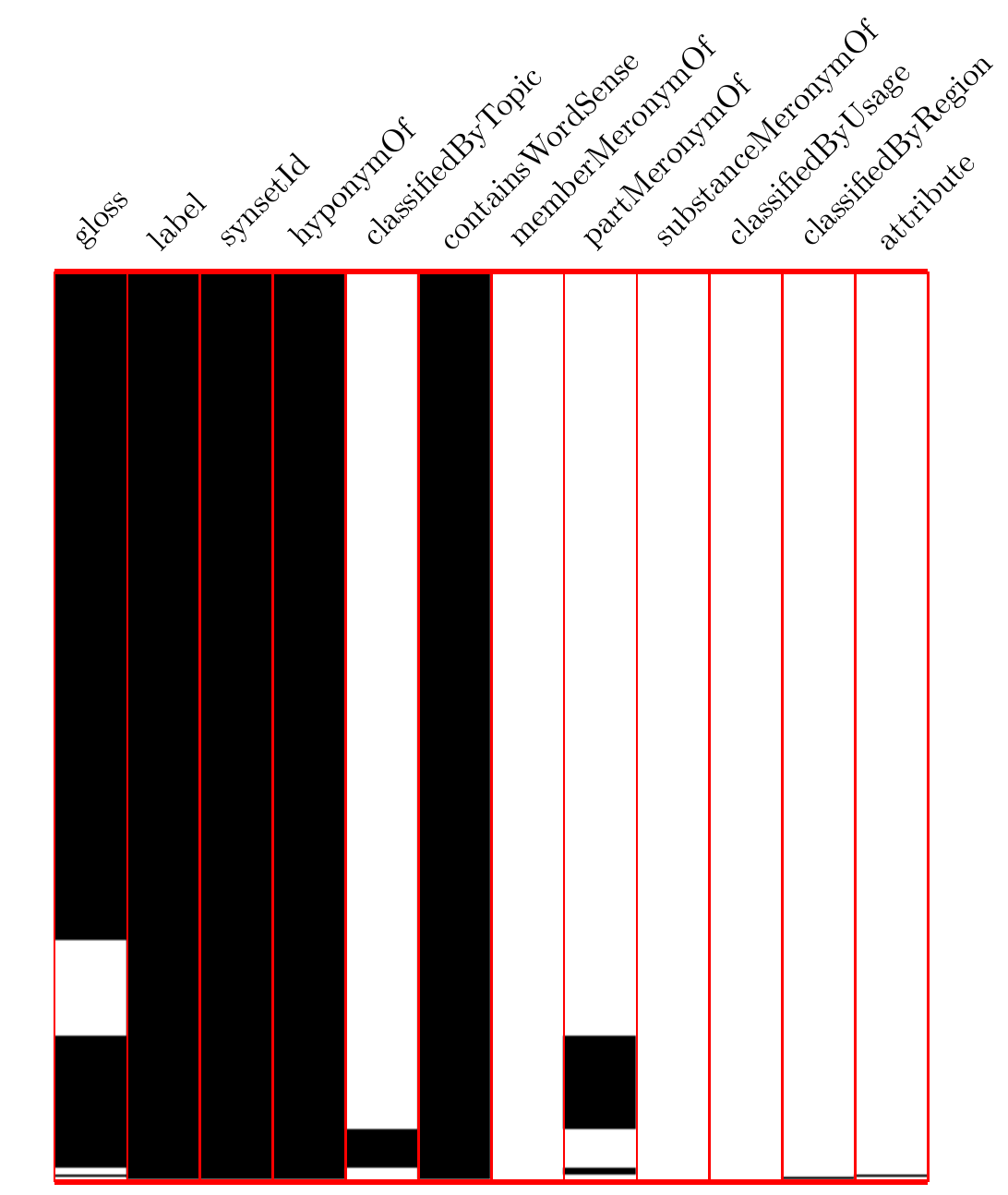}
    \caption{Using the $\sigma_{\Cov}$ function, the left sort has 14,938 subjects and 35 signatures, $\sigma_{\Cov} = 0.55$, $\sigma_{\Sim} = 0.93$. The right sort has 64,751 subjects and 18 signatures, $\sigma_{\Cov} = 0.56$, $\sigma_{\Sim} = 0.95$.}
  \label{fig:wordnet_nouns_k2_bestTheta_cov}\end{subfigure}

  \begin{subfigure}[b]{\columnwidth}
    \centering
    \includegraphics[trim=5cm 13.2cm 5.5cm 4cm, clip=true, width=4cm]{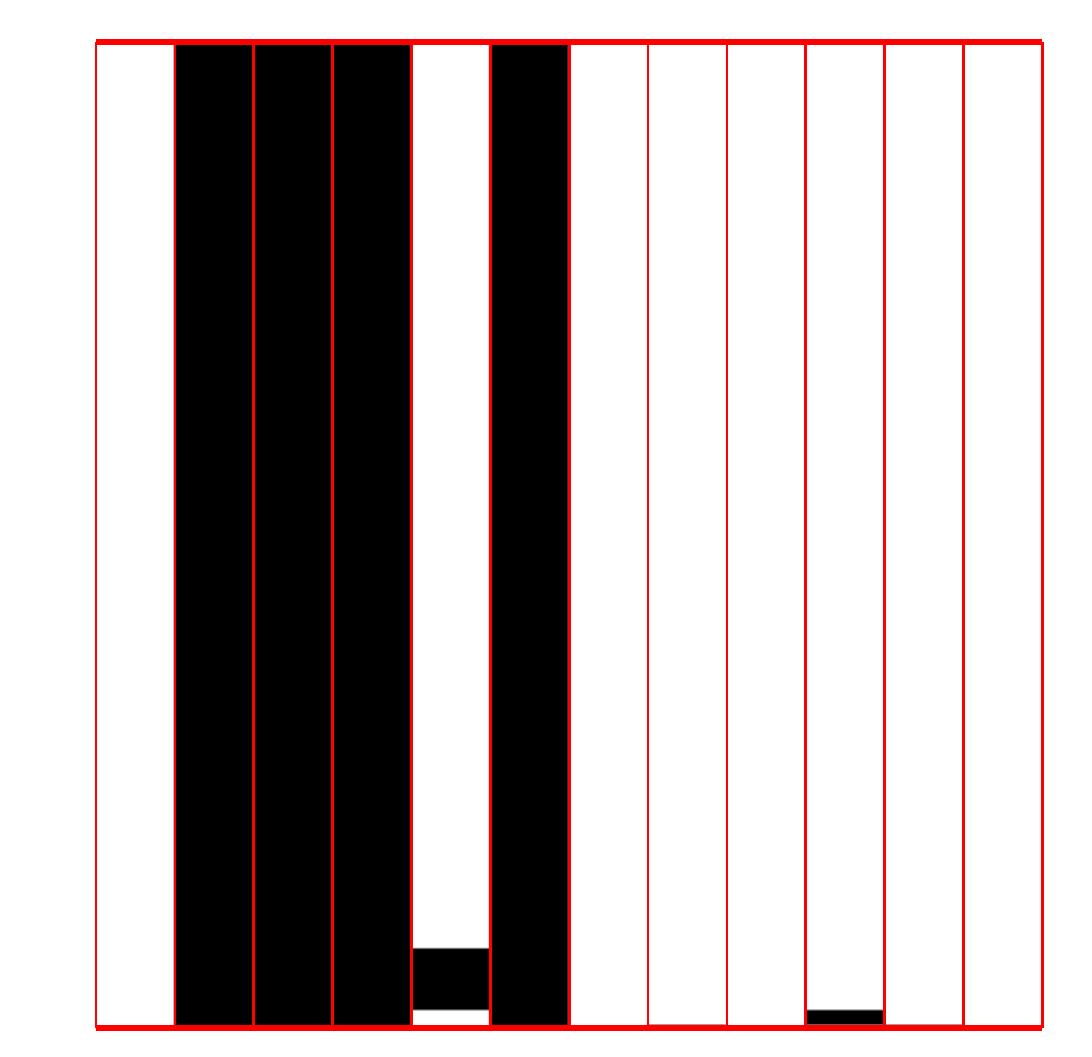}
    \includegraphics[trim=5cm 13.2cm 5.5cm 4cm, clip=true, width=4cm]{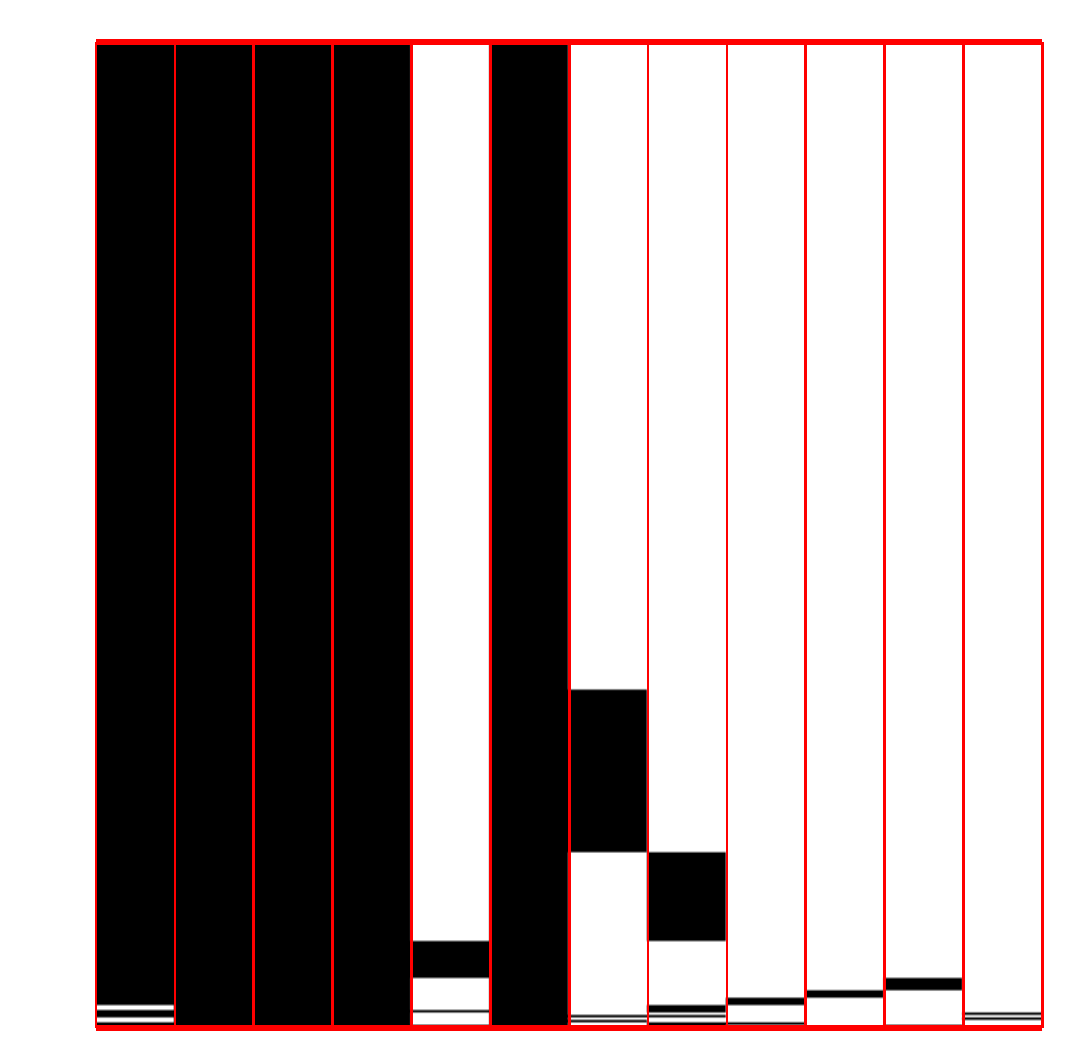}
    \caption{Using the $\sigma_{\Sim}$ function, the left sort has 7,311 subjects and 13 signatures, $\sigma_{\Cov} = 0.34$, and $\sigma_{\Sim} = 0.98$. The right sort has 72,378 subjects and 40 signatures, $\sigma_{\Cov} = 0.45$, and $\sigma_{\Sim} = 0.94$. } 
    %Note: This timed out at 1 hr. Has log file!}
  \label{fig:wordnet_nouns_k2_bestTheta_sim}\end{subfigure}

  \caption{WordNet Nouns split into $k = 2$ implicit sorts, using two different structuredness functions: (a) $\sigma_{\Cov}$, and (b) $\sigma_{\Sim}$.}

\label{fig:wordnet_nouns_k2_bestTheta}\end{figure}

Figure \ref{fig:wordnet_nouns_k2_bestTheta_cov} shows the result for $\sigma_{\Cov}$. The largest difference between both sorts is that the left sort mostly consists of subjects which have the {\small \texttt{memberMeronymOf}} property (the 7th property). The improvement in the structuredness of these two sorts is very small in comparison to the original dataset (from 0.44 to 0.55), suggesting that $k = 2$ is not enough to discriminate sub-sorts in this dataset, and with this rule. This is due to the presence of many of signatures which represent very few subjects, and have different sets of properties. Here, all ILP instances were solved in under 1 s.

Figure \ref{fig:wordnet_nouns_k2_bestTheta_sim} shows the result for
$\sigma_{\Sim}$. The difference between the two sorts is {\small
\texttt{gloss}}, which is absent in the left sort. The placement of the smaller
signatures does not seem to follow any pattern, as the $\Sim$ function is
not sensitive to their presence. Although the structuredness is high in this
case, the improvement is small, since the original dataset is
highly structured with respect to $\sigma_{\Sim}$ anyway. A discussion is in
order with respect to the running times.  Recall that the ILP instances are
solved for increasing values of $\theta$ (the increment being $0.01$). For all
values of $\theta$ lower than 0.95 each ILP instance is solved in less than 5
s. For the value $\theta = 0.95$ however (the first value for which there
is no solution), after 75 hours of running time, the ILP solver was not able to
find a solution or prove the system infeasible. Although there is an enourmous
asymmetry between the ease of finding a solution and the difficulty of proving
an instance infeasible, in every instance a higher threshold solution is found,
in which case it is reasonable to let the user specify a maximum running time
and keep the best solution found so far.

%================================================================
%  Figure: wordnet k2
%================================================================

%================================================================
%  Subsubsection: wordnet theta
%================================================================
\subsubsection{A lowest $k$ sort refinement for fixed $\theta$}

As with the previous experimental setup, Nouns proves more difficult to solve. For the $\sigma_{\Cov}$ we set the usual threshold of 0.9, however, since the structuredness value of Wordnet Nouns under the $\sigma_{\Sim}$ function is 0.93 originally, this exersize would be trivial if $\theta$ is 0.9. For that reason, in this last case we fix the threshold at 0.98.

Figure \ref{fig:wordnet_nouns_theta09_bestK_cov} shows the first 10 sorts of the $k = 31$ solution for $\sigma_{\Cov}$. The sheer amount of sorts needed is a indication that WordNet Nouns already represents a highly structured sort. The sorts in many cases correspond to individual signatures, which are the smallest sets of identically structured entities. In general, it is probably not of interest for a user or database administrator to be presented with an sort refinement with so many sorts. This setup was the longest running, at an average 7 hours running time per ILP instance. This large number is another indication of the difficulty of partitioning a dataset with highly uniform entities.

Figure \ref{fig:wordnet_nouns_theta09_bestK_sim} shows the solution for $\sigma_{\Sim}$, which is for $k = 4$. As with the $k = 2$ case, there is a sort which does not include the \texttt{gloss} property. The general pattern of this sort refinement, however, is that the four largest signatures are each placed in their own sort. Beyond that, the presence of the smaller signatures does not greatly affect the structuredness value (runtime: apx.~15 min.).

It is to be expected that a highly structured RDF graph like WordNet Nouns will not be a prime candidate for discovering refinements of the sort, which is confirmed by these experiments.

%================================================================
%  Figure: wordnet theta09
%================================================================
\newcommand{\wnWidth}{1.5cm}
\newcommand{\wnSpacing}{1.5mm}
\newcommand{\wnVSpacing}{1.0mm}
\begin{figure}[t]
\centering

  \begin{subfigure}[b]{\columnwidth}
    \centering
    \includegraphics[trim=5.8cm 13.5cm 5.9cm 4.3cm, clip=true, width=\wnWidth]{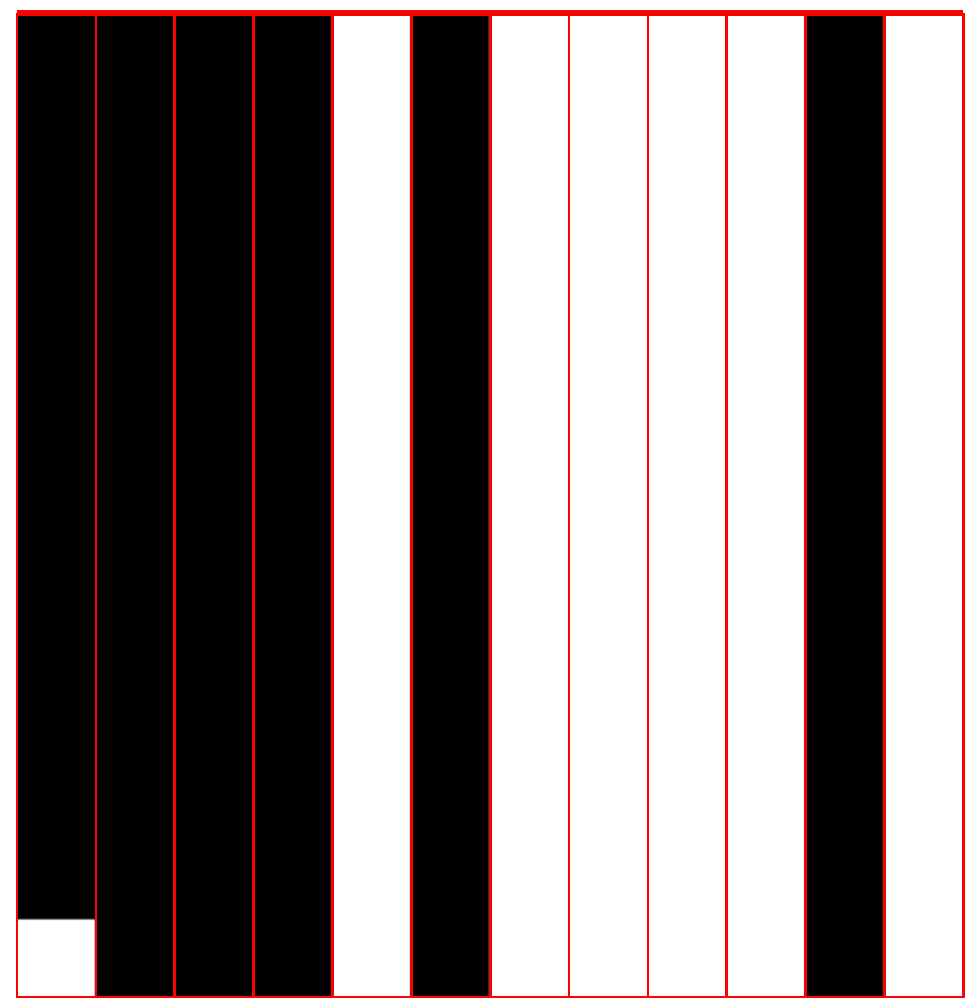}\hspace{\wnSpacing}
    \includegraphics[trim=5.8cm 13.5cm 5.9cm 4.3cm, clip=true, width=\wnWidth]{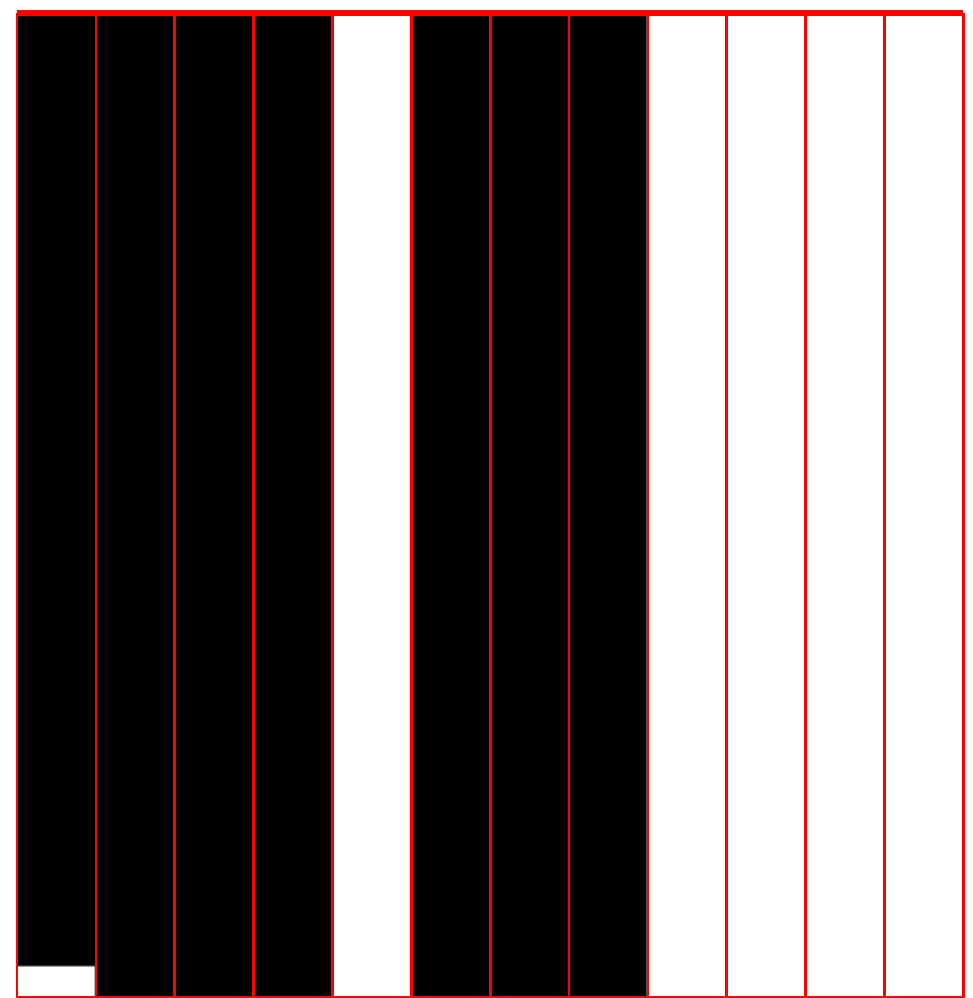}\hspace{\wnSpacing}
    \includegraphics[trim=5.8cm 13.5cm 5.9cm 4.3cm, clip=true, width=\wnWidth]{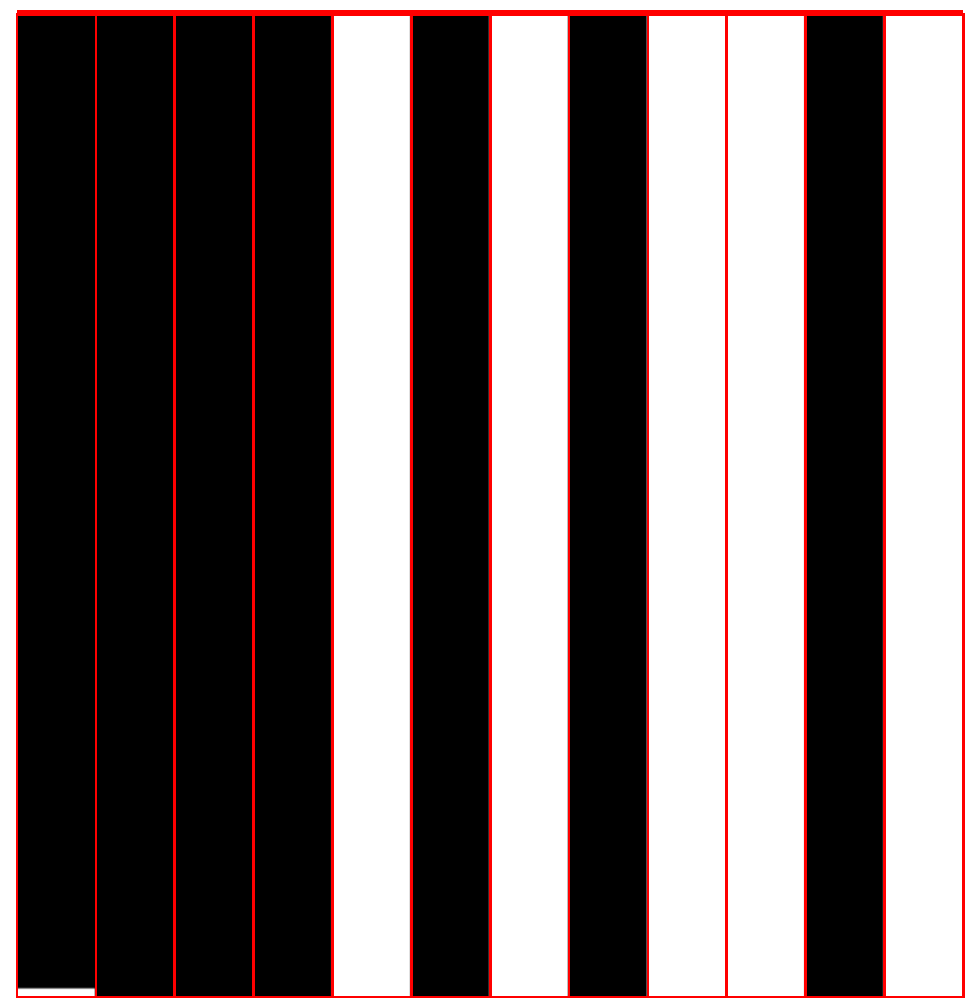}\hspace{\wnSpacing}
    \includegraphics[trim=5.8cm 13.5cm 5.9cm 4.3cm, clip=true, width=\wnWidth]{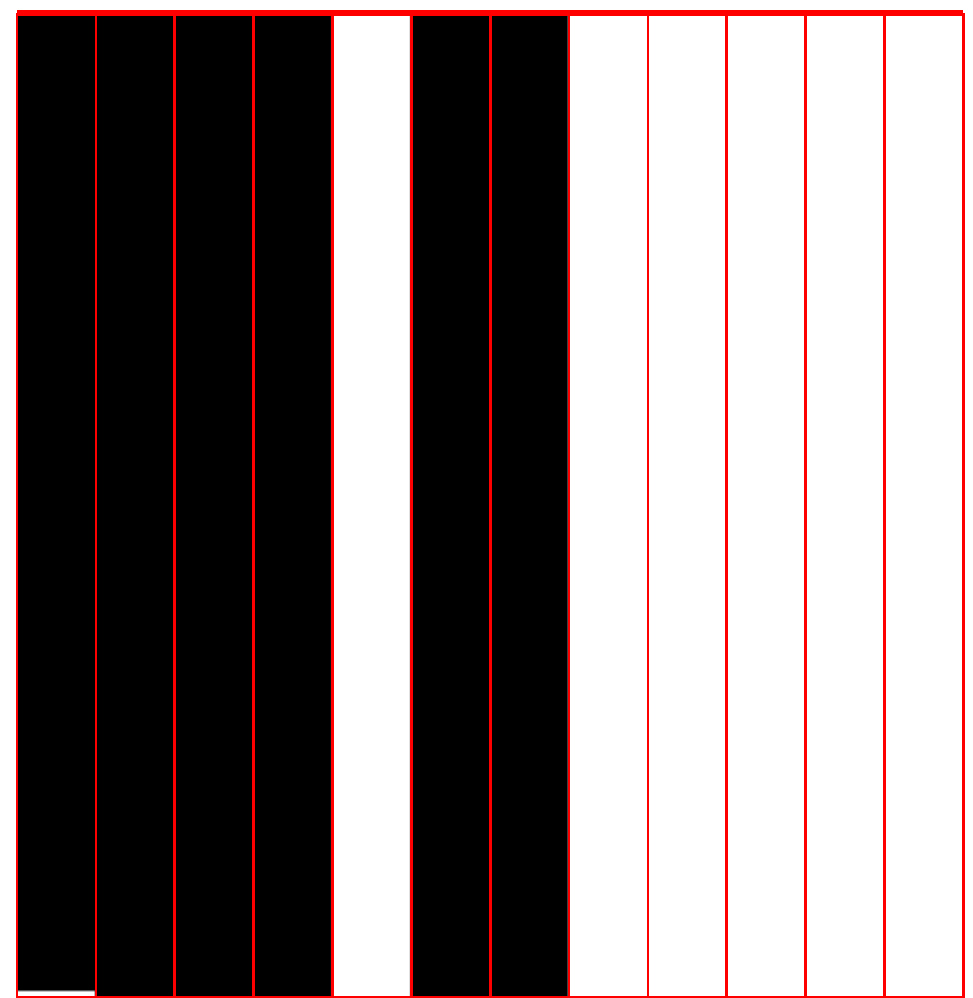}\\\vspace{\wnVSpacing}

    \includegraphics[trim=5.8cm 13.5cm 5.9cm 4.3cm, clip=true, width=\wnWidth]{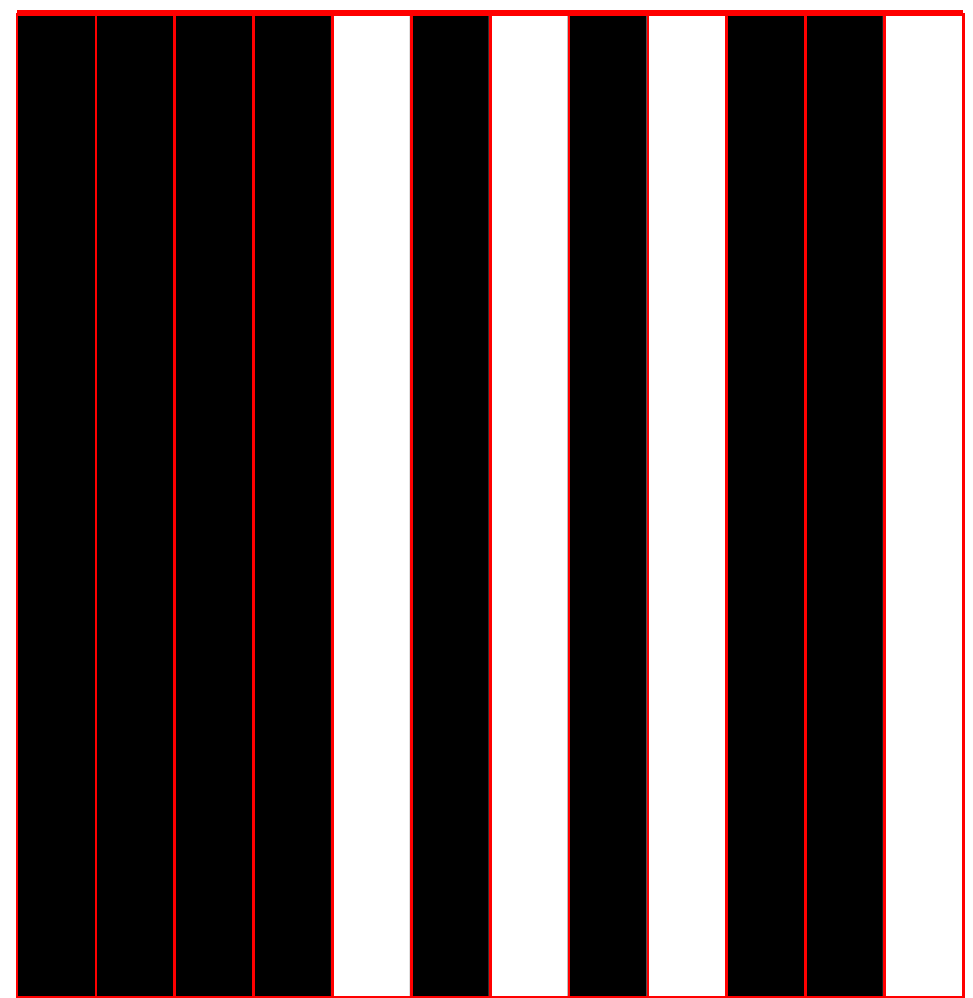}\hspace{\wnSpacing}
    \includegraphics[trim=5.8cm 13.5cm 5.9cm 4.0cm, clip=true, width=\wnWidth]{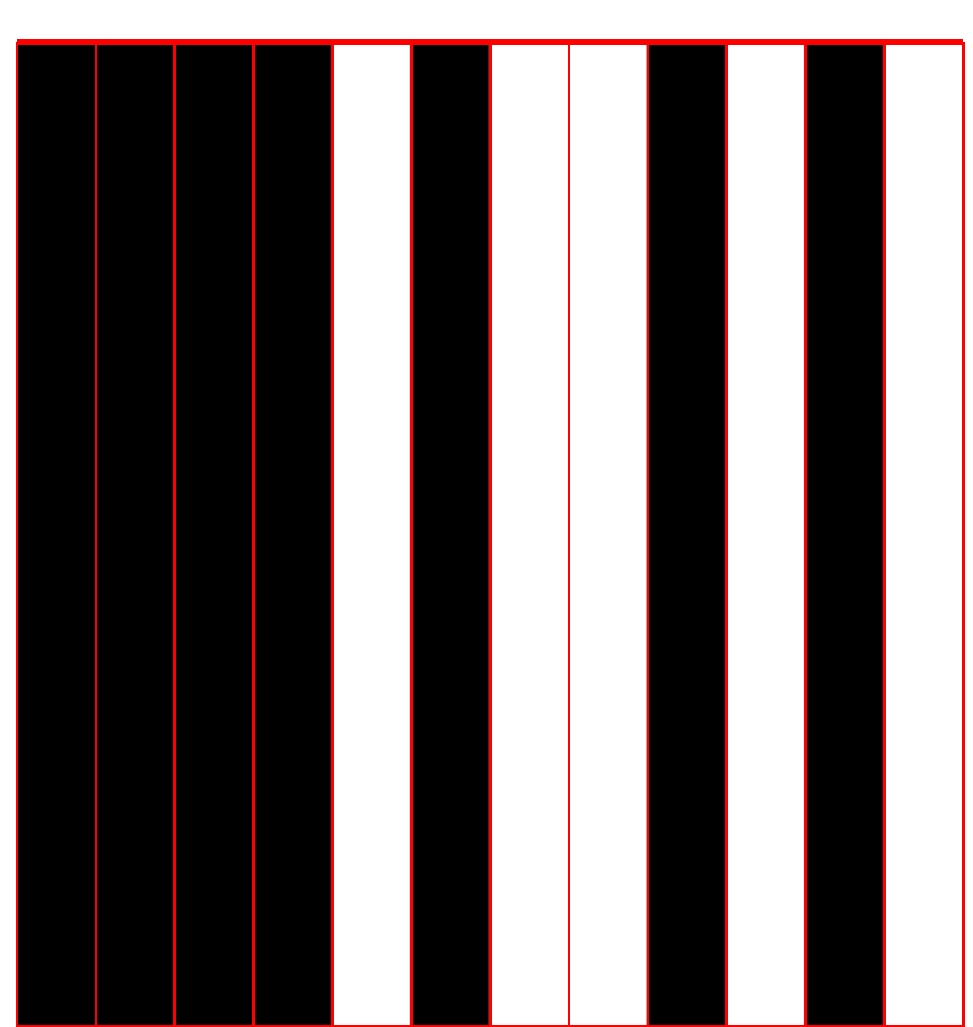}\hspace{\wnSpacing}
    \includegraphics[trim=5.8cm 13.5cm 5.9cm 4.0cm, clip=true, width=\wnWidth]{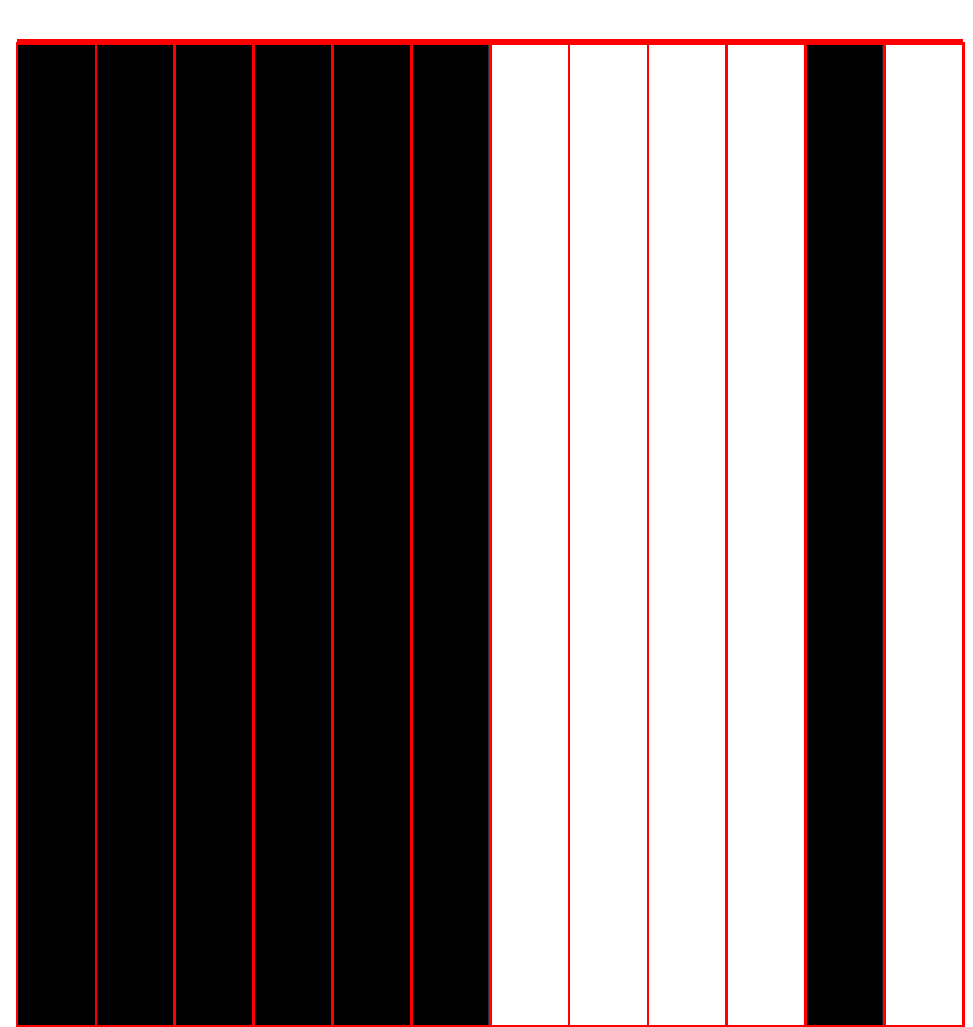}\hspace{\wnSpacing}
    \includegraphics[trim=5.8cm 13.5cm 5.9cm 4.0cm, clip=true, width=\wnWidth]{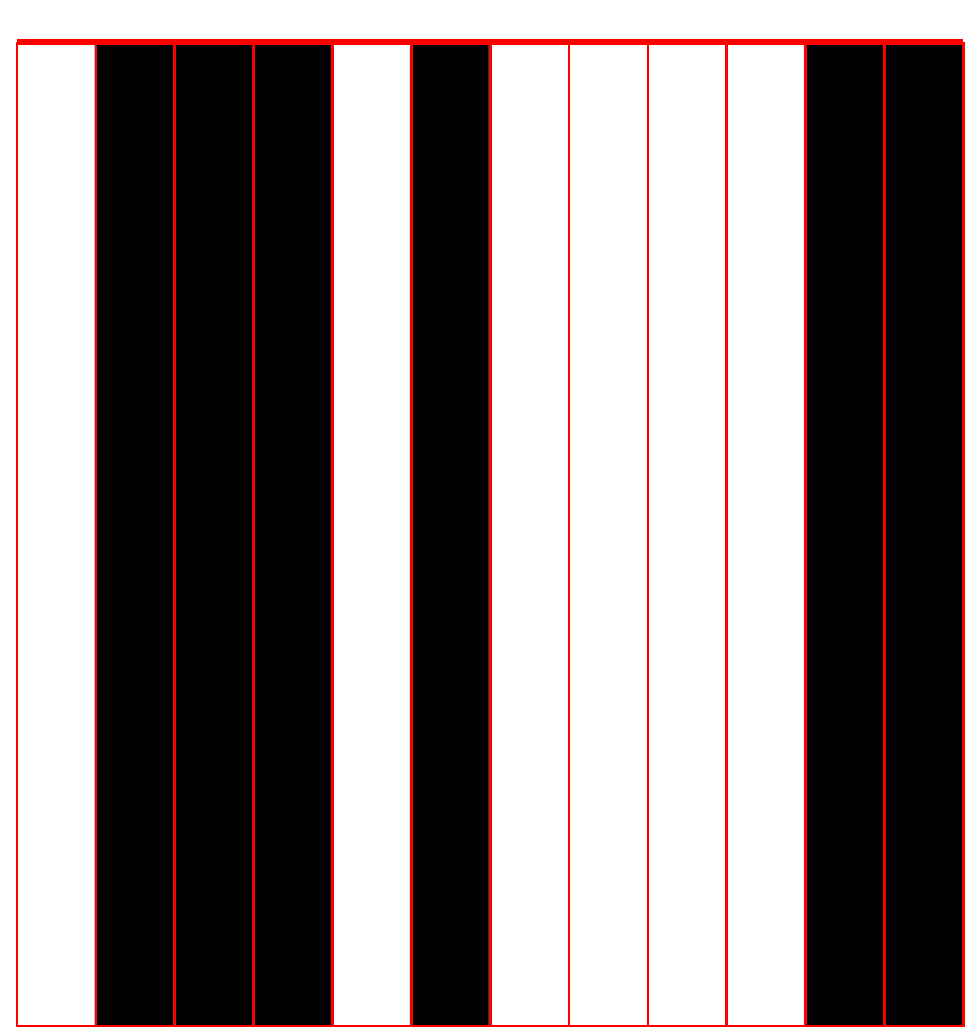}\\\vspace{\wnVSpacing}

    \includegraphics[trim=5.8cm 13.5cm 5.9cm 4.0cm, clip=true, width=\wnWidth]{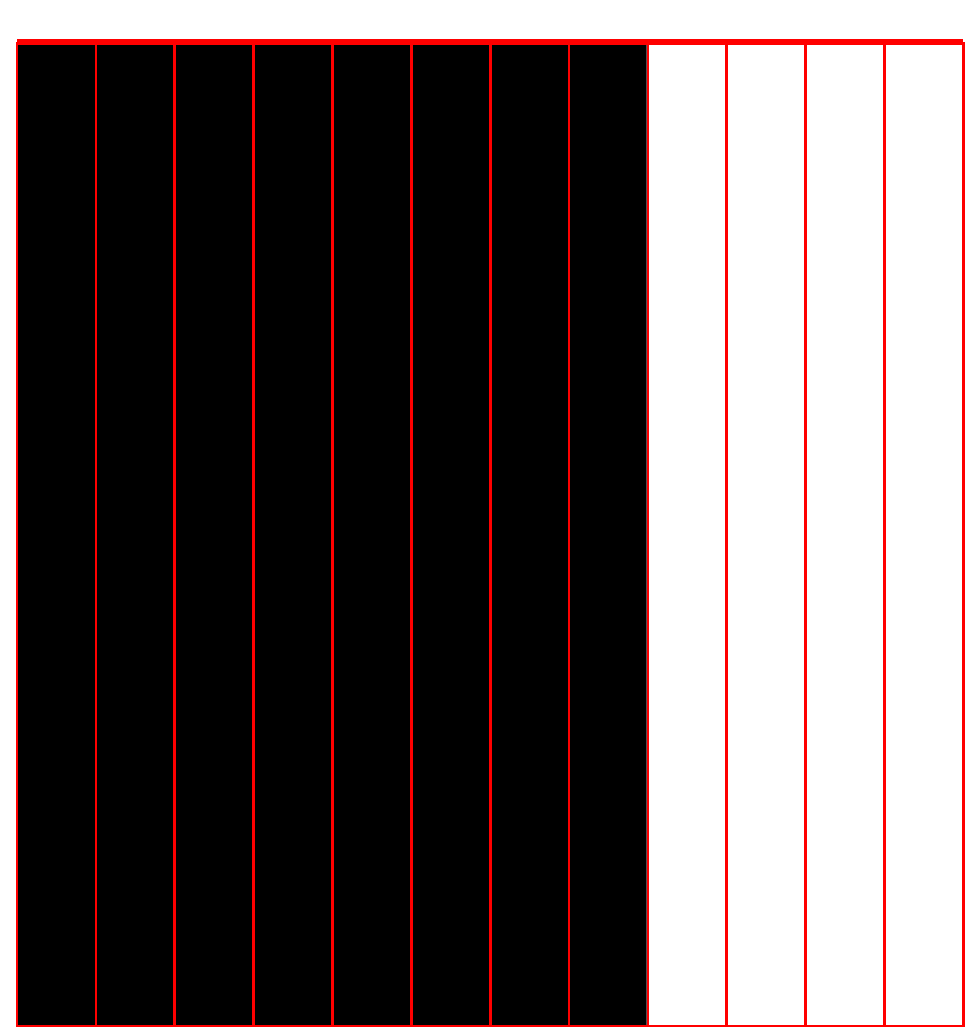}\hspace{\wnSpacing}
    \includegraphics[trim=5.8cm 13.5cm 5.9cm 4.0cm, clip=true, width=\wnWidth]{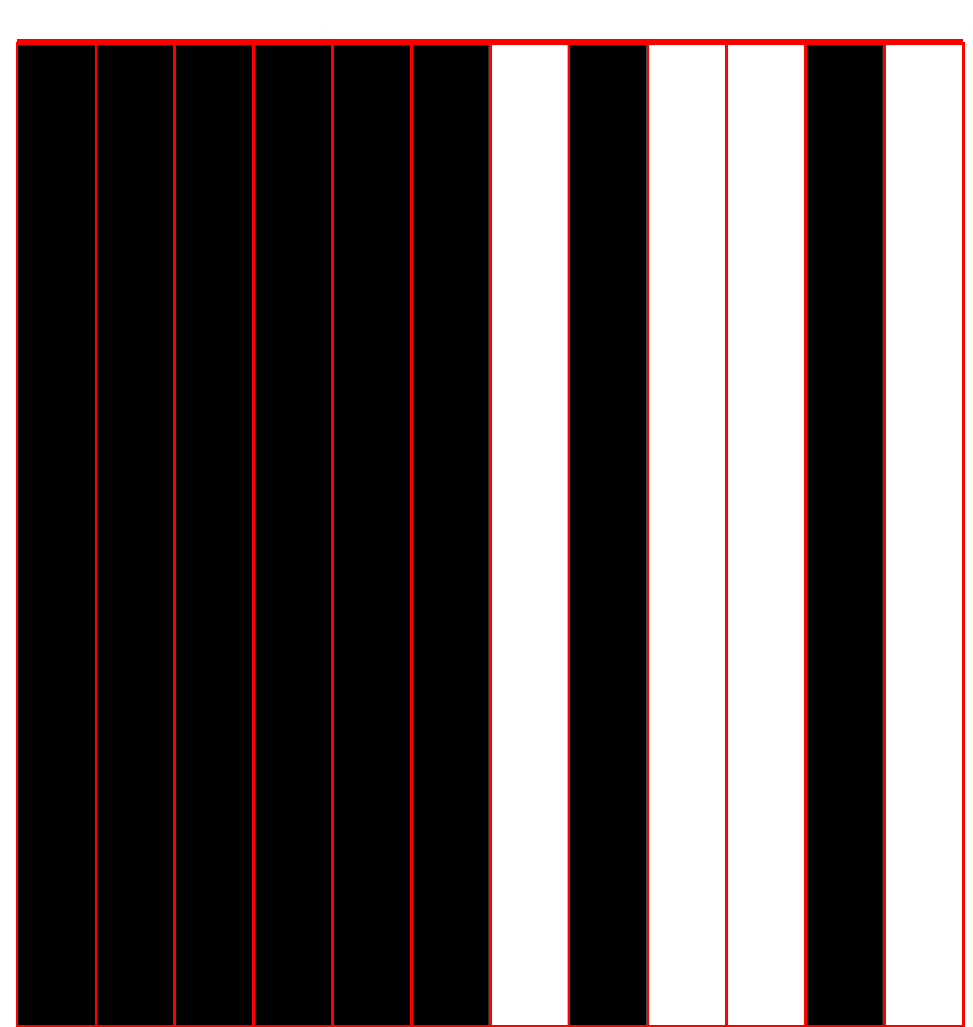}\hspace{\wnSpacing}
    \includegraphics[trim=5.8cm 13.5cm 5.9cm 4.0cm, clip=true, width=\wnWidth]{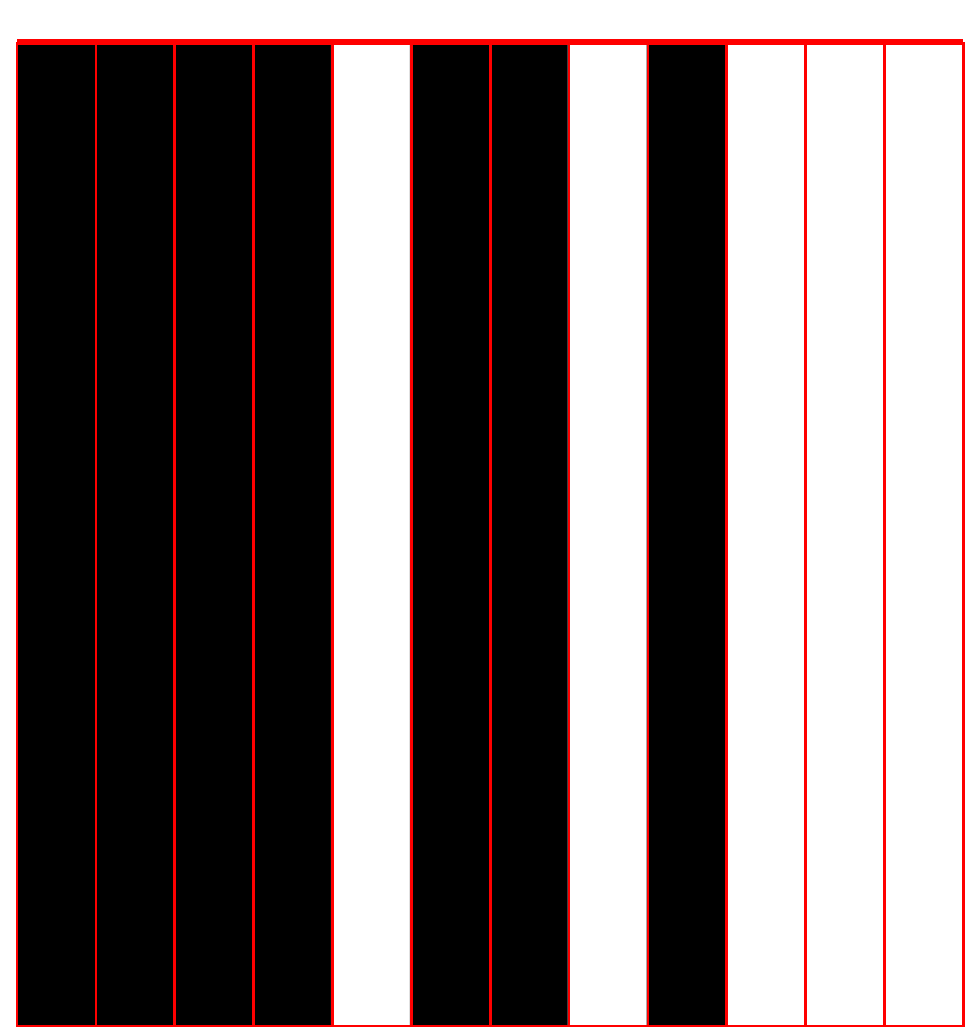}\hspace{\wnSpacing}
    \includegraphics[trim=5.8cm 13.5cm 5.9cm 4.0cm, clip=true, width=\wnWidth]{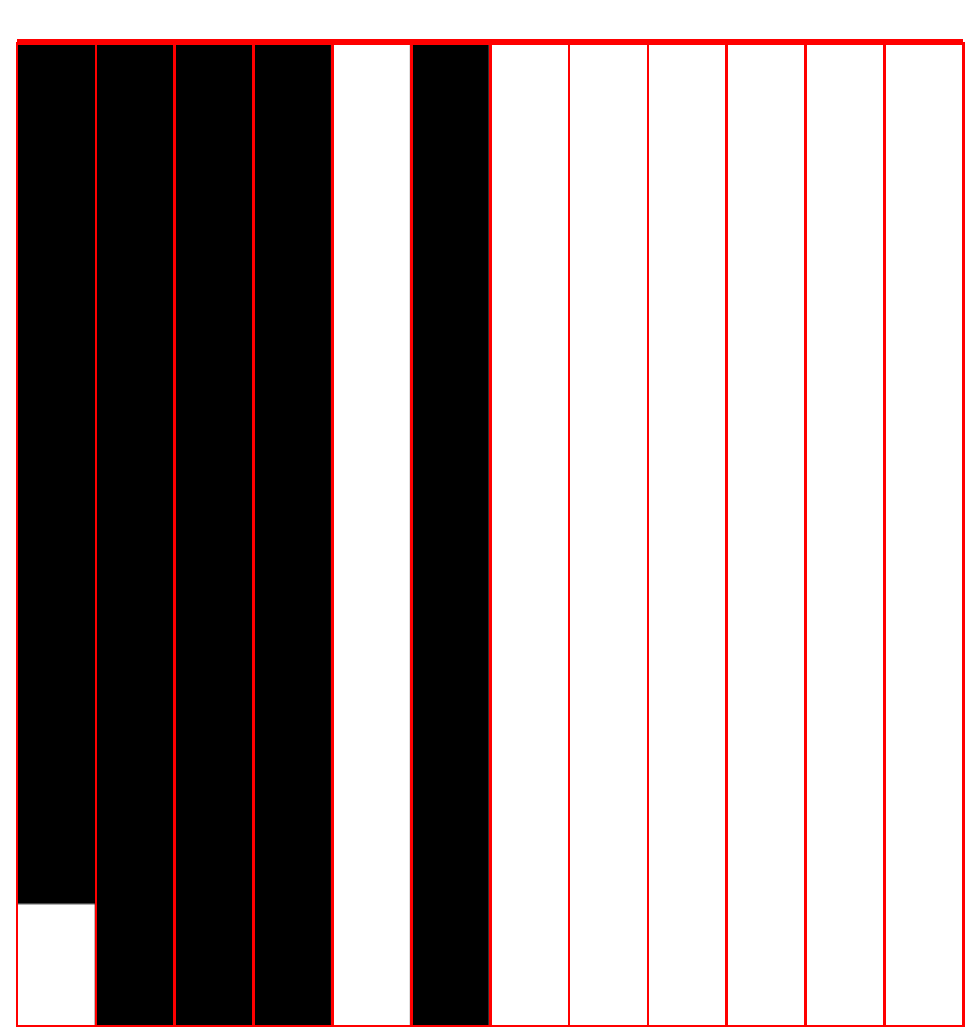}

    \caption{WordNet Nouns split into $k = 31$ implicit sorts, using the $\sigma_{\Cov}$ function and a threshold of $\theta = 0.9$. Only the first 12 sorts are shown here.}
  \label{fig:wordnet_nouns_theta09_bestK_cov}\end{subfigure}

  \begin{subfigure}[b]{\columnwidth}
    \centering
    \includegraphics[trim=5.8cm 13.5cm 5.9cm 3.0cm, clip=true, width=\wnWidth]{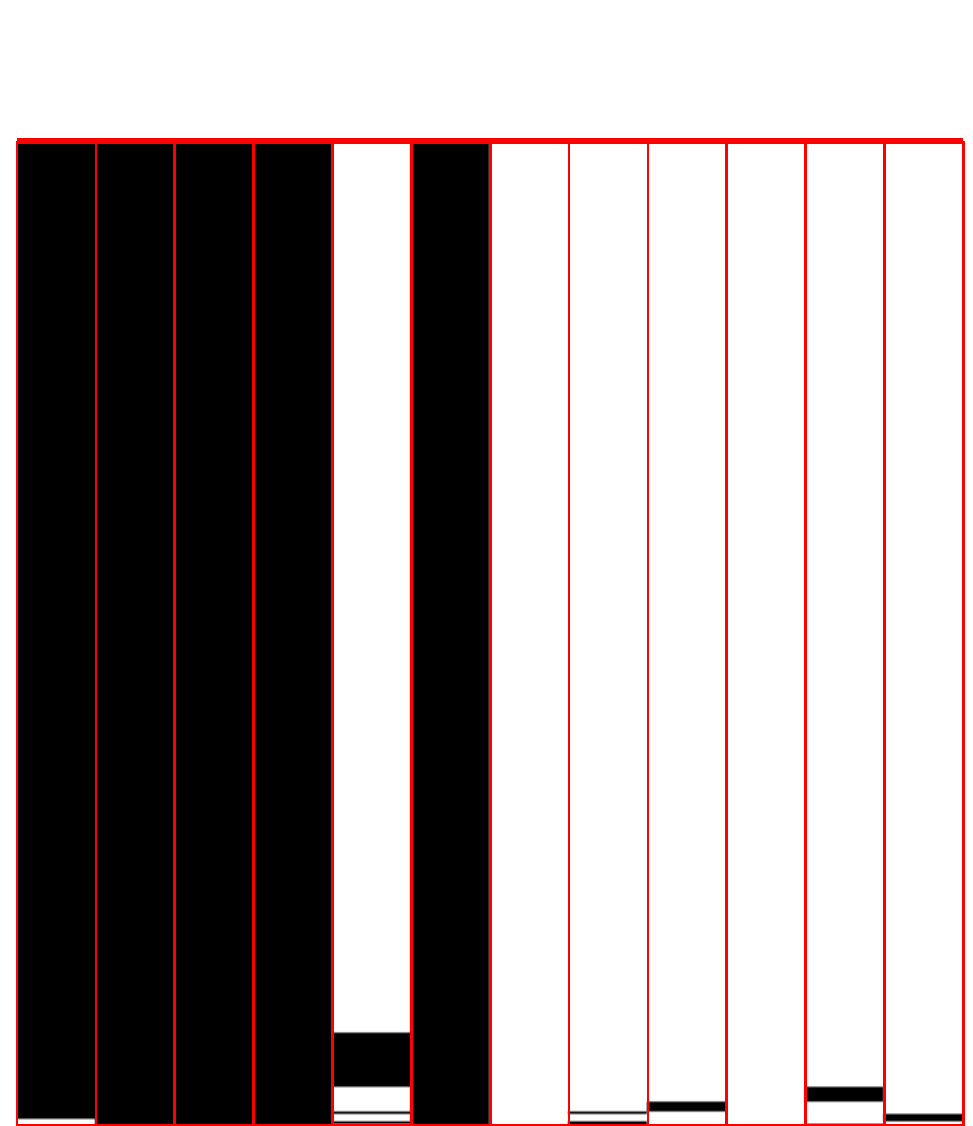}\hspace{\wnSpacing}
    \includegraphics[trim=5.8cm 13.5cm 5.9cm 3.0cm, clip=true, width=\wnWidth]{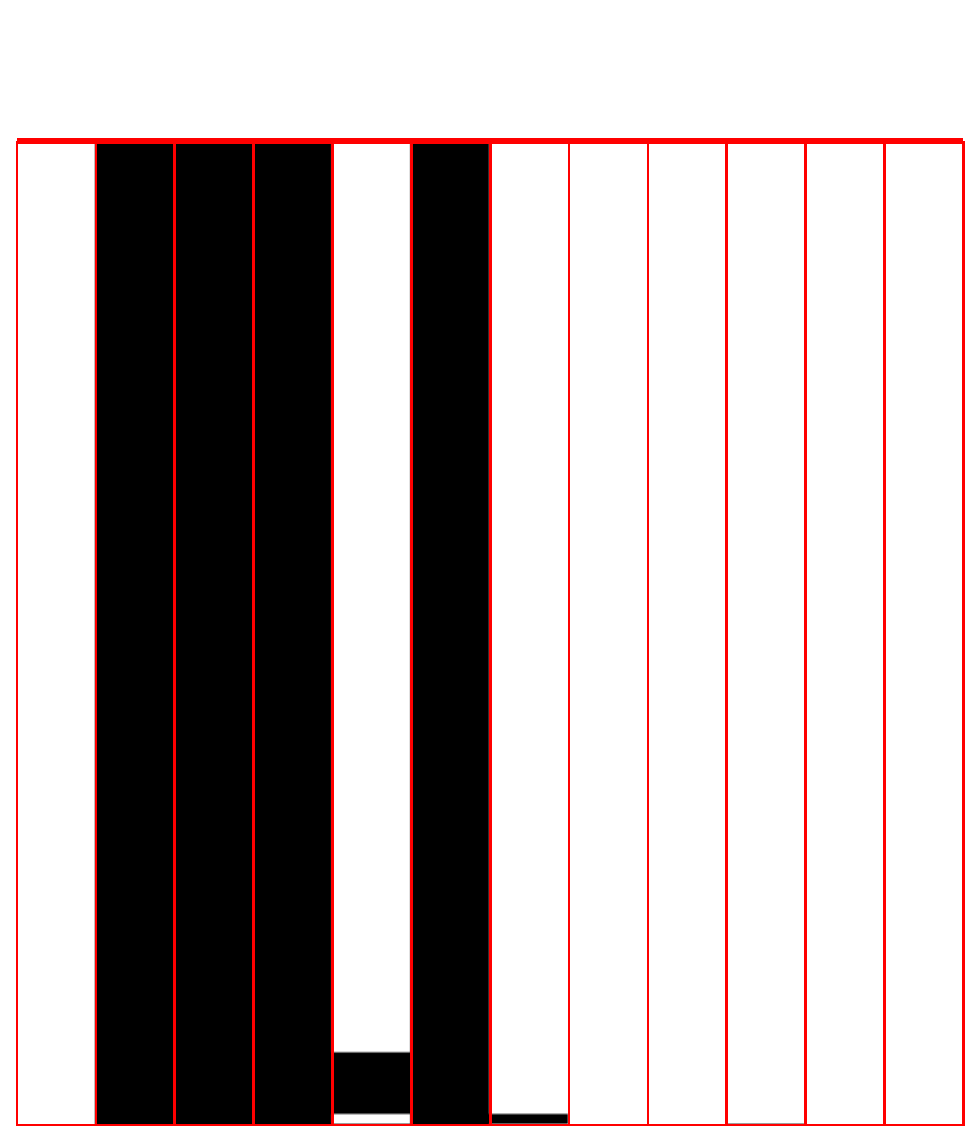}\hspace{\wnSpacing}
    \includegraphics[trim=5.8cm 13.5cm 5.9cm 3.0cm, clip=true, width=\wnWidth]{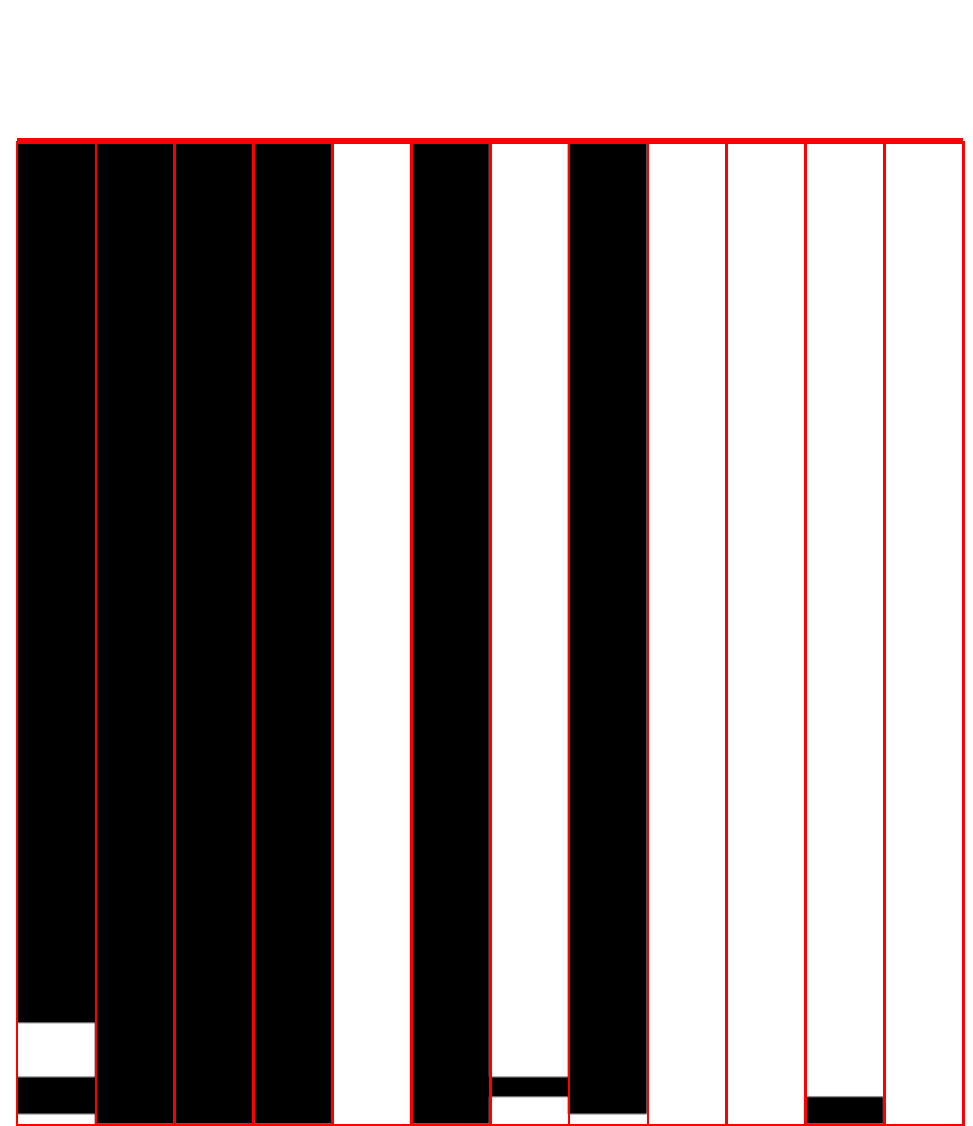}\hspace{\wnSpacing}
    \includegraphics[trim=5.8cm 13.5cm 5.9cm 3.0cm, clip=true, width=\wnWidth]{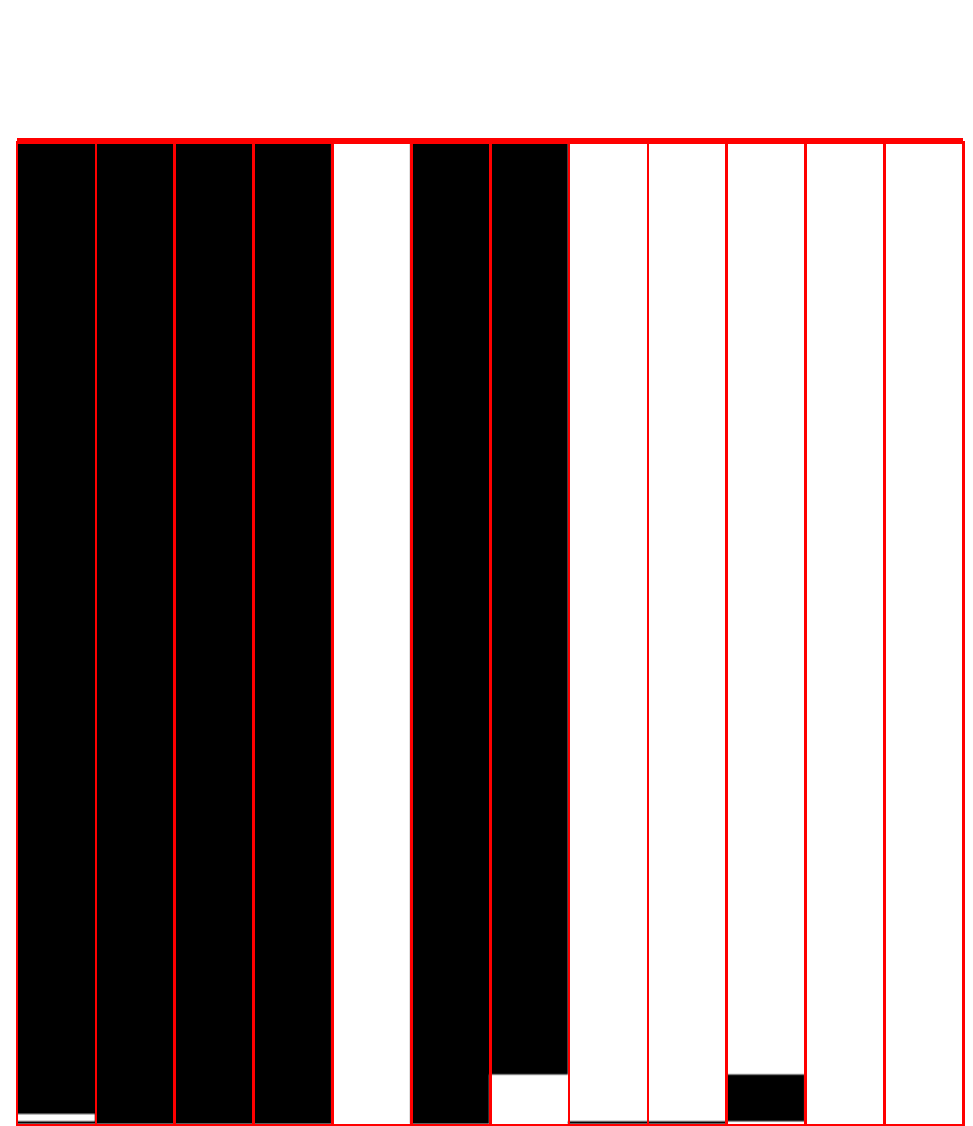}

%Set 0: 
%  7037 subjects
%  7 signatures
%  cov = 0.6068210885320449
%  sim = 0.9881510370174102
%Set 1: 
%  12449 subjects
%  34 signatures
%  cov = 0.4999129782847351
%  sim = 0.9803864807638205
%Set 2: 
%  52880 subjects
%  6 signatures
%  cov = 0.565393343419062
%  sim = 0.9801169847823905
%Set 3: 
%  7323 subjects
%  6 signatures
%  cov = 0.45325989652085513
%  sim = 0.9813148538409249

    \caption{WordNet Nouns split into $k = 4$ implicit sorts, using the $\sigma_{\Sim}$ function. The threshold of this sort refinement is $\theta = 0.98$. The sizes of the sorts range from 52,880 subjects (the third sort) to 7,037 subjects (the first sort).}
  \label{fig:wordnet_nouns_theta09_bestK_sim}\end{subfigure}

\caption{WordNet Nouns partitioned into the lowest $k$ with a fixed threshold.}
\label{fig:wordnet_nouns_theta09_bestK}\end{figure}

%================================================================
%  SCALING ANALYSIS
%================================================================
\subsection{Scalability Analysis}

To study how the ILP-based solution scales, we turn to the knowledge base YAGO. This proves to be a useful dataset for this study, as it contains apx.~380,000 explicit sorts extracted from several sources, including Wikipedia. From YAGO, a sample of apx.~500 sorts is randomly selected (as most explicit sorts in YAGO are small, we manually included larger sorts in the sample). The sample contains sorts with sizes ranging from $\sim$100 to $\sim$10$^5$ subjects, from $1$ to $\sim$350 signatures, and from $\sim$10 to $\sim$40 properties.

\begin{figure}[t]
\centering

%  \begin{subfigure}[b]{\columnwidth}
%    \centering
%    \includegraphics[width=5.0cm]{runtime_vs_subjs.pdf}
%    \caption{A plot of the runtime as a function of the number of subjects in a sort. There does not appear to be a dependency.}
%  \label{fig:runtime_vs_subjs}\end{subfigure}

  \begin{subfigure}[b]{\columnwidth}
%    \centering
    \includegraphics[width=4.5cm]{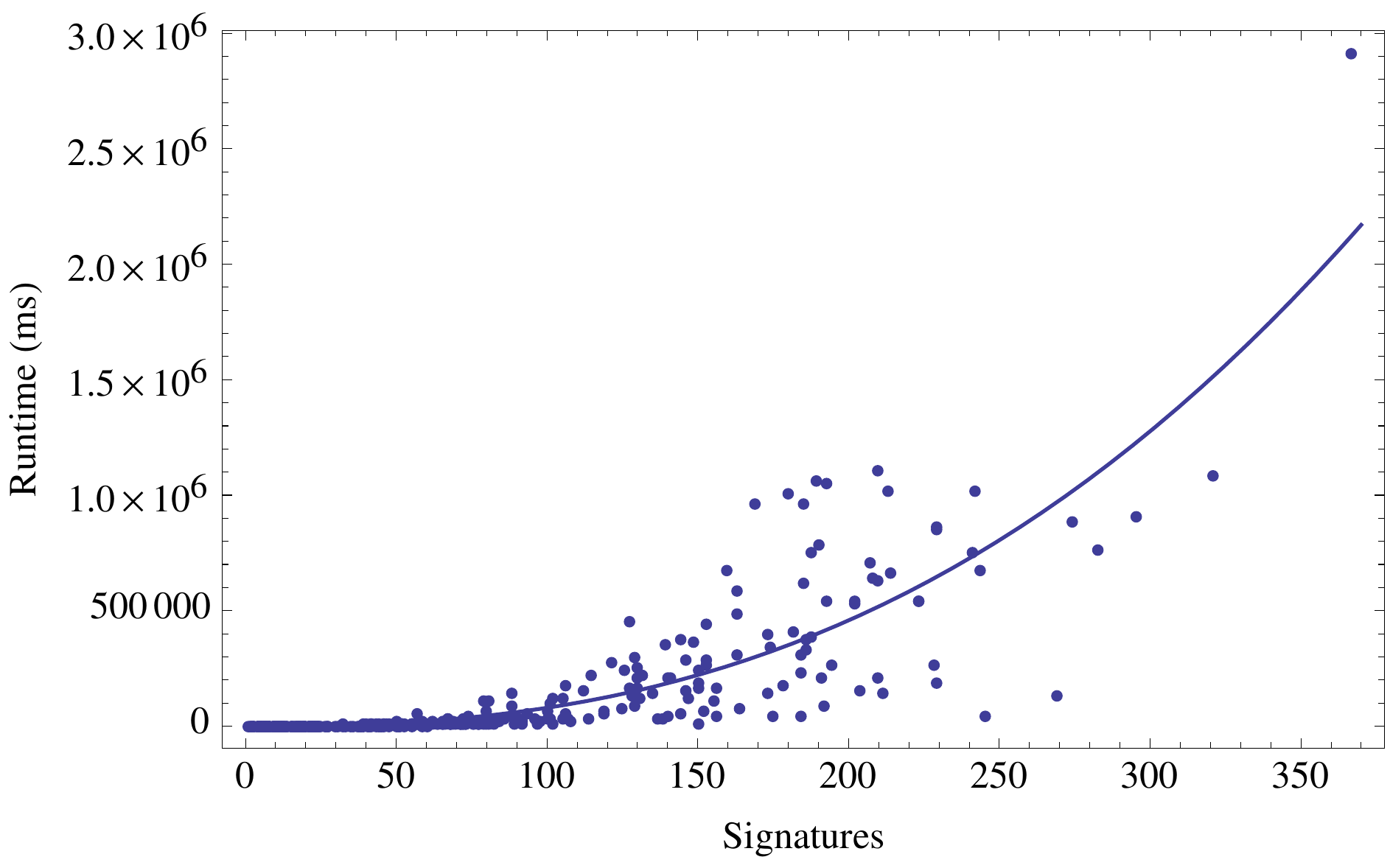}~\raisebox{2mm}{\includegraphics[width=3.8cm]{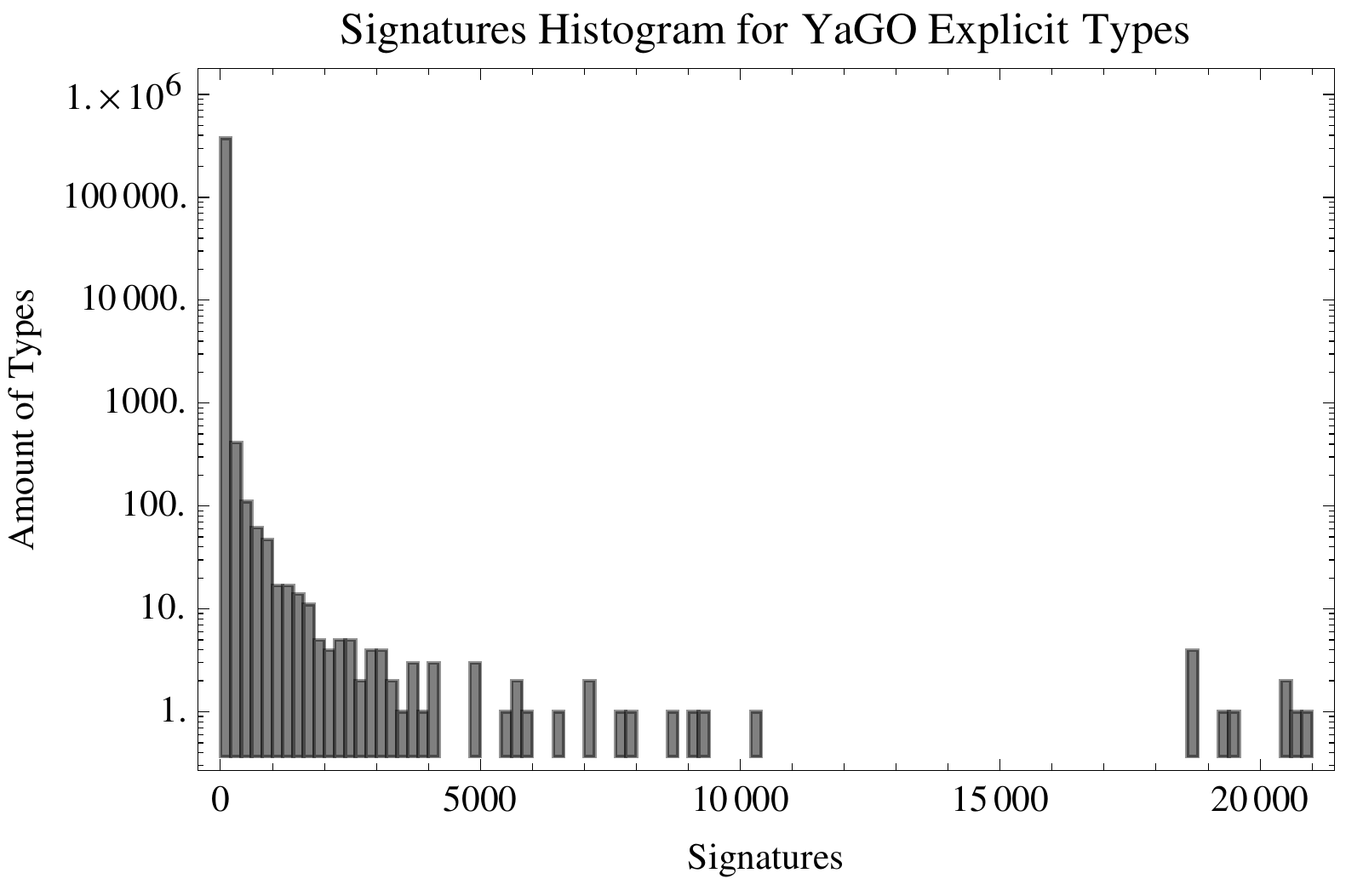}}
    \caption{Left: A plot of the runtime as a function of the number of signatures in a sort. The best fit is shown, and corresponds to $f(s) \approx x^{2.53}$ ($R^2 = 0.72$). Right: A logarithmic histogram of the number of signatures in YAGO explicit sorts. Note that 99.9\% of sorts have less than 350 signatures.}
  \label{fig:runtime_vs_sigs}\end{subfigure} %f(x) = -82.61 + 0.69 x^{2.53}

  \begin{subfigure}[b]{\columnwidth}
%    \centering
    \includegraphics[width=4.5cm]{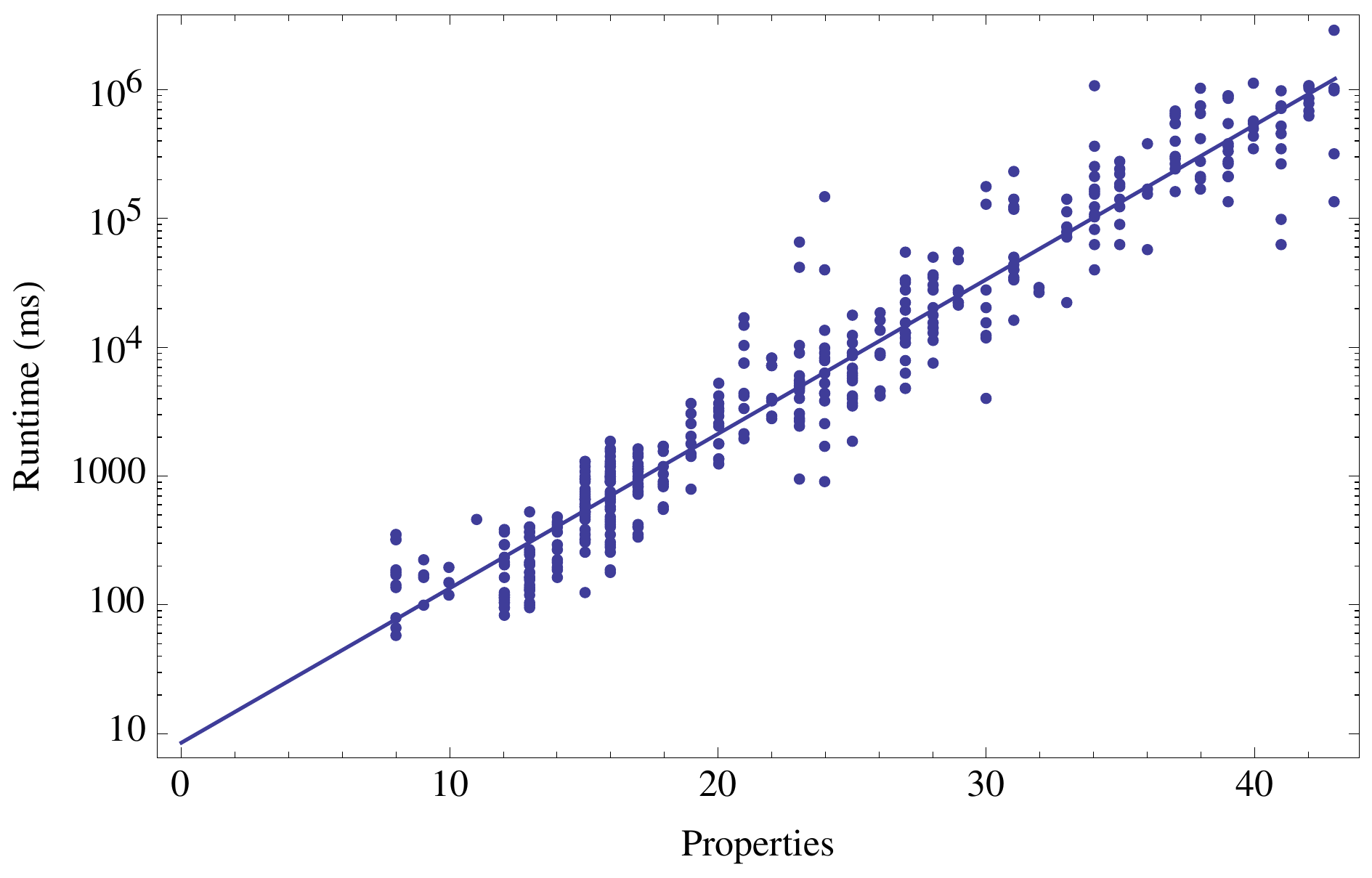} ~\raisebox{2mm}{\includegraphics[width=3.8cm]{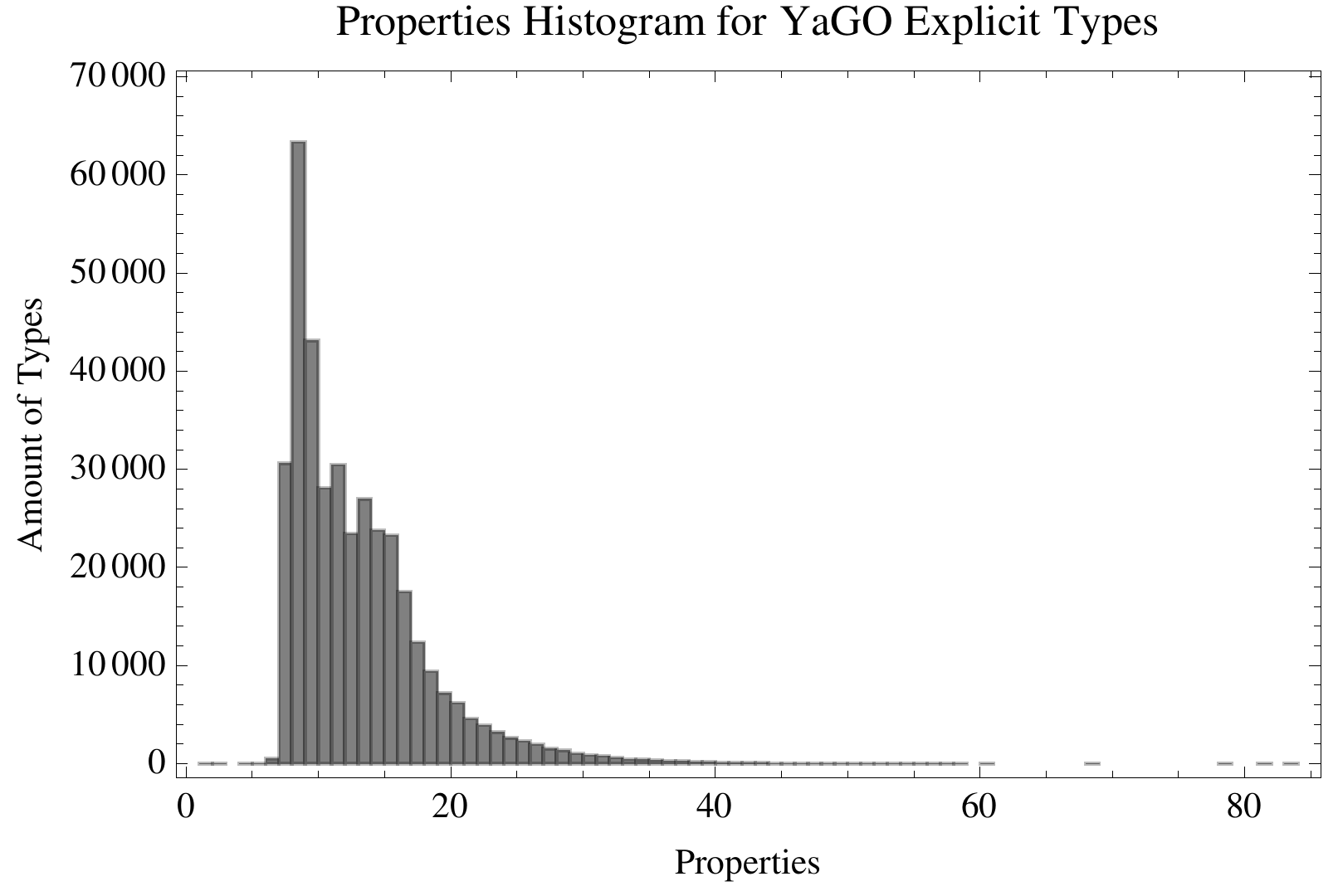}}
    \caption{Left: A logarithmic plot of the runtime as a function of the number of properties in a sort. The best fit is shown, and corresponds to $f(p) \approx e^{0.28 p}$ ($R^2 = 0.61$). Right: A histogram of the number of properties in YAGO explicit sorts. Note that 99.8\% of sorts have less than 40 properties.}
  \label{fig:runtime_vs_props_logplot}\end{subfigure} %f(x) = 8.52 e^{0.28 x}

  \caption{For a sample of YAGO explicit sorts, a highest $\theta$ sort refinement for $k = 2$ is solved.}
\label{fig:runtime_plots}\end{figure}

As can be seen in the histograms of figure \ref{fig:runtime_plots}, there exist larger and more complex explicit sorts in YAGO than those sampled (e.g.~sorts with $\sim$20,000 signatures or $\sim$80 properties). However, the same histograms show that 99.9\% of YAGO explicit sorts have under 350 signatures, and 99.8\% have under 40 properties, and thus the sample used is representative of the whole. We also note that, in a practical setting, the amount of properties can be effectively reduced \emph{with the language itself} by designing a rule which restricts the structuredness function to a set of predefined properties.

For each sort in the sample, a highest $\theta$ sort refinement for $k = 2$ is solved, and the total time taken to solve the ILP instances is recorded (we may solve multiple ILP instances, each for larger values of $\theta$, until an infeasible instance is found). See figure \ref{fig:runtime_plots}.

Due to lack of space, a plot for the runtime as a function of the number of subjects in a sort is not shown, but the result is as expected: the runtime of the ILP-based solution does not depend on the amount of subjects in an explicit sort; the difficulty of the problem depends on the structure of the dataset, and not its size.

In contrast, figures \ref{fig:runtime_vs_sigs} and \ref{fig:runtime_vs_props_logplot} both reveal clear dependencies of the runtime on the number of signatures and properties in a sort.

In the first case, the best polynomial fit shows that runtime is $\mathcal{O}(s^{2.5})$, where $s$ is the number of signatures (the best fit is obtained by minimising the sum of squared errors for all data points). This result limits the feasibility of using the ILP-based solution for more complicated explicit sorts. However, sorts with more than 350 signatures are very rare. In the second case, the best exponential fit shows that runtime is $\mathcal{O}(e^{0.28 p})$, where $p$ is the number of properties. Although an exponential dependency can be worrisome, the situation is similar to that for signatures.

%\begin{figure}[t]
%\centering
%  \includegraphics[width=5.0cm]{yagoSubjsHist.pdf}
%  \includegraphics[width=5.0cm]{yagoSigsHist.pdf}
%  \includegraphics[width=5.0cm]{yagoPropsHist.pdf}
%
%  \caption{YAGO has apx.~380,000 explicit sorts. For these, histograms for amount of subjects, amount of signatures, and amount of properties are shown, respectively.}
%\label{fig:yagoHistograms}\end{figure}

%================================================================
%  SEMANTIC CORRECTNESS
%================================================================
\subsection{Semantic Correctness}

A final practical question is whether our ILP-based solution can reproduce existing differences between sorts. To address this, we have chosen two explicit sorts from YAGO: Drug Companies and Sultans. All triples whose subject is declared to be of sort Drug Company or Sultan are included in one mixed dataset. We then solve a highest $\theta$ sort refinement for fixed $k = 2$ problem on this mixed dataset and compare the resulting sort refinement with the actual separation between Drug Companies and Sultans.
\begin{center}
\begin{tabular}{r|c|c}
                           & Is Drug Company & Is Sultan \\\hline
Classified as Drug Company & 27 & 17 \\\hline
Classified as Sultan       & 0 & 23
\end{tabular}
\end{center}
In order to describe the quality of these results, we interpret this experiment as the binary classification problem of Drug Companies. Thus, Drug Companies become the positive cases, whereas Sultans become negative cases. With this,
%If this is interpreted as the problem of classifying signatures as corresponding to a Drug Company or not, then
the resulting accuracy of the result is $74.6\%$, the precision is $61.4\%$, and the recall is $100\%$. These results can be improved by considering a modified $\Cov$ rule which ignores properties defined in the syntax of RDF (\texttt{type}, \texttt{sameAs}, \texttt{subClassOf}, and \texttt{label}), which can be achieved by adding four conjuncts of the form $\prop(c) \neq u$ to the antecedent of $\Cov$. This results in $82.1\%$ accuracy, $69.2\%$ precision, and $100\%$ recall.
%
%\begin{center}
%\begin{tabular}{r|c|c}
%                           & Is Drug Company & Is Sultan \\\hline
%Classified as Drug Company & 27 & 12 \\\hline
%Classified as Sultan       & 0 & 28
%\end{tabular}
%\end{center}

It must be noted, however, that this experiment relies on the assumption that the initial explicit sorts are well differentiated to begin with, which is precisely the assumption we doubt in this paper.

%================================================================
%================================================================

\section{Related Work}

Our work is most related to efforts that mine RDF data to discover frequently co-occurring property sets that can be stored in separate so-called 'property tables'.  A recent example of this approach is exemplified in \cite{Lee13}, ``\emph{Attribute Clustering by Table Load}'' where the authors consider the problem of partitioning the properties of an RDF graph into \emph{clusters}. Each cluster defines the columns of a property table in which each row will represent a subject. A cluster is valid insofar as the \emph{table load factor} remains above a threshold. The table load factor Lee et.~al.~defined is equivalent to the coverage value defined in \cite{DKSU11} ($\Cov$ metric as per the notation of this paper). Their approach, however, differs from ours in the following way: while they seek to partition the properties of an RDF graph for the purpose of generating property tables, we seek to discover sets of subjects which, when considered together as an RDF graph, result in a highly structured relational database. The sub-sorts generated by us may use overlapping sets of properties.

Similarly, \cite{Ding03application-specificschema}  and \cite{Levandoski:2009:RDS:1586636.1586985} use frequent item set sequences \cite{Agrawal:1995:MSP:645480.655281} data mining techniques to discover, in a RDF dataset, properties that are frequently defined together for the same subject (e.g., first name, last name, address, etc.). Such properties represent good candidates to be stored together in  a property table.  Although the goal of \cite{Ding03application-specificschema} and \cite{Levandoski:2009:RDS:1586636.1586985} is to improve performance by designing a customized database schema to store a RDF dataset,  a property table can also be viewed as a refined sort whose set of instances consists of all resources specifying at least one of the properties of the table.   In \cite{Ding03application-specificschema} and \cite{Levandoski:2009:RDS:1586636.1586985}, various important parameters controlling the sort refinement approach  are chosen in an ad-hoc manner (e.g.,  in \cite{Ding03application-specificschema} the  minimum support used is chosen after manually inspecting partial results produced by an handful of minimum support  values, and, in \cite{Levandoski:2009:RDS:1586636.1586985}, it is explicit specified by the user); whereas, in our approach, key parameters (e.g., $k$ and $\theta$) are  selected in a principled way to reach an optimal value of a user defined structuredness metric.  

Other than the property tables area, our work can be positioned in the broader context of inductive methods to acquiring or refining schema-level knowledge for RDF data \cite{Volker:2011:SSI:2008892.2008904,d'Amato:2010:ILS:2019445.2019452,DBLP:phd/de/Lehmann2010b,DelteilFD01:1,maedche2002clustering,foaflearning}. Prior works have typically relied on statistical or logic programming approaches to discover ontological relations between sorts and properties. However, to the best of our knowledge, our work presents the first principled approach to refine the sort by altering the assignment of resources to a refined set of sorts in order to improve some user defined measure of structuredness. 

In the area of knowledge discovery in general, the work by Yao \cite{Vinh:2010:ITM:1953011.1953024} offers a nice overview of several information-theoretic measures for knowledge discovery, including, attribute entropy and mutual information. A common characteristic of all these measures is that they focus on the particular values of attributes (in our case, predicates) and attempt to discover relationships between values of the same attribute, or relationships between values of different attributes. As is obvious from Section 3, our work focuses on discovering relationships between entities (and their respective schemas) and therefore we are only interested in the presence (or absence) of predicates for particular attributes for a given entity, therefore ignoring the concrete values stored there. Hence the our measures are orthogonal to those discussed by Yao \cite{Vinh:2010:ITM:1953011.1953024}.

%================================================================
%================================================================

\section{Conclusions}

We have presented a framework within which it is possible to study the structuredness of RDF graphs using measures which are tailored to the needs of the user or database administrator. This framework includes a formal language for expressing structuredness rules which associate a structuredness value to each RDF graph. We then consider the problem of discovering a partitioning of the entities of an RDF graph into subset which have high structuredness with respect to a specific structuredness function chosen by the user. Although this problem is intractable in general, we define an ILP instance capable of solving this problem within reasonable time limits using commercially available ILP solvers.

We have used our framework to study two real world RDF datasets, namely DBpedia Persons and WordNet Nouns, the former depending on a publicly editable web source and therefore containing data which does not clearly conform to its schema, and the latter corresponding to a highly uniform set of dictionary entries. In both cases the results obtained were meaningful and intuitive. We also studied the scalability of the ILP-based solution on a sample of explicit sorts extracted from the knowledge base YAGO, showing that our solution is practical for all but a small minority of existing sorts.

The obvious next goal is to better understand the expressiveness of structuredness rules, and to explore subsets of our language with possibly lower computational complexity. In particular, the NP-hardness of the decision problem has been proven for a rule which uses disjunction in the consequent; if this were to be disallowed, it might be possible to lower the complexity of the problem. Finally, it would also be interesting to explore the existence of rules for which a high structuredness value can predict good performance for certain classes of queries.

%================================================================
%================================================================

\bibliographystyle{abbrv}
\bibliography{bibliography}

%================================================================
%================================================================
\newpage
\onecolumn

\appendix

%================================================================
% Section: reduction from 3-coloring
%================================================================
\section{Reducing 3-Coloring to the problem}

Recall the definition of the decision problem, where $r$ is a rule:

\begin{center}
\framebox{
\begin{tabular}{rp{11cm}}
{\bf Problem}: & $\eits(r)$\\
{\bf Input}:   &  An RDF graph $D$, a rational number $\theta$ such that $0 \leq \theta \leq 1$, and a positive integer $k$.\\
{\bf Output}:  & \textbf{true} if there exists a $\sigma_{r}$-sort refinement $\mathcal{T}$ of $D$ with threshold $\theta$ that contains at most $k$ implicit sorts, and \textbf{false} otherwise.
\end{tabular}}
\end{center}

We will now prove that $\eits(r_0)$ is NP-complete for $\theta = 1$ and $k = 3$, where $r_0$ is the rule defined in equation \ref{eq:def-r0}. For this, we will first prove that the general problem $\eits(r)$ is in NP. Then we will show that fixing $\theta = 1$ and $k = 3$ and using rule $r_0$ yields a version of $\eits(r)$ that is NP-hard.

\subsection{$\eits(r)$ is in NP}

Given an RDF graph $D$, a rational number $\theta$ such that $0 \leq \theta \leq 1$, and a positive integer $k$, a Non-Deterministic Turing Machine (NDTM) must guess $T_1, \ldots, T_l$ with $l \leq k$. The NDTM must then verify that $T_1, \ldots, T_l$ form a sort refinement with threshold $\theta$. For each $T_i$, $i \in [1, l]$, the structuredness can be determined in the following way: Consider that $T_i$ contains $S_i$ subjects (rows, if represented as a matrix) and $P_i$ properties (columns). Let $v$ be the number of variables in $r_0$. Since rule $r_0$ is fixed, $v$ is also fixed. There are $(S_i P_i)^v$ different possible assignments of the variables in $r_0$, which is polynomial in the size of $D$.

For each assignment, the NDTM must check if (i) it satisfies the antecedent of $r_0$ and (ii) if it satisfies the antecedent and the consequent of $r_0$, together. Since both checks are similar, we will focus on checking if the antecedent of $r_0$ is satisfied by an assignment $\rho$.

The antecedent of $r_0$ will consist of a number of equalities (or inequalities) bounded by the size of $r_0$ (which is itself fixed). If $c_a$ and $c_b$ are variables in $r_0$, each equality will be of the form $x = y$, where $x$ may be $c_a$, $\subj(c_a)$, $\prop(c_a)$, or $\val(c_a)$, and $y$ will be defined accordingly, obeying the syntax of the rules (defined previously). The NDTM can encode the assignment $\rho$ in the following way: each of the $v$ variables can be encoded using $\log(v)$ bits, and for each variable it must store the subject and property it is assigned to, using $\log(S_i)$ and $\log(P_i)$ bits, respectively. Since $S_i$ and $P_i$ are both bounded by the size $|D|$ of $D$, the assignment $\rho$ can be encoded in space which is logarithmic in $|D|$.

To determine the value of a variable, the NDTM must find the subject-property pair in the encoded version of the RDF graph $D$. This lookup will take time at most linear in $|D|$. The comparison itself can be done polynomial time.

The entire verification process will take time which is polynomial in the size of the RDF graph $D$, although it depends non-trivially on the size of the rule $r$. Therefore, $\eits(r)$ is in NP.

\subsection{$\eits(r_0)$ with $k = 1$ and $\theta = 1$ is NP-hard}

To prove that $\eits(r_0,1,3)$ is NP-hard, we will use a reduction from 3-\textsc{Colorability}, which is the following decision problem: given an undirected graph without loops (self-edges) $G = (V,E)$, decide if there exists a 3-coloring of $G$ (i.e. a function $f: V \rightarrow \{ 1, 2, 3 \}$ such that for all pairs of nodes $u, v \in V$, if $(u,v) \in E$ then $f(u) \neq f(v)$).

Consider a non-directed graph $G$ with $n$ nodes, defined by the $n \times n$ adjacency matrix $A_G$. We will construct an RDF graph $D_G$ defined by its accompanying matrix $M_G = M(D_G)$ so that $G$ is 3-colorable if and only if there exists a $\sigma_{r_0}$-sort refinement $\mathcal{T}$ of $D$ with threshold 1, consisting of at most $3$ implicit sorts.

First, we construct the accompanying matrix $M_G$ of $D_G$ by blocks, in the following way:

\begin{equation*}
	M_G = \begin{pmatrix}
		0_{n \times 1} & 0_{n \times 1} & 1_{n \times 1} & D_{n \times n} & D_{n \times n} \\
		0_{n \times 1} & 1_{n \times 1} & 1_{n \times 1} & D_{n \times n} & D_{n \times n} \\
		1_{n \times 1} & 0_{n \times 1} & 1_{n \times 1} & D_{n \times n} & D_{n \times n} \\
		1_{n \times 1} & 1_{n \times 1} & 0_{n \times 1} & D_{n \times n} & \bar{A}_G
	\end{pmatrix}_{(4n) \times (2n + 3)}
\end{equation*}

Here, $D_{n \times n}$ is a $n \times n$ unit matrix (i.e.~with 1's in the diagonal and 0's everywhere else), $0_{n \times 1}$ is a single column of $n$ zeroes, and $1_{n \times 1}$ is a single column of $n$ ones. The block $\bar{A}_G$ contains the complement of the adjacency matrix $A_G$ ($\bar{A}_G[i,j] = 1 - A_G[i,j]$).

The first $3n$ rows will be referred to as the \emph{upper section} of $M_G$, while the last $n$ rows will be referred to as the \emph{lower section} of $M_G$. The upper section consists of three sets of auxiliary rows, which are identical in every column except the first and second. The first 2 columns will be called \textbf{sp1} and \textbf{sp2}, respectively, the third column will be called the \textbf{idp} column, the next $n$ columns will be referred to as the \emph{left column set} and the last $n$ rows as the \emph{right column set}. In this way, the complemented adjacency matrix $\bar{A}_G$ is contained in the right column set, in the lower section of $M_G$.

Every row of $M_G$ represents a subject and every column represents a property, however, we will not explicitly mention the names of any subjects or properties, except for the first three properties: \textbf{sp1}, \textbf{sp2}, and \textbf{idp}.

\begin{example}
{\em Consider the following simple graph $G$, its adjacency matrix $A_G$ and the complemented matrix $\bar{A}_G$:
\begin{center}
\begin{tikzpicture}[->,>=stealth',shorten >=1pt,node distance=2cm,main node/.style={circle,draw}]

  \node[main node] (1) {1};
  \node[main node] (2) [below left of=1] {2};
  \node[main node] (3) [below right of=1] {3};

  \path[->]
    (1) edge (2)
    (2) edge (1);
\end{tikzpicture}
\end{center}
\begin{align*}
	A_G = \begin{pmatrix}
		0 & 1 & 0 \\
		1 & 0 & 0 \\
		0 & 0 & 0 \\
	\end{pmatrix}, \qquad
	\bar{A}_G = \begin{pmatrix}
		1 & 0 & 1 \\
		0 & 1 & 1 \\
		1 & 1 & 1 \\
	\end{pmatrix} \quad
\end{align*}

We construct an RDF graph $D_G$ as described above. $D_G$ is an RDF graph containing RDF triples, which we will not show explicitly, as only the accompanying matrix is necessary for the reduction. For increased visibility, zeroes are omitted from matrix $M_G$, everywhere except in the $\bar{A}_G$ block.

\hspace{5.5cm}$\overbrace{\phantom{111111111}}^{\text{left column set}}$$\overbrace{\phantom{111111111}}^{\text{right column set}}$

\vspace{-4mm}\setlength{\columnsep}{0cm}
\begin{multicols}{2}
\hfill $M_G = \begin{pmatrix}[cc|c|ccc|ccc]
		  &   & 1 & 1 &   &   & 1 &   &   \\
		  &   & 1 &   & 1 &   &   & 1 &   \\
		  &   & 1 &   &   & 1 &   &   & 1 \\\hline
		  & 1 & 1 & 1 &   &   & 1 &   &   \\
		  & 1 & 1 &   & 1 &   &   & 1 &   \\
		  & 1 & 1 &   &   & 1 &   &   & 1 \\\hline
		1 &   & 1 & 1 &   &   & 1 &   &   \\
		1 &   & 1 &   & 1 &   &   & 1 &   \\
		1 &   & 1 &   &   & 1 &   &   & 1 \\\hline
		1 & 1 &   & 1 &   &   & 1 & 0 & 1 \\
		1 & 1 &   &   & 1 &   & 0 & 1 & 1 \\
		1 & 1 &   &   &   & 1 & 1 & 1 & 1 \\
	\end{pmatrix}$

\noindent $\left. \phantom{\begin{pmatrix}a\\a\\a\\a\\a\\a\\a\\a\\a\end{pmatrix}} \right\} \text{upper section}$ \\
$\left. \phantom{\begin{pmatrix}a\\a\\a\end{pmatrix}} \right\} \text{lower section}$ \\
\end{multicols}

$\qed$
}
\end{example}

The next step is to introduce the fixed rule $r_0$ to be used. The variables of $r_0$ are $x, c_1, c_2, y, d_1, d_2, z, e, u, f_1$, and $f_2$, and the rule itself is:
\begin{equation}\label{eq:def-r0}
\left(
\begin{array}{l}
\prop(c_1) \neq \mbox{`\textbf{sp1}'} \ws \prop(c_1) \neq \mbox{`\textbf{sp2}'} \ws \\
\prop(c_2) \neq \mbox{`\textbf{sp1}'} \ws \prop(c_2) \neq \mbox{`\textbf{sp2}'} \ws \\
\prop(d_1) \neq \mbox{`\textbf{sp1}'} \ws \prop(d_1) \neq \mbox{`\textbf{sp2}'} \ws \\
\prop(d_2) \neq \mbox{`\textbf{sp1}'} \ws \prop(d_2) \neq \mbox{`\textbf{sp2}'} \ws \\
%\prop(z) \neq \mbox{`\textbf{sp1}'} \ws \prop(z) \neq \mbox{`\textbf{sp2}'} \ws \\
\prop(e) \neq \mbox{`\textbf{sp1}'} \ws \prop(e) \neq \mbox{`\textbf{sp2}'} \ws \\
\prop(f_1) \neq \mbox{`\textbf{sp1}'} \ws \prop(f_1) \neq \mbox{`\textbf{sp2}'} \ws \\
\prop(f_2) \neq \mbox{`\textbf{sp1}'} \ws \prop(f_2) \neq \mbox{`\textbf{sp2}'} \ws \\
\prop(x) = \mbox{`\textbf{idp}'} \ws \val(x) = 1 \ws \\
c_1 \neq x \ws \subj(c_1) = \subj(x) \ws \val(c_1) = 1 \ws \\
c_2 \neq x \ws \subj(c_2) = \subj(x) \ws \val(c_2) = 1 \ws \\
c_1 \neq c_2 \ws \\
\prop(y) = \mbox{`\textbf{idp}'} \ws \val(y) = 0 \ws \\
\subj(d_1) = \subj(y) \ws \prop(d_1) = \prop(c_1) \ws \\
\subj(d_2) = \subj(y) \ws \prop(d_2) = \prop(c_2) \ws \\
\prop(z) = \mbox{`\textbf{idp}'} \ws \subj(z) = \subj(e) \ws \\
\prop(e) = \prop(c_1) \ws e \neq c_1 \ws \val(e) = 1 \ws \\
\prop(u) = \mbox{`\textbf{idp}'} \ws \val(u) = 0 \ws \\
\subj(u) = \subj(f_1) \ws \prop(f_1) = \prop(c_1) \ws \\
\subj(u) = \subj(f_2) \ws \prop(f_2) = \prop(c_2) \ws \\
\val(f_1) = 1 \ws \val(f_2) = 1 \\
\end{array}
\right)
\qquad\mapsto\qquad ( \val(d_1) = 1 \;\vee\; \val(d_2) = 1 ) \ws \val(z) = 0
\end{equation}

\noindent Rule $r_0$ is constructed as follows:

\begin{enumerate}
	\item The subexpression
	\begin{align*}
	&\prop(c_1) \neq \mbox{`\textbf{sp1}'} \ws \prop(c_1) \neq \mbox{`\textbf{sp2}'} \ws \prop(c_2) \neq \mbox{`\textbf{sp1}'} \ws \prop(c_2) \neq \mbox{`\textbf{sp2}'} \ws \\
	&\prop(d_1) \neq \mbox{`\textbf{sp1}'} \ws \prop(d_1) \neq \mbox{`\textbf{sp2}'} \ws \prop(d_2) \neq \mbox{`\textbf{sp1}'} \ws \prop(d_2) \neq \mbox{`\textbf{sp2}'} \ws \\
%	&\prop(z) \neq \mbox{`\textbf{sp1}'} \ws \prop(z) \neq \mbox{`\textbf{sp2}'} \ws
	&\prop(e) \neq \mbox{`\textbf{sp1}'} \ws \prop(e) \neq \mbox{`\textbf{sp2}'} \ws \\
	&\prop(f_1) \neq \mbox{`\textbf{sp1}'} \ws \prop(f_1) \neq \mbox{`\textbf{sp2}'} \ws \prop(f_2) \neq \mbox{`\textbf{sp1}'} \ws \prop(f_2) \neq \mbox{`\textbf{sp2}'}
	\end{align*}
	ensures that no variables can be assigned to columns \textbf{sp1} or \textbf{sp2} (i.e.~the set of total cases will only consider assignments where the variables are mapped to cells whose columns are not \textbf{sp1} or \textbf{sp2}). It is not necessary to include variables $x$, $y$, $z$ and $u$ here, as their columns are fixed.
	\item The subexpression $\prop(x) = \mbox{`\textbf{idp}'} \ws \val(x) = 1$ forces variable $x$ to point to a cell in the upper section of the \textbf{idp} column.
	
	\item The subexpression $c_1 \neq x \ws \subj(c_1) = \subj(x) \ws \val(c_1) = 1 \ws c_2 \neq x \ws \subj(c_2) = \subj(x) \ws \val(c_2) = 1 \ws c_1 \neq c_2$ defines variables $c_1$ and $c_2$, which share their row with $x$. Both must point to an element of the diagonal of the $D_{n \times n}$ blocks (since their value must be 1) and since $x$, $c_1$, and $c_2$ are all distinct, $c_1$ and $c_2$ must each point to a different $D_{n \times n}$ block. From here on, and without loss of generality, we will assume $c_1$ points to the left column set and $c_2$ points to the right column set.

	\item The subexpression $\prop(y) = \mbox{`\textbf{idp}'} \ws \val(y) = 0 \ws \subj(d_1) = \subj(y) \ws \prop(d_1) = \prop(c_1) \ws \subj(d_2) = \subj(y)$ $\ws \prop(d_2) = \prop(c_2)$ defines variables $y$, $d_1$, and $d_2$. Variable $y$ will point to the lower section of the \textbf{idp} column, and $d_1$ will point to a cell whose row and column are determined by $y$ and $c_1$, respectively. On the other hand, $d_2$ will point to a cell whose row and column are determined by $y_2$ and $c_2$, respectively. With this configuration (and the assumption given in the previous item), $d_1$ will point to the $D_{n \times n}$ in the left column set, lower section, and $d_2$ will point to the $\bar{A}_G$ block.
	
	\item The subexpression $\prop(z) = \mbox{`\textbf{idp}'} \wedge \subj(z) = \subj(e) \wedge \prop(e) = \prop(c_1) \wedge e \neq c_1 \wedge \val(e) = 1$ defines $z$, which points to the \textbf{idp} column, and $e$, whose position is fixed by the row of $z$ and column of $c_1$. Note that the rows of $z$ and $e$ are not fixed by the antecedent of the rule.
	
	\item The subexpression $\prop(u) = \mbox{`\textbf{idp}'} \ws \val(u) = 0 \ws \subj(u) = \subj(f_1) \ws \prop(f_1) = \prop(c_1) \ws \subj(u) = \subj(f_2) \ws \prop(f_2) = \prop(c_2)$ defines $u$, which points to the lower section of the \textbf{idp} column, $f_1$ is fixed by the row of $u$ and the column of $c_1$, and $f_2$ is analogous to $f_1$.
	
	\item The subexpression $\val(f_1) = 1 \ws \val(f_2) = 1$ ensures that the values of $f_1$ and $f_2$ are 1 (in an assignment that is to be included in the set of total cases).
	
	\item The subexpression $( \val(d_1) = 1 \;\vee\; \val(d_2) = 1 ) \ws \val(z) = 0$ constitutes the consequent of the rule and states the additional conditions that must be met by an assignment $\rho$ for it to be included in the set of favorable cases.
\end{enumerate}

\begin{example}
{\em We can visualize the positioning of the variables assigned to cells in $M_G$:

\begin{align*}
\begin{pmatrix}[cc|c|cccc|cccc]
   &   &    &   &     &   &   &   &     &   &   \\
   &   &  x &   & c_1 &   &   &   & c_2 &   &   \\
   &   &    &   &     &   &   &   &     &   &   \\
   &   &  z &   & e   &   &   &   &     &   &   \\
   &   &    &   &     &   &   &   &     &   &   \\\hline
   &   &  y &   & d_1 &   &   &   & d_2 &   &   \\
   &   &    &   &     &   &   &   &     &   &   \\
   &   &  u &   & f_1 &   &   &   & f_2 &   &   \\
\end{pmatrix} \qquad
\end{align*}

The $(x, c_1, c_2)$ trio must be located in the upper section as explained previously (items 2 and 3). The $(y, d_1, d_2)$ trio and the $(u, f_1, f_2)$ trio must be located in the lower section (items 4 and 6, respectively). The $(z, e)$ duo is shown in the upper section, although it may be assigned to the lower section also (item 5).

$\qed$
}
\end{example}

Now that RDF graph $D_G$ and rule $r_0$ have been defined, we will give an intuition as to how the existence of an implicit $\sigma_{r_0}$-sort refinement $\mathcal{T}$ of $D_G$ with threshold 1 implies that graph $G$ is 3-colorable.

An implicit $\sigma_{r_0}$-sort $T$ of $D_G$ with threshold 1 can be understood as a subset of the rows of $M_G$, which themselves form an RDF graph which we will call $D_T$, with accompanying matrix $M_T = M(D_T)$. We must be careful, however, with the following condition: an implicit $\sigma_{r_0}$-sort $T$ must be closed under signatures, that is, if a row is present in $M_T$, then all other identical rows must be included in $M_T$ as well. This condition has been made trivial with the creation of columns $\textbf{sp1}$ and $\textbf{sp2}$, whose sole purpose is to ensure that there are no two identical rows in matrix $M_G$. Since each row of matrix $M_G$ has its own unique signature, we have no need to discuss signatures for this reduction.

The problem of deciding if there exists an implicit $\sigma_{r_0}$-sort refinement $\mathcal{T}$ of $D_G$ with threshold 1 and with at most 3 sorts is the problem of partitioning the rows of $M_G$ into at most 3 implicit $\sigma_{r_0}$-sorts $T_1, T_2$, and $T_3$ such that each $T_i$ satisfies $\sigma_{r_0}(T_i) = 1$ (i.e. the value of the structuredness of $T_i$ is 1 when using $r_0$).

\begin{example}\label{example3}
{\em Given matrix $M_G$ of our working example, we show a possible partitioning of rows:
\begin{align*}
	M_{T_1} = \begin{pmatrix}[cc|c|ccc|ccc]
		0 & 0 & 1 & 1 &   &   & 1 &   &   \\
		0 & 0 & 1 &   & 1 &   &   & 1 &   \\
		0 & 0 & 1 &   &   & 1 &   &   & 1 \\\hline
		1 & 1 &   & 1 &   &   & 1 & 0 & 1 \\
		1 & 1 &   &   &   & 1 & 1 & 1 & 1 \\
	\end{pmatrix}, \quad
	M_{T_2} = \begin{pmatrix}[cc|c|ccc|ccc]
		0 & 1 & 1 & 1 &   &   & 1 &   &   \\
		0 & 1 & 1 &   & 1 &   &   & 1 &   \\
		0 & 1 & 1 &   &   & 1 &   &   & 1 \\\hline
		1 & 1 &   &   & 1 &   & 0 & 1 & 1 \\
	\end{pmatrix}
	M_{T_3} = \begin{pmatrix}[cc|c|ccc|ccc]
		1 & 0 & 1 & 1 &   &   & 1 &   &   \\
		1 & 0 & 1 &   & 1 &   &   & 1 &   \\
		1 & 0 & 1 &   &   & 1 &   &   & 1 \\
	\end{pmatrix}
\end{align*}

Note that $M_{T_1}$ includes two rows in the lower section which represent the set $\{1,3\}$ of nodes of $G$ and, analogously, $M_{T_2}$ includes rows which represent the set $\{2\}$. In this case, each implicit $\sigma_{r_0}$-sort $T_i$ has included one set of auxiliary rows from the upper section of matrix $M_G$. Also, the first sort, $T_1$, has included the first and the third row from the lower section of $M_G$, while sort $T_2$ has included the second row from the lower section of $M_G$. Sort $T_3$ has not incuded any additional rows. We will later see that this choice of sorts represents a possible partitioning of the nodes of graph $G$ into independent sets.

$\qed$
}
\end{example}

We will now see how an implicit $\sigma_{r_0}$-sort refinement $\mathcal{T}$ of $D_G$ with threshold 1 partitions the rows of $M_G$ in such a way that each sort represents an independent set of graph $G$. Given an implicit sort $T$, we will explain how the rows from the lower section which have been included represent nodes of graph $G$. These nodes form an independent set in $G$ if and only if the structuredness of sort $T$ under the structuredness function $\sigma_{r_0}$ is 1.

For every pair of nodes included in $T$, $r_0$ must check that they are not connected, using the adjacency submatrix $\bar{A}_G$. The simplest way to do this would be to select a row from the lower section to represent the first node, a column from the right column set to represent the second node, and to check that the appropriate cell in $\bar{A}_G$ is set to 1 (recall that $\bar{A}_G$ is the complement of the adjacency matrix $A_G$ of graph $G$). However, when building the implicit sort $T$ for each row, all columns are included (i.e.~full rows are included). The effect of this is that, while the rows included in $M_T$ only represent a subset of the nodes in $G$, all nodes of $G$ are represented in the columns of $M_T$. Rule $r_0$ must first ensure that only nodes included in $T$ will be compared.

We now briefly review the role of the different variables in $r_0$. The $(z,e)$ pair serves to ensure that only one copy of the auxiliary rows is present in each implicit sort. To illustrate this, consider that, in a given assignment $\rho$, the triple $(x, c_1, c_2)$ will occupy an auxiliary row. If this auxiliary row is duplicated in $T$, then it is possible to assign $(z,e)$ to the duplicate auxiliary row. This assignment will satisfy the antecedent of $r_0$, but will not satisfy the consequent, since $\val(z) = 0$ will not hold. This will cause the structuredness of $T$ to be less than 1 (see example).

\begin{example}\label{example4}
{\em Using the same working example, consider the following subset $T$ of $D_G$ where two copies of an auxiliary row have been included (more precisely, the two copies of the auxiliary rows which are included are equal in every column except \textbf{sp1} and \textbf{sp2}):
\begin{align*}
	M_{T} = \begin{pmatrix}[cc|c|ccc|ccc]
		0 & 1 & 1 & 1 &   &   & 1 &   &   \\
		0 & 0 & 1 & 1 &   &   & 1 &   &   \\
		0 & 0 & 1 &   & 1 &   &   & 1 &   \\
		0 & 0 & 1 &   &   & 1 &   &   & 1 \\\hline
		1 & 1 &   & 1 &   &   & 1 & 0 & 1 \\
		1 & 1 &   &   &   & 1 & 1 & 1 & 1 \\
	\end{pmatrix}, \quad
\end{align*}

A possible assignment $\rho$ of the variables is shown as superscripts (note that only the variables relevant to the example are shown):
\begin{align*}
	\begin{pmatrix}[ll|l|lll|lll]
		0 & 1 & 1^{z} & 1^{e} &   &   & 1 &   &   \\\hline
		0 & 0 & 1^{x} & 1^{c_1} &   &   & 1^{c_2} &   &   \\
		0 & 0 & 1 &   & 1 &   &   & 1 &   \\
		0 & 0 & 1 &   &   & 1 &   &   & 1 \\\hline
		1 & 1 & 0 & 1 &   &   & 1 & 0 & 1 \\
		1 & 1 &   &   &   & 1 & 1 & 1 & 1 \\
	\end{pmatrix}
\end{align*}

The assignment shown will cause the structuredness to be less than 1. This is because the assignment satisfies the antecedent but not the consequent, as $\val(z) = 0$ is false. This example illustrates how variables $(z,e)$ serve to ensure only one set of auxiliary rows can be included in a sort.

%\begin{center}
%\begin{tabular}{| p{2mm} | p{2mm} || p{8mm} || p{8mm} | p{8mm} | p{8mm} || p{8mm} | p{8mm} | p{8mm} |}\hline
% 0 & 1 & $1^{z}$   & $1^{e}$       &   &   & 1             &   &   \\\hline
% 0 & 0 & $1^{x}$   & $1^{c_1}$     &   &   & $1^{c_2}$     &   &   \\\hline
% 0 & 0 & 1         &               & 1 &   &               & 1 &   \\\hline
% 0 & 0 & 1         &               &   & 1 &               &   & 1 \\\hline\hline
% 1 & 1 & $0^{y,u}$ & $1^{d_1,f_1}$ & 0 &   & $1^{d_2,f_2}$ & 0 & 1 \\\hline
% 1 & 1 & 0         &               &   & 1 & 1             & 1 & 1 \\\hline
%\end{tabular}
%\end{center}

$\qed$
}
\end{example}

Next, the $(u, f_1, f_2)$ triple ensures that $c_1$ (and, by association, $c_2$) is assigned to a column of $M_T$ which represents a node of $G$ that is also represented by a row of $M_T$. To understand this better, recall that a sort $T$ of $D_G$ is represented by its matrix $M_T$, which is itself built by selecting a subset of the rows of $M_G$. When including a row from the lower section of $M_G$ in $M_T$, we are also selecting the appropriate node of $G$. In this way, the rows of the lower section of $M_T$ represent a certain subset of the nodes in $G$. This is not true for the columns in $M_T$, which are all included indiscriminatedly. We will use variables $(u, f_1, f_2)$ to only consider only the columns which represent the appropriate nodes.

%In an assignment which obeys the restriction for $(u,f_1,f_2)$, if $u$ points to the row which represents node $n$, then either $f_1$ or $f_2$ must point to the cell in the same row, in the column representing node $n$, for its value to be 1, while the last variable ($f_2$ or $f_1$, respectively) will point to the cell $(n,n)$ of the adjacency submatrix (which will always contain a 1, since $n$ is not connected to $n$ in graph $G$).

\begin{example}
{\em Consider matrix $M_{T_1}$ as was defined in example \ref{example3}:
\begin{align*}
	M_{T_1} = \begin{pmatrix}[ll|l|lll|lll]
		0 & 0 & 1 & 1 &   &   & 1 &   &   \\
		0 & 0 & 1 &   & 1 &   &   & 1 &   \\
		0 & 0 & 1 &   &   & 1 &   &   & 1 \\\hline
		1 & 1 & 0 & 1 &   &   & 1 & 0 & 1 \\
		1 & 1 &   &   &   & 1 & 1 & 1 & 1 \\
	\end{pmatrix}
\end{align*}

Because of the inclusion of the first and third rows from the lower section of matrix $M_G$, sort $T_1$ represents the subset $V_1 = \{1, 3\}$ of the nodes of $G$.

Consider the following assignment $\rho$ of the variables in $r_0$ (shown as superscripts):
\begin{align*}
	\begin{pmatrix}[ll|l|lll|lll]
		0 & 0 & 1 & 1 &   &   & 1 &   &   \\
		0 & 0 & 1^{x} &   & 1^{c_1} &   &   & 1^{c_2} &   \\
		0 & 0 & 1 &   &   & 1 &   &   & 1 \\\hline
		1 & 1 &   & 1 &   &   & 1 & 0 & 1 \\
		1 & 1 & 0^{u} &   & 0^{f_1} & 1 & 1 & 1^{f_2} & 1 \\
	\end{pmatrix}
\end{align*}

Here, $c_1$ is assigned to the column which represents node 2 of $G$. This is undesired, since node 2 is not present in $V_1$. This assignment will not be included in the set of total cases because $\val(f_1) = 0$. Furthermore, it is not possible to build an assignment which assigns $c_1$ to the column representing node 2 and which also satisfies the antecedent, since there is no 1 valued cell to be found in the lower section of that column.

In contrast, the following assignment does satisfy the antecedent:
\begin{align*}
	\begin{pmatrix}[ll|l|lll|lll]
		0 & 0 & 1 & 1 &   &   & 1 &   &   \\
		0 & 0 & 1 &   & 1 &   &   & 1 &   \\
		0 & 0 & 1^{x} &   &   & 1^{c_1} &   &   & 1^{c_2} \\\hline
		1 & 1 &   & 1 &   &   & 1 & 0 & 1 \\
		1 & 1 & 0^{u} &   &   & 1^{f_1} & 1 & 1 & 1^{f_2} \\
	\end{pmatrix}
\end{align*}

As a final comment, note that there are two possibilities for discarding a variable assignment $\rho$: (i) it may be that the variable assignment does not satisfy the antecedent of rule $r_0$, in which case it will never be counted (as is the case in this example), or (ii) it may be that a variable assignment, if valid in a subgraph, will cause the structuredness to be less than 1, as it does satisfy the antecedent but not the consequent (as is the case in example \ref{example4}).

$\qed$
}
\end{example}

Finally, the trio $(y,d_1,d_2)$ serve to assure that the node represented by the row of $y$ and the node represented by the column of $c_1$ (and $c_2$) are not connected in $G$. The $y$ variable is free to be assigned to any cell whose row is in the lower section of $M_T$ and whose column is \textbf{idp} (i.e.~any cell representing a node which is included in $T$). Also, $d_1$ and $d_2$ are already assigned cells whose columns represent a node in $N_T$ (possibly different to the node represented by the row of $y$). While one of $d_1$, $d_2$ point to the lower section, left column set, the other of the two points to the adjacency submatrix. In the consequent of $r_0$, it would be enough to ask that the $d_i$ variable which points to the adjacency matrix point to a 1 valued cell. However, it is not possible to distinguish $d_1$ from $d_2$, therefore, both are checked in a symmetrical fashion, with the subexpression $\val(d_1) = 1 \vee \val(d_2) = 1$. Momentarily, let us assume that $d_1$ points to the left column set and $d_2$ points to $\bar{A}_G$ and let $n$ be the node represented by $y$ and $m$ be the node represented by the column of $c_1$. In this case, if $\val(d_1) = 1$, then $n = m$ and $d_2$ will necessarily point to the cell $(n,n)$ of $\bar{A}_G$, which will always contain a 1 (recall that $G$ does not have self-edges). If $\val(d_1) = 0$, then $d_2$ will point to the cell $(n,m)$ of $\bar{A}_G$. This last cell must contain a 1 for the assignment to satisfy the consequent (meaning $n$ and $m$ are not connected).

\begin{example}
{\em Using the same matrix $M_{T_1}$ once again, consider the following assignment $\rho$ of the variables in $r_0$ (shown as superscripts):
\begin{align*}
	\begin{pmatrix}[ll|l|lll|lll]
		0 & 0 & 1 & 1 &   &   & 1 &   &   \\
		0 & 0 & 1 &   & 1 &   &   & 1 &   \\
		0 & 0 & 1^{x} &   &   & 1^{c_1} &   &   & 1^{c_2} \\\hline
		1 & 1 & 0^{y} & 1 &   & 0^{d_1} & 1 & 0 & 1^{d_2} \\
		1 & 1 & 0 &   &   & 1 & 1 & 1 & 1 \\
	\end{pmatrix}
\end{align*}

Here, the row of $y$ represents node 1 and the column of $c_1$ represents node 3. This assigment satisfies the antecedent of $r_0$. As such, both nodes are in $V_1$, so we must now check if they are connected in $G$. This is done with the subexpression $\val(d_1) = 1 \;\vee\; \val(d_2) = 1$ of the consequent of $r_0$. Here, the consequent is also satisfied by $\rho$, since $\val(d_2) = 1$, which tells us that nodes $1$ and $3$ are not connected in $G$.

$\qed$
}
\end{example}

We will now prove that graph $G$ is 3-colorable if and only if RDF graph $D_G$ has a $\sigma_{r_0}$-sort refinement $\mathcal{T}$ with threshold 1 consisting of at most 3 implicit sorts.

\subsubsection{$G$ is 3-colorable if and only if the sort refinement exists}

If the graph $G = (V,E)$ is 3-colorable, then let $f: V \rightarrow \{ 1, 2, 3 \}$ be the coloring function, with the property that for every pair of nodes $u, v \in V$, if $(u,v) \in E$ then $f(v) \neq f(u)$. This property can also be expressed with the adjacency matrix $A_G$: for every pair $i, j \in \{ 1, \ldots, n \}$, if $A_G[i,j] = 1$ then $f(i) \neq f(j)$. Let $n$ be the number of nodes in $G$.

The RDF graph $D_G$, constructed as was previously specified, has three sets of $n$ auxiliary rows (upper section) and one set of $n$ rows (lower section). The right column set of the lower section contains the complemented adjacency matrix $\bar{A}_G$.

We construct a sort refinement with three sorts, $T_1$, $T_2$ and $T_3$, as follows. The first set of auxiliary rows is assigned to $T_1$, the second set of auxiliary rows is assigned to $T_2$ and the third set of auxiliary rows is assigned to $T_3$. The last set of rows range from $3n+1$ to $3n + n$. For every row $3n + i$ with $i \in [1,n]$, we assign this row to the sort $T_{f(i)}$. That is, the row $3n + i$ is assigned to the sort given by the color of the corresponding node in $G$.

We now argue that the sort refinement $\mathcal{T} = \{ T_1, T_2, T_3 \}$ has threshold 1 using rule $r_0$. For this, consider a assignment $\rho$ of the variables $(x, c_1, c_2, y_1, d_1, d_2, z, e, u, f_1, f_2)$ in $r_0$ to the cells in the sort $T_1$. To be included in the set of total cases, the assignment must satisfy the antecedent of $r_0$. We will now consider the restrictions produced by this fact (without loss of generality, we assume that the rows of $M_{T_1}$ are ordered in the same fashion as presented previously).

Considering the subexpression $\prop(x) = \mbox{`\textbf{idp}'} \wedge \val(x) = 1$ of $r_0$, let $\rho(x) = (i_x, 3)$, where $i_x \in [1,n]$. That is, the variable $x$ is assigned to a cell in row $i_x$ and column 3 (recall that column $\textbf{idp}$ is the third column).

With the subexpression $c_1 \neq x \wedge \subj(c_1) = \subj(x) \wedge \val(c_1) = 1 \wedge c_2 \neq x \wedge \subj(c_2) = \subj(x) \wedge \val(c_2) = 1 \wedge c_1 \neq c_2$ we can restrict the assignment of $c_1$ and $c_2$. Without loss of generality, assume that:
\begin{equation*}
\rho(c_1) = (i_x, 3 + i_x) \quad\text{and}\quad \rho(c_2) = (i_x, 3 + n + i_x)
\end{equation*}
Both $c_1$ and $c_2$ are assigned to the same row as $x$ and furthermore, since the auxiliary rows contain diagonal matrices ($D_{n \times n}$), the row $i_x$ restricts the possible columns to only two, given by the cells with value 1.

Given that the subexpression $\prop(y) = \mbox{`\textbf{idp}'} \wedge \val(y) = 0 \wedge \subj(d_1) = \subj(y) \wedge \prop(d_1) = \prop(c_1) \wedge \subj(d_2) = \subj(y) \wedge \prop(d_2) = \prop(c_2)$ holds, we have:
% $\prop(y) = \mbox{`\textbf{idp}'} \wedge \val(y) = 0 \wedge \subj(y) = \subj(d_1) \wedge \prop(d_1) = \prop(c_1)$
%
\begin{equation*}
\rho(y) = (n + i_{y}, 3), \quad \rho(d_1) = (n + i_{y}, 3 + i_x), \quad\text{and}\quad \rho(d_2) = (n + i_{y}, 3 + n + i_x).
\end{equation*}

\noindent where $i_{y}$ ranges from 1 to the number of nodes with color 1.

Next, the subexpression $\prop(z) = \mbox{`\textbf{idp}'} \wedge \subj(z) = \subj(e) \wedge \prop(e) = \prop(c_1) \wedge e \neq c_1 \wedge \val(e) = 1$ sets:
\begin{equation*}
\rho(e) = (n + \alpha_{i_x}, 3 + i_x) \quad\text{and}\quad \rho(z) = (n + \alpha_{i_x}, 3).
\end{equation*}

Since the value of the cell assigned to $e$ must be 1, it must share columns with $c_1$, and $e \neq c_1$, we must assign $e$ to a cell of the diagonal of the lower section, left column set. However, note that this block is incomplete: only the rows corresponding to nodes in color 1 have been included. If there is no node which provides a value 1 on column  $3 + i_x$, then this assignment cannot be considered as a total case. We therefore assume that this node is present. In that case, we define $n + \alpha_{i_x}$ to be the row at which node $i_x$ has been placed. Furthermore, if this subset is empty (i.e.~no nodes of $G$ have been included) then there will be no total cases. In this case, by definition the structuredness is assigned value 1.

The last subexpression of the antecedent is $\prop(u) = \mbox{`\textbf{idp}'} \wedge \val(u) = 0 \wedge \subj(u) = \subj(f_1) \wedge \prop(f_1) = \prop(c_1) \wedge \subj(u) = \subj(f_2) \wedge \prop(f_2) = \prop(c_2) \wedge \val(f_1) = 1 \wedge \val(f_2) = 1$. Since the variables $u$, $f_1$, and $f_2$ do not appear in the consequent, this subexpression acts only as a restriction to the total cases. It allows us to write:
\begin{equation*}
\rho(u) = (n + i_u, 3), \quad \rho(f_1) = (n + i_u, 3 + i_x), \quad\text{and}\quad \rho(f_2) = (n + i_u, 3 + n + i_x)
\end{equation*}

\noindent This restriction has the effect of only considering, in the set of total cases, the assigments where column $3 + n + i_x$ (and column $3 + i_x$ also) refers to a node of $G$ which has been included in color 1. This is because the value of the cell assigned to $f_1$ must be 1. Variable $f_2$ will be restricted to the cell which represents the edge $(i_x, i_x)$ of the graph $G$, which, since there are no self-edges, will be 1 (when complemented).

We now turn our attention to the consequent of $r_0$. Given an assignment $\rho$ of the variables in $r_0$ as seen before, we shall show that it will also satisfy the consequent.

Consider the subexpression $\val(d_1) = 1 \vee \val(d_2) = 1$. Variable $d_2$ will be assigned to a cell of the complemented adacency matrix. Since column $3 + n + i_x$ of $d_2$ is fixed, $d_2$ will point to a cell which corresponds to the edge $(i_{y}, i'_x)$ of $G$, where we define $i'_x$ to be the node represented by row $i_x$ (recall that the index $i_x$ actually ranges from 1 to the number of nodes included in the subset). If it is the case that $i'_x = i_y$, then  $\val(d_1) = 1$ will hold. Now, we know node $i'_x$ is included in color 1. Since color 1 is an independent set, the complemented adjacency must contain a 1 in the corresponding cell. Therefore, the value of the cell assigned to $d_2$ will be 1.

The subexpression $\val(z) = 0$ will be true because we have included exactly one copy of the $n$ auxiliary rows in $T_1$. More precisely, since the column of $e$ is fixed and $e \neq c_1$, $e$ must point to the only other cell in that column which has value 1, for the value of $z$ to be 0. If any auxiliary row had been included twice, then $z$ could be assigned to a cell with value 1.

We have shown that every assignment which satisfies the antecedent (is a total case) also satisfies the consequent (is a favorable case). Therefore, the structuredness value for color 1 is 1. The same reasoning can be applied to $T_2$ and $T_3$. Therefore, the sort refinement constructed has threshold 1.

The proof of the other direction (if the sort refinement exists then $G$ is 3-colorable) is analogous and will not be shown explicitly here.

\end{document}